  \documentclass[prc,amsmath,amsfonts,showpacs,eqsecnum,floatfix,twocolumn]{revtex4}

 \usepackage{graphicx}  
 \usepackage{pstricks}  

\begin{document}
\hyphenation{Rijken}
\hyphenation{Nijmegen}
 
\title{
    Extended-soft-core Baryon-Baryon Model \\   
    II. Hyperon-Nucleon Interaction }
\author{Th.A.\ Rijken}
\affiliation{ Institute of Mathematics, Astrophysics, and Particle Physics \\
 Radboud University, Nijmegen, The Netherlands}               
%\affiliation{Institute for Theoretical Physics, University of Nijmegen,
%         Nijmegen, The Netherlands}
\author{Y.\ Yamamoto}
\affiliation{Physics Section, Tsuru University, Tsuru, Yamanashi 402-8555, Japan}


\date{version of: \today}
 
\begin{abstract}                                       
The {\it YN} results are presented from the Extended-soft-core (ESC) interactions.
They consist of local- and non-local-potentials due to (i) One-boson-exchanges 
(OBE), which are the members of nonets of pseudoscalar-, vector-, scalar-, and
axial-mesons, (ii) Diffractive exchanges, (iii) Two pseudoscalar exchange (PS-PS),
and (iv) Meson-Pair-exchange (MPE). Both the OBE- and Pair-vertices are regulated 
by gaussian form factors producing potentials with a soft behavior near the origin.
The assignment of the cut-off masses for the BBM-vertices is dependent on the 
SU(3)-classification of the exchanged mesons for OBE, and a similar scheme 
for MPE.        

%Besides the results for the fit to the scattering data, which
%defines the model largely, also the application to hypernuclear
%systems is described using the G-matrix method.

The particular version of the ESC-model, called ESC04 \cite{Rij04a}, describes 
nucleon-nucleon ({\it NN}) and hyperon-nucleon ({\it YN}) in a unified way using broken 
SU(3)-symmetry. Novel ingredients are the inclusion of (i) the axial-vector
meson potentials, (ii) a zero in the scalar- and axial-vector meson form factors. 
These innovations
made it possible for the first time to keep the parameters of the model closely 
to the predictions of the $^3P_0$ quark-antiquark creation (QPC) model. This is 
also the case for the $F/(F+D)$-ratio's. Furthermore, the introduction of the zero
helped to avoid the occurrence of unwanted bound states.
 
Broken SU(3)-symmetry serves  to connect the {\it NN} and the {\it YN}
channels, which leaves after fitting {\it NN} only a few free parameters
for the determination of the {\it YN}-interactions. In particular,
the meson-baryon coupling constants are calculated via SU(3)
using the coupling constants of the {\it NN}-analysis as input.
Here, as a novel feature, we allow for medium strong flavor-symmetry-breaking (FSB) 
of the coupling constants, using the $^{3}P_{0}$-model with a Gell-Mann-Okubo   
hypercharge breaking for the BBM-coupling.
We obtained very good fits for ESC-model with and without FSB. 
The charge-symmetry-breaking (CSB) in the $\Lambda p$ and $\Lambda n$ channels, which is 
an SU(2) isospin breaking, is included in the OBE-, TME-, and MPE-potentials.

We describe simultaneous fits to the {\it NN}- and the {\it YN}- scattering data, using 
different options for the ESC-model. 
For the selected 4233 {\it NN}-data with energies $0 \le T_{lab} \leq 350$ MeV, we 
typically reached a $\chi^2/N_{\rm data}=1.22$. 
For the usual set of 35 {\it YN}-data and 3 $\Sigma^+p$ cross-sections from a recent 
KEK-experiment E289 \cite{Kanda05} we obtained $\chi^{2}/{\it YN}_{\rm data} \approx 0.63$. 
In particular, we
were able to fit the precise experimental datum $r_{R}=0.468 \pm 0.010$ for
the inelastic capture ratio at rest rather well.      

%The model is used to calculate various properties of nuclear- and 
%hyperonic matter, using the G-matrix method.
The four versions (a,b,c,d) of ESC04 presented in this paper, give different 
results for hypernuclei. The reported G-matrix calculations are performed for
{\it YN} ($\Lambda N$, $\Sigma N$, $\Xi N$) pairs in nuclear matter. The obtained
well depths ($U_\Lambda$, $U_\Sigma$, $U_\Xi$) reveal distinct features of 
ESC04a-d. The $\Lambda\Lambda$-interactions are demonstrated to be consistent
with the observed data of $^{\ 6}_{\Lambda \Lambda}$He. The possible three-body
effects are investigated by considering phenomenologically the changes of the
vector-meson masses.

 \end{abstract}

\maketitle

\section{Introduction}
\label{sec:1}

%---------------------------------------------------------------------------------
%---------------------------------------------------------------------------------
%General things on the ESC-model:
This is the second in a series of papers where we present the recent results 
obtained with the Extended-Soft-Core model, henceforth referred to as ESC04, 
model for nucleon-nucleon ({\it NN}), hyperon-nucleon ({\it YN}), and hyperon-hyperon ({\it YY}). 
This paper treats the {\it N\!N}- and {\it Y\!N}(S=-1)-systems.
In \cite{Rij04a}, in the following referred to as I, 
many formal aspects have been described
or discussed rather extensively. Therefore, in this paper we will concentrate on 
items that are in particularly important in {\it YN}, and that were not treated in 
\cite{Rij04a}. In paper III \cite{Rij04c} the $S=-2$ channels will be described.

In \cite{MRS89} and \cite{RSY99}  
it has been shown the a soft-core (SC) one-boson-exchange (OBE) model, based on
regge-pole theory \cite{Rij85}, provides a satisfactory description of many aspects 
of the nucleon-nucleon ({\it N\!N}) and  hyperon-nucleon ({\it Y\!N}) channels.           

Since for {\it N\!N} the ESC-model has given a big step forward in the detailed 
description, one may expect that a simultaneous and unified treatment of the
{\it N\!N} and {\it Y\!N} channels, using broken SU(3), will give a very realistic
model for the baryon-baryon interactions. (In this paper by SU(3) is meant 
always SU(3)-flavor.)

In all previous work of the Nijmegen group, the exploration of 
(broken) SU(3)-symmetry connects the {\it N\!N} and the {\it Y\!N}
channels, leaving after fitting ${\it N\!N}$ only a few free parameters
to be determined in the {\it YN}-interactions. The latter is important 
in view of the scarce experimental {\it Y\!N}-data.  In particular,
the baryon-baryon-meson BBM coupling constants are calculated via SU(3)
using the coupling constants of the {\it N\!N}-analysis as input.
The first versions of the ESC-model, referred to as ESC00 \cite{Rij00,Rij01}, 
worked along the same procedures.
The aim of the ESC00 work was to demonstrate 
the ability of the ESC-model to realize a very good description of the 
{\it N\!N}- and {\it Y\!N}-data, i.e. a low $\chi^2$. Therefore, we left much 
freedom to the parameters.
However, in the ESC04 version presented here, we focus on the improvement
of the physics of the model by restricting the coupling constants in the 
BBM-vertices by the predictions of the quark-model (QM) in the form of 
the $^3P_0$ antiquark-quark pair creation model (QPC) \cite{Mic69}.
Also, all $\alpha=F/(F+D)$-ratios are taken close to the QPC-model predictions
for the BBM- and the
{\it B\!B}-Pair-vertices. An exception is made here for the pseudoscalar $\alpha_{PV}$ and
the $\alpha_V^m$ ratios. The first because it is interesting to see how close
or different it becomes as compared to the same ratio in weak interactions, where
$\alpha_{PV}=0.355$. The second, because it proved to be an important regulator for 
the S-wave spin-dependence of the $\Lambda N$-interaction \cite{RSY99}.

In \cite{MRS89} the 
magnetic ratio $\alpha_{V}^{m}$ for the vector-mesons was fixed to its
SU(6)-value, but the spin-spin interaction needed a correction. 
This because the S-wave spin-spin interaction in the $\Lambda N$-channels 
became later well known from hypernuclear systems \cite{Yam90,Mot95,Has95}.
In \cite{RSY99}, to improve the spin-spin interaction,
we left $\alpha_{V}^{m}$ free and made fits for different values
of this parameter. It turned out that in this way we indeed can construct 
soft-core {\it Y\!N}-models
which encompass a range of scattering length's in the $^{1}S_{0}$ and the
$^{3}S_{1}$ $\Lambda N$ channels. 
From the NSC97a-f models the sensitivity with respect to $\alpha^m_V$ is evident.

%---------------------------------------------------------------------------------
%Medium-strong SU(3)-breaking:
SU(3)-symmetry and the QPC-model give strong constraints on the coupling 
parameters. In order to keep some more flexibility in distinguishing the {\it N\!N}- and
the {\it Y\!N}(S=-1)-channels, similarly to the NSC97 models \cite{RSY99}, 
we allow for medium strong breaking of the
coupling constants, employing again the $^{3}P_{0}$-model with a Gell-Mann-Okubo   
hypercharge breaking for the BBM-coupling. This leads to a universal scheme
for SU(3)-breaking of the coupling constants for all meson nonets, in terms 
of a single extra parameter. 

To summarize the different sources of SU(3)-breaking, we include
%\begin{itemize}
\noindent (i) using the physical masses of the mesons and baryons
in the potentials and Schr\"{o}dinger equation,               
\noindent (ii) allowing for meson-mixing within a nonet
\mbox{$(\eta-\eta',\ \omega-\phi,\ \varepsilon-f'_{0}$},
\noindent (iii) including Charge Symmetry Breaking (CSB) \cite{Dal64} 
due to $\Lambda\Sigma$-mixing, which for example introduces a 
one-pion-exchange (OPE) potential in the $\Lambda N$ channel,
\noindent (iv) taking into account the Coulomb-interaction.
%\end{itemize}

The electromagnetic SU(2)-breaking \cite{Dal64}, called Charge-Symmetry-Breaking (CSB),
in the $\Lambda p$ and $\Lambda n$ channels is included, not only for the BBM- but 
also for the {\it BB}-Pair-couplings.

%---------------------------------------------------------------------------------
%gaussian form factors:
The BBM-vertices are described by coupling constants and
form factors, which correspond to the regge residues at high
energies \cite{Rij85}. The form factors are taken
to be of the gaussian-type, like the residue functions in many
regge-pole models for high energy scattering.
Note that also in (nonrelativistic) quark models (QM's)
a gaussian behavior of the form factors is most natural.
These form factors evidently
guarantee a soft behavior of the potentials in configuration
space at small distances.
 
In \cite{RSY99} the assignment of the cut-off parameters in the form factors 
was made for the individual
baryon-baryon-meson (BBM) vertices, constrained by broken SU(3)-symmetry.
This in distinction to the first attempt to construct soft-core interaction \cite{MRS89},
where cut-offs were assigned per baryon-baryon SU(3)-irrep. The latter scheme
we consider now not natural and we use here the same scheme as in \cite{RSY99}.
Moreover, this way we obtain immediately full predictive power for the $S=-2$ etc. 
baryon-baryon channels, e.g. $\Lambda\Lambda, \Xi N$-channels which involve the singlet 
$\{1\}$-irrep that does not occur in the {\it N\!N} and {\it Y\!N} channels. 

The dynamics of the ESC04 model has been described and discussed in paper I \cite{Rij04a}, 
and it is sufficient to refer to this here for all types of exchanges that are included.
Nevertheless, some more remarks on the scalar mesons are appropriate. An extensive 
discussion of the situation with respect to the scalar mesons is given in \cite{RSY99}.
The question whether the $J^{PC}=0^{++}$-mesons are of the Dalitz-type ($Q\bar{Q}$) or 
of the Jaffe-type ($Q^2\bar{Q}^2$) is not yet decided. In the coupling to the baryons,
we assume here that the basic process is described by the QPC-model \cite{Mic69}. 
It has been shown in paper I that this seems rather successful, justifying this 
assumption.

%---------------------------------------------------------------------------------

With a combined treatment of the $N\!N$ and $Y\!N$ channels we aim at
a high quality description of the baryon-baryon interactions.
By high quality we understand here a ${\it YN}$-fit with
low $\chi^{2}$ and such that,
while keeping the constraints forced on the potentials by the ${\it NN}$-fit,
the free parameters with a clear physical significance, like e.g. the
$F/(F+D)$-ratio's $\alpha_{PV}$ and $\alpha_{V}^{m}$, assume realistic values. 

Such a combined study of all baryon-baryon interactions, 
and especially {\it N\!N} and {\it Y\!N}, is desirable if one wants:  
\begin{itemize}
\item  To study the assumption of broken SU(3)-symmetry. For example we
       want to investigate the properties of the scalar mesons
       ($\varepsilon(760)$, $f_{0}(975)$, $a_{0}(980)$, $\kappa(1000)$).
       We note that especially the status of the scalar nonet is at
       present not established yet.
\item  To determination of $F/(F+D)$-ratio's.
\item  To extract, in spite of the scarce experimental ${\it YN}$-data,
 information about scattering lengths, effective
 ranges, the existence of resonances etc.
\item  To provide realistic baryon-baryon potentials, which can be 
       applied in few-body computations, nuclear- and hyperonic matter
       studies. 
\item  To extend the theoretical description to the $\Lambda\Lambda$
       and $\Xi N$ channels, where experiments may be realized in the
       foreseeable future.
\end{itemize}
 
%In this paper we treat in detail
%the following ${\it YN}$-reactions:
%\begin{itemize}
%\item[(i)] The coupled channels $\Lambda p \Rightarrow \Lambda p, 
% \Sigma^{+} n, \Sigma^{0} p$;
%\item[(ii)] The coupled
%channels $\Sigma^{-}p \Rightarrow \Sigma^{-}p, \Sigma^{0}n, \Lambda n$;
%\item[(iii)] The single channel $\Sigma^{+}p \Rightarrow \Sigma^{+}p$.
%\end{itemize}

% Different versions of the ESC04-model:
In the construction of the ESC-models there are two important options: 
\begin{enumerate}
\item[(i)] First, there is the choice of the pv- or the ps-coupling for the pseudoscalar
mesons, or some mixture, regulated by the $a_{PV}$-parameter, of these. 
This choice affects some $1/M^2$-terms in the 
ps-ps-exchange potentials. 
\item[(ii)] Second, yes or no medium strong symmetry-breaking of the couplings, 
regulated by a $\Delta_{FSB}$-parameter.
\end{enumerate}
We have accordingly produced four different solutions, fitting simultaneously the 
${\it NN} \oplus {\it YN}$-data, which are referred to as follows: 
ESC04a$(\Delta_{FSB} \neq 0, a_{PV}=0.5)$, 
ESC04b$(\Delta_{FSB} \neq 0, a_{PV}=1.0)$, 
ESC04c$(\Delta_{FSB} = 0, a_{PV}=0.5)$, 
ESC04d$(\Delta_{FSB} = 0, a_{PV}=1.0)$.  
Here, $a_{PV}=1.0$ and $a_{PV}=0.0$ means pure pseudo-vector respectively purely 
pseudoscalar coupling. 
It appears that there are notable differences
between these models, in particularly their properties for matter, e.g. 
 well-dephs $U_\Lambda, U_{\Sigma}$, and $U_\Xi$, are rather distinct.\\
We will display and discuss in this paper only the results for the ESC04a-model
in detail. With the exception of the G-matrix results, 
we will be very brief on ESC04b-d and will compare these models only very globally.
So, in the following, by ESC04 is meant ESC04a, unless specified otherwise.\\

As in all Nijmegen models, the Coulomb interaction is included exactly, for which we 
solve the multichannel
Schr\"{o}dinger equation on the physical particle basis.
The nuclear potentials are calculated on the isospin basis,
in order to limit the number of different form factors.
This means that we include only the so-called 'medium strong' 
SU(3)-breaking in the potentials.
 
%----------------------------------------------------------------------
%contents:
The contents of this paper are as follows. In section II we describe 
the $S=-1$ $Y\!N$-channels on the isospin and particle basis, and
the use of the multi-channel Schr\"{o}dinger equation is mentioned.
The potentials in momentum and configuration space are defined by
referring to the description given in paper I.
The BBM-couplings are discussed both in the $3\times 3$-matrix and
the cartesian-octet representation. The SU(3)-couplings of the 
OBE- and TME-graphs are given in a form suitable for 
a digital evaluation.\\
In section III the meson-pair interaction Hamiltonians are given in
the context of SU(3). Expressions for the meson-pair-exchange (MPE)
graphs are given, again in an immediately programmable form.
In section IV the medium-strong breaking of SU(3)-symmetry of the 
coupling constants is described. The QPC-model is employed for the 
development of a universal scheme for this breaking.
Here also the detailed prescription for the handling of the cut-off
parameters is given, in particularly for the cases of meson-mixing.\\
In section V the simultaneous ${\it NN} \oplus {\it YN}$ fitting procedure is reviewed.
In section VI the results for the coupling constants and $F/(F+D)$-ratios
for OBE and MPE are given. 
They are discussed and compared with the predictions of the 
QPC-model. Here, also the values of the $BBM$-couplings are displayed
for pseudoscalar, vector, scalar, and axial-vector mesons.\\
In section VII the {\it NN}-results from the combined ${\it NN} \oplus {\it YN}$-fit,  
model ESC04a, henceforth called ESC04, are discussed 
and compared with the results of paper I, referred to ESC04({\it NN}).
In section VIII we discuss the fit to the {\it Y\!N} scattering data from the 
combined ${\it NN} \oplus {\it YN}$-fit.
In section IX we compare very briefly the models ESC04a-d.\\            
In section X, the hypernuclear properties of ESC04a-d are studied through
the G-matrix calculations for {\it YN} ($\Lambda N$, $\Sigma N$, $\Xi N$)
and their partial-wave contributions. Here, the implications of possible
three-body effects for the nuclear saturation and baryon well-dephs 
are discussed. Also, the $\Lambda \Lambda$ interactions in ESC04a-d are 
demonstrated to be consistent with the observed data of $^{\ 6}_{\Lambda\Lambda}$He.
In section XI we finish by a final discussion and draw some conclusions.

\section{Channels, Potentials, and SU(3)-symmetry}
\label{sec:2}

\subsection{Channels and Potentials}     
\label{sec:2a}
In this paper we consider the hyperon-nucleon reactions with $S=-1$
\begin{equation}
 Y(p_a,s_a)+N(p_b,s_b) \rightarrow Y(p'_a,s'_a)+N(p'_b,s'_b)             
\label{eq:2.1}\end{equation}
Like in Ref.'s~\cite{MRS89,RSY99} we will also refer to $Y$ and $Y'$
as particles 1 and 3, and to $N$ and $N'$ as particles 2 and 4. For the 
kinematics and the definition of the amplitudes, we refer to paper I 
\cite{Rij04a} of this series. Similar material can be found in \cite{MRS89}.
Also, in paper I the derivation of the Lippmann-Schwinger equation 
in the context of the relativistic two-body equation is described.

On the physical particle basis, there are four charge channels:
\begin{eqnarray}
   q=+2:\ \  && \Sigma^+p\rightarrow\Sigma^+p,         \nonumber\\
   q=+1:\ \  && (\Lambda p, \Sigma^+n, \Sigma^0p)\rightarrow
             (\Lambda p, \Sigma^+n, \Sigma^0p),        \nonumber\\
   q=\ \ 0:\ \ && (\Lambda n, \Sigma^0n, \Sigma^-p)\rightarrow
             (\Lambda n, \Sigma^0n, \Sigma^-p),        \nonumber\\
   q=-1:\ \  && \Sigma^-n\rightarrow\Sigma^-n.
\label{eq:2.2}\end{eqnarray}

Like in \cite{MRS89,RSY99}, the potentials are calculated on the isospin basis.
For $S=-1$ hyperon-nucleon systems there are only two isospin channels:
(i) $ I={\textstyle\frac{1}{2}}:\ \ (\Lambda N,\Sigma N\rightarrow
                                  \Lambda N,\Sigma N)$, and              
(ii) $ I={\textstyle\frac{3}{2}}:\ \ \Sigma N\rightarrow\Sigma N$. 

Obviously, the potential on the particle basis for the $q=2$ and
$q=-1$ channels are given by the $I={\textstyle\frac{3}{2}}$
$\Sigma N$ potential on the isospin basis. For $q=1$ and $q=0$, the
potentials are related to the potentials on the isospin basis by an
isospin rotation. Using a notation where we only list the hyperons
[$V_{\Lambda\Sigma^+}=(\Lambda p|V|\Sigma^+n)$, etc.], we find for $q=1$

%------------------------------------------------------------------------------
 \begin{widetext}
%\onecolumngrid                                          
 \begin{equation}
  \left(\begin{array}{ccc}
    V_{\Lambda\Lambda}  & V_{\Lambda\Sigma^+}  & V_{\Lambda\Sigma^0} \\[2mm]
    V_{\Sigma^+\Lambda} & V_{\Sigma^+\Sigma^+} & V_{\Sigma^+\Sigma^0}\\[2mm]
    V_{\Sigma^0\Lambda} & V_{\Sigma^0\Sigma^+} & V_{\Sigma^0\Sigma^0}
        \end{array}\right) =
  \left(\begin{array}{ccc}
    V_{\Lambda\Lambda}
    &  \sqrt{\textstyle\frac{2}{3}}V_{\Lambda\Sigma}
    & -\sqrt{\textstyle\frac{1}{3}}V_{\Lambda\Sigma} \\[2mm]
    \sqrt{\textstyle\frac{2}{3}}V_{\Sigma\Lambda}
    &  {\textstyle\frac{2}{3}}V_{\Sigma\Sigma}({\textstyle\frac{1}{2}})
      +{\textstyle\frac{1}{3}}V_{\Sigma\Sigma}({\textstyle\frac{3}{2}})
    &  {\textstyle\frac{1}{3}}\sqrt{2}\left[
           V_{\Sigma\Sigma}({\textstyle\frac{3}{2}})
          -V_{\Sigma\Sigma}({\textstyle\frac{1}{2}})\right] \\[2mm]
    -\sqrt{\textstyle\frac{1}{3}}V_{\Sigma\Lambda}
    &  {\textstyle\frac{1}{3}}\sqrt{2}\left[
           V_{\Sigma\Sigma}({\textstyle\frac{3}{2}})
          -V_{\Sigma\Sigma}({\textstyle\frac{1}{2}})\right]
    &  {\textstyle\frac{1}{3}}V_{\Sigma\Sigma}({\textstyle\frac{1}{2}})
      +{\textstyle\frac{2}{3}}V_{\Sigma\Sigma}({\textstyle\frac{3}{2}})
        \end{array}\right)\ ,
\label{eq:2.3} \end{equation}
while for $q=0$ we find
 \begin{equation}
  \left(\begin{array}{ccc}
    V_{\Lambda\Lambda}  & V_{\Lambda\Sigma^0}  & V_{\Lambda\Sigma^-} \\[2mm]
    V_{\Sigma^0\Lambda} & V_{\Sigma^0\Sigma^0} & V_{\Sigma^0\Sigma^-}\\[2mm]
    V_{\Sigma^-\Lambda} & V_{\Sigma^-\Sigma^0} & V_{\Sigma^-\Sigma^-}
        \end{array}\right) =
  \left(\begin{array}{ccc}
    V_{\Lambda\Lambda}
    &  \sqrt{\textstyle\frac{1}{3}}V_{\Lambda\Sigma}
    & -\sqrt{\textstyle\frac{2}{3}}V_{\Lambda\Sigma} \\[2mm]
    \sqrt{\textstyle\frac{1}{3}}V_{\Sigma\Lambda}
    &  {\textstyle\frac{1}{3}}V_{\Sigma\Sigma}({\textstyle\frac{1}{2}})
      +{\textstyle\frac{2}{3}}V_{\Sigma\Sigma}({\textstyle\frac{3}{2}})
    &  {\textstyle\frac{1}{3}}\sqrt{2}\left[
           V_{\Sigma\Sigma}({\textstyle\frac{3}{2}})
          -V_{\Sigma\Sigma}({\textstyle\frac{1}{2}})\right] \\[2mm]
    -\sqrt{\textstyle\frac{2}{3}}V_{\Sigma\Lambda}
    &  {\textstyle\frac{1}{3}}\sqrt{2}\left[
           V_{\Sigma\Sigma}({\textstyle\frac{3}{2}})
          -V_{\Sigma\Sigma}({\textstyle\frac{1}{2}})\right]
    &  {\textstyle\frac{2}{3}}V_{\Sigma\Sigma}({\textstyle\frac{1}{2}})
      +{\textstyle\frac{1}{3}}V_{\Sigma\Sigma}({\textstyle\frac{3}{2}})
        \end{array}\right)\ .
\label{eq:2.4} \end{equation}
%\twocolumngrid                                          
 \end{widetext}
%------------------------------------------------------------------------------

%The isospin factors for the various OBE-exchanges in the two
%isospin channels are given in Table~\ref{tabisofac}. We use the
%pseudoscalar mesons as a specific example, and $P$ is the exchange
%operator alluded to in the previous section. We also include the
%coupling of the $\Lambda$-hyperon to the neutral pion, which is
%non-zero due to $\Lambda$-$\Sigma^0$ mixing, as was discussed earlier.
%However, this matrix element is {\it only\/} included for the
%potentials on the particle basis.

For the kinematics of the reactions and the various thresholds, see \cite{RSY99}.
In this work we do not solve the Lippmann-Schwinger equation, but the 
multi-channel Schr\"{o}dinger equation in configuration space, completely 
analogous to \cite{MRS89}.
The multichannel Schr\"odinger equation for the configuration-space
potential is derived from the Lippmann-Schwinger equation through
the standard Fourier transform, and the equation for the radial
wave function is found to be of the form~\cite{MRS89}
\begin{equation}
   u^{\prime\prime}_{l,j}+(p_i^2\delta_{i,j}-A_{i,j})u_{l,j}
       -B_{i,j}u'_{l,j}=0,           
\label{eq:2.5}\end{equation}
where $A_{i,j}$ contains the potential, nonlocal contributions, and
the centrifugal barrier, while $B_{i,j}$ is only present when non-local
contributions are included. 
The solution in the presence of open and closed channels is given,
for example, in Ref.~\cite{Nag73}.
The inclusion of the Coulomb interaction in the
configuration-space equation is well known and included in the evaluation
of the scattering matrix.

The momentum space and configuration space potentials for the ESC04-model
have been described in paper I \cite{Rij04a} for baryon-baryon in general.
Therefore, they apply also to hyperon-nucleon and we can refer for that
part of the potential to paper I.  
Also in the ESC-model, the potentials are of such a form that they are exactly 
equivalent in both momentum space and configuration space. 
The treatment of the mass differences among the baryons are handled exactly 
similar as is done in \cite{MRS89,RSY99}. Also, exchange potentials related to
strange meson exchanhe $K, K^*$ etc. , can be found in these references. 

The baryon mass differences in the intermediate states for TME- and MPE-
potentials has been neglected for {\it YN}-scattering. This, although possible
in principle, becomes rather laborious and is not expected to change the 
characteristics of the baryon-baryon potentials much.

\subsection{BBM-couplings in SU(3), Matrix-representation}
\label{sec:2b}
In previous work of the Nijmegen group, e.g. \cite{MRS89} and \cite{RSY99},
the treatment of SU(3) has been given in detail for the BBM interaction
Lagrangians and the coupling coefficients of the OBE-graphs. However,
for the ESC-models we also need the coupling coefficients for the TME- and the
MPE-graphs. Since there are many more TME- and MPE-graphs than OBE-graphs, an
computerized computation is desirable. For that purpose we found the so-called
'cartesian-octet'-representation quite useful. Therefore, we give an exposition of this 
representation, its connection with the matrix representation used in our previous
work, and the formulation of the coupling coefficients used in the 
automatic computation. 

In the matrix representation, the eight $J^P={\textstyle\frac{1}{2}}^+$ baryons 
are described by a traceless matrix, see e.g. \cite{Swa63},                 
\begin{equation}
  B = \left( \begin{array}{ccc}
      {\displaystyle\frac{\Sigma^{0}}{\sqrt{2}}+\frac{\Lambda}{\sqrt{6}}}
               &  \Sigma^{+}  &  p  \\[2mm]
      \Sigma^{-} & {\displaystyle-\frac{\Sigma^{0}}{\sqrt{2}}
                   +\frac{\Lambda}{\sqrt{6}}}  &  n \\[2mm]
      -\Xi^{-} & \Xi^{0} &  {\displaystyle-\frac{2\Lambda}{\sqrt{6}}}
             \end{array} \right).
\label{eq:2.6}\end{equation}
Similarly, the
various meson nonets (we take the pseudoscalar mesons with $J^P=0^+$
as an example) are represented by 
\begin{equation}
     P=P_{\{1\}}+P_{\{8\}},
\label{eq:2.7}\end{equation}
where the singlet matrix $P_{\{1\}}$ has elements $\eta_0/\sqrt{3}$
on the diagonal, and the octet matrix $P_{\{8\}}$ is given by
\begin{equation}
   P_{\{8\} } = \left( \begin{array}{ccc}
      {\displaystyle\frac{\pi^{0}}{\sqrt{2}}+\frac{\eta_{8}}{\sqrt{6}}}
             & \pi^{+}  &  K^{+}  \\[2mm]
      \pi^{-} & {\displaystyle-\frac{\pi^{0}}{\sqrt{2}}
         +\frac{\eta_{8}}{\sqrt{6}}}  &   K^{0} \\[2mm]
      K^{-}  &  \overline{{K}^{0}}
             &  {\displaystyle-\frac{2\eta_{8}}{\sqrt{6}}}
             \end{array} \right).
\label{eq:2.8}\end{equation}
Exploiting the SU(3)-invariant combinations, see e.g. \cite{RSY99,Swa63},
 $ \left[\overline{B}BP\right]_{F} $, $ \left[\overline{B}BP\right]_{D} $ ,      
 and $ \left[\overline{B}BP\right]_{S} $ , the SU(3)-invariant BBP-interaction
Lagrangian can be written as \cite{Swa63}
\begin{eqnarray}
   {\cal L}_{I} &=& -g_{8}\sqrt{2}\left\{
     \alpha\left[\overline{B}BP\right]_{F}+
     (1-\alpha)\left[\overline{B}BP\right]_{D}\right\}\, 
   \nonumber\\ && - \,
     g_{1}{\textstyle\sqrt{\frac{1}{3}}}
     \left[\overline{B}BP\right]_{S},            
\label{eq:2.9}\end{eqnarray}
where $g_8$ and $g_1$ are the singlet and octet couplings, 
$\alpha$ is known as the $F/(F+D)$ ratio, and the square-root
factors are introduced for later convenience.

The convention used for the isospin doublets is
\begin{eqnarray}
&&  N=\left(\begin{array}{c} p \\ n \end{array} \right), \ \ \
  \Xi=\left(\begin{array}{c} \Xi^{0} \\ \Xi^{-} \end{array} \right), 
 \nonumber\\ 
&&  K=\left(\begin{array}{c} K^{+} \\ K^{0} \end{array} \right),
  \ \ \   K_{c}=\left(\begin{array}{c} \overline{K^{0}} \\
               -K^{-} \end{array} \right),        
\label{eq:2.10}\end{eqnarray}
and for the isovectors in the SU(2)$_I$-tensor notation $(a,b=1,2)$
\begin{eqnarray}
 \pi^a_b &=& \left(\begin{array}{cc} 
 \frac{\pi^0}{\sqrt{2}} & \pi^+ \\ \pi^- & -\frac{\pi^0}{\sqrt{2}} 
 \end{array}\right) , 
 \Sigma^a_b  =  \left(\begin{array}{cc} 
 \frac{\Sigma^0}{\sqrt{2}} & \Sigma^+ \\ \Sigma^- & -\frac{\Sigma^0}{\sqrt{2}} 
 \end{array}\right)\  
\label{eq:2.11}\end{eqnarray}
 where we have chosen  the phases of the isovector fields 
 such~\cite{Swa63} that
\begin{equation}
  \mbox{\boldmath $\Sigma$}\!\cdot\!\mbox{\boldmath $\pi$} 
  = \sum_{a,b=1}^3 \Sigma^a_b\ \pi^b_a 
  = \Sigma^{+}\pi^{-} +\Sigma^{0}\pi^{0}+\Sigma^{-}\pi^{+}.
\label{eq:2.12}\end{equation}
The expression of the interaction Lagrangian (\ref{eq:2.9}) in terms of the 
isospin singlets (I=0), doublets (I=1/2), and triplets (I=1), 
is given e.g. in \cite{RSY99}. Also, the $BBM$-couplings of the octet-members are
given in terms of $g_8$ and $\alpha=F/(F+D)$. See \cite{RSY99}, equations 
 (2.10)-(2.14).

\subsection{Cartesian-octet Representation}                             
\label{sec:2c}
\begin{table}
\caption{Octet Representation Baryons and Mesons.}               
\begin{tabular}{ccccccc} & & & & & & \\ \hline\hline & & & & & & \\
  $\Sigma^{+}$ &=& $\frac{1}{\sqrt{2}}(\psi_{1}-i \psi_{2})$&\hspace*{5mm}& 
  $\pi^{+}$ &=& $\frac{1}{\sqrt{2}}(\phi_{1}-i \phi_{2})$ \\ && \\
  $\Sigma^{-}$ &=& $\frac{1}{\sqrt{2}}(\psi_{1}+i \psi_{2})$&\hspace*{5mm}& 
  $\pi^{-}$ &=& $\frac{1}{\sqrt{2}}(\phi_{1}+i \phi_{2})$ \\ && \\
  $\Sigma^{0}$ &=& $\psi_{3}$&\hspace*{5mm}& $\pi^{0}$&=& $\phi_{3}$\\
  && \\
  $ p        $ &=& $\frac{1}{\sqrt{2}}(\psi_{4}-i \psi_{5})$&\hspace*{5mm}& 
  $  K^{+}$ &=& $\frac{1}{\sqrt{2}}(\phi_{4}-i \phi_{5})$ \\ && \\
  $ n        $ &=& $\frac{1}{\sqrt{2}}(\psi_{6}-i \psi_{7})$&\hspace*{5mm}& 
  $  K^{0}$ &=& $\frac{1}{\sqrt{2}}(\phi_{6}-i \phi_{7})$ \\ && \\
  $ \Xi^{-}$ &=& $\frac{1}{\sqrt{2}}(\psi_{4}+i \psi_{5})$&\hspace*{5mm}& 
  $  K^{-}$ &=& $\frac{1}{\sqrt{2}}(\phi_{4}+i \phi_{5})$ \\ && \\
  $ \Xi^{0}  $ &=& $\frac{1}{\sqrt{2}}(\psi_{6}+i \psi_{7})$&\hspace*{5mm}& 
  $  \bar{K}^{0}$ &=& $\frac{1}{\sqrt{2}}(\phi_{6}+i \phi_{7})$ \\ && \\
  $\Lambda   $ &=& $\psi_{8}$&\hspace*{5mm}& $\eta   $&=& $\phi_{8}$\\
     & & & & & & \\
\hline\hline
\end{tabular}
\label{taboctet1} 
\end{table}
The annihilation operators corresponding to the baryon and pseudoscalar
SU(3) octet-representation $\{8\}$ are given in Table~\ref{taboctet1}.
Here we used the cartesian octet fields. For baryons these are denoted
by $\psi_i(i=1,2, \ldots,8)$, and for the pseudoscalar mesons by
$\phi_i (i=1,2,\ldots,8)$ \cite{Car66,Mar69,Swa63}.
The particle states are created by these operators are given in 
Table~\ref{taboctet2} \cite{Swa63}.
\begin{table}
\caption{Octet Particle States.}               
\begin{tabular}{ccrcccr} & & & & & & \\ \hline\hline & & & & & & \\
$|\pi^+\rangle$ &=& $-\pi^{+ \dagger}|0\rangle $ & \hspace*{5mm}&
$|\Sigma^+\rangle$ &=&$ -\Sigma^{+ \dagger} |0\rangle$ \\ && \\
$|\pi^-\rangle$ &=& \hspace{2mm}$ \pi^{- \dagger}|0\rangle$ &\hspace*{5mm}&
$|\Sigma^+\rangle$ &=& \hspace{2mm}$ \Sigma^{- \dagger} |0\rangle$ \\ && \\
$|\pi^0\rangle$ &=& \hspace{2mm} $\pi^{0 \dagger}|0\rangle $ &\hspace*{5mm}&
$|\Sigma^0\rangle$ &=& \hspace{2mm} $\Sigma^{0 \dagger} |0\rangle$ \\ && \\
$| K^+\rangle$ &=& \hspace{2mm} $K^{+ \dagger}|0\rangle$ &\hspace*{5mm}&
$| p \rangle$ &=& \hspace{2mm} $p^{ \dagger} |0\rangle$ \\ && \\
$| K^0\rangle$ &=& \hspace{2mm} $K^{0 \dagger}|0\rangle$ &\hspace*{5mm}&
$| n \rangle$ &=& \hspace{2mm} $n^{ \dagger} |0\rangle$ \\ && \\
$| K^-\rangle$ &=& \hspace{2mm} $K^{- \dagger}|0\rangle$ &\hspace*{5mm}&
$|\Xi^-\rangle$ &=& \hspace{2mm} $\Xi^{- \dagger} |0\rangle$ \\ && \\
$| \bar{K}^0\rangle$ &=& \hspace{2mm} $\bar{K}^{0 \dagger}|0\rangle$ &\hspace*{5mm}&
$|\Xi^0\rangle$ &=& \hspace{2mm} $\Xi^{0 \dagger} |0\rangle$ \\ && \\
$| \eta_8\rangle$ &=& \hspace{2mm} $\eta_8^{\dagger}|0\rangle$ &\hspace*{5mm}&
$|\Lambda\rangle$ &=& \hspace{2mm} $\Lambda^{\dagger} |0\rangle$ \\ 
     & & & & & & \\
\hline\hline
\end{tabular}
\label{taboctet2} 
\end{table}
Similar expressions hold for the vector, axial-vector, and scalar mesons.
The connection between the matrix-representation (\ref{eq:2.6}) and the 
cartesian-octet representation is 
\begin{eqnarray}
&& \hspace{-8mm} B^a_b = \frac{1}{\sqrt{2}} \sum_{i=1}^8 (\lambda_i)_{ab} \psi_i\ ,
 \psi_i =  \frac{1}{\sqrt{2}} \sum_{a,b=1}^3 (\lambda_i)_{ab} B^a_b\  
\label{eq:2.13}\end{eqnarray}
where $\lambda_i, i=1,8$ are the Gell-Mann matrices \cite{Swa63}, and 
where the indices $(a,b=1,2,3)$. The same expression holds for $P^a_b$ of
(\ref{eq:2.8}) in terms of the $\phi_i$'s. The SU(3)-invariants in the 
cartesian-octet representation read
\begin{subequations}
\begin{eqnarray}
 \left[\overline{B}BP\right]_{F} &=&  \sum_{i,j,k=1}^8 f_{ijk}\left[\overline{\psi}_i 
 \psi_j\right]\ \phi_k\ , \\
 \left[\overline{B}BP\right]_{D} &=&  \sum_{i,j,k=1}^8 d_{ijk}\left[\overline{\psi}_i 
 \psi_j\right]\ \phi_k\ , \\
 \left[\overline{B}BP\right]_{S} &=&  \sum_{i,j=1}^8 \delta_{ji}\
 \left[\overline{\psi}_i \psi_j\right]\ \phi_9\ ,            
\label{eq:2.14}\end{eqnarray}
\end{subequations}
where $f_{ijk}$ are the totally anti-symmetric SU(3)-structure constants,  
$d_{ijk}$ are the totally symmetric constants, and $\phi_9$ denotes the unitary singlet.
They are given by the following
commutators and anti-commutators
\begin{eqnarray}
 \left[\lambda_i,\lambda_j\right] = 2i f_{ijk}\ \lambda_k, 
 \left\{\lambda_i,\lambda_j\right\} = \frac{4}{3}\delta_{ij}+ 2 d_{ijk}\ \lambda_k\ . 
\label{eq:2.15}\end{eqnarray}

\noindent The baryon-baryon matrix elements can now be computed using the 
cartesian octet states
\begin{equation}
 \langle B_{3},B_{4} | M | B_{1},B_{2}\rangle =
 C^{*}_{3j} C^{*}_{4n}\ M(j,n; i,m)\ C_{1 i} C_{2 m}\ ,
\label{eq:2.16}\end{equation}
where C-coefficients relate the particle states to the cartesian states,
see Table~\ref{taboctet2}, and $M(j,n;i,m)$ depends on the structure of the graph.
Below, we work out the $M$-operator for OBE-, TME-, and MPE-graphs in the
cartesian-octet representation. Then, the physical two-baryon matrix elements in 
(\ref{eq:2.16}) can be obtained easily.

\subsection{ Computations for OBE-, TME-graphs SU(3)-factors}
\label{sec:2d}
\begin{figure}   
% {\includegraphics[3.2in,8.5in][11in,10in]{ynfig/su3.obe.ps}}
  {\includegraphics[2.50in,9.0in][10.30in,10.5in]{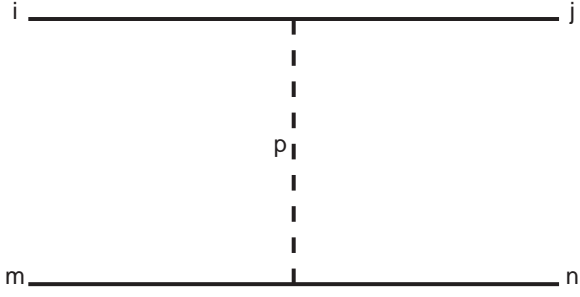}}
 \caption{Octet representation indices OBE-graphs. 
         The solid lines denote baryons with labels $i,m,j,n$.
         The dashed line with label p refers to the
         bosons: pseudoscalar, vector, axial-vector, or scalar mesons.}
 \label{fig.su3.obe}
  \end{figure}

$\bullet$ {\bf One-Boson-Exchange}:
The SU(3) matrix element for the OBE-graph Fig.~\ref{fig.su3.obe} is given
by 
\begin{equation}
      M_{obe}(j,n;i,m) = \sum_{p}^{\prime}         
      H_{1}^{(a)}(j,i,p)\ H_{2}^{(a)}(n,m,p)\ ,                    
\label{eq:2.17}\end{equation}
where $a=P,V,A,S$ and
\begin{eqnarray}
 H_{a}(j,i,p)&=&\left[2g^{(a)}_{8}\left\{i\alpha f_{jip}+(1-\alpha) d_{jip}
 \right\}\right. \nonumber\\ && \nonumber\\
     & & \left. \hspace{0.5cm} + g^{(a)}_{1} \delta_{ji}\delta_{p9}\right]
\label{eq:2.18}\end{eqnarray}
The summation over $p$ determines which mesons contribute to (\ref{eq:2.18}), 
and the prime indicates that one may restrict this summation in order to 
pick out a particular meson. This is in general necessary because within 
an SU(3) nonet the mesons have different masses, and we need their couplings
separately for a proper calculation of the potentials.

To illustrate this method of computation we consider $\pi$-exchange in 
$\Sigma N \rightarrow \Sigma N$. We have
\begin{eqnarray}
&& \langle \Sigma^+ n| M_\pi |\Sigma^+ n \rangle = \frac{1}{4}
 \sum_{i,j,m,n=1}^8 \sum_{p=1}^3 \langle \psi_1-i\psi_2|\psi_j\rangle 
 \cdot \nonumber\\ && \times
 \langle \psi_6-i\psi_7|\psi_n\rangle 
 \langle\psi_j\psi_n| M_\pi | \psi_i \psi_m \rangle
 \cdot \nonumber\\ && \times
 \langle \psi_i |\psi_1-i\psi_2\rangle \langle \psi_m|\psi_6-i\psi_7\rangle 
 \nonumber\\ && =\frac{1}{4}
 \sum_{i,j,m,n=1}^8 \sum_{p=1}^3 
 (\delta_{1j}+i\delta_{2j}) (\delta_{6n}+i\delta_{7n})\cdot \nonumber\\ &&
 \times(\delta_{1i}-i\delta_{2i}) (\delta_{6n}-i\delta_{7n}) 
 \langle\psi_j\psi_n| M_\pi | \psi_i \psi_m \rangle \nonumber\\ &&
 =\frac{1}{4} \sum_{i,j=1}^2 \sum_{m,n=6}^7\sum_{p=1}^3 
 Z(j,i)\ Z(n-j,m-i)\cdot \nonumber\\ && \times
 \left\{g^{(P)}_8\left[-i\alpha_{PV} f_{jip}+(1-\alpha_{PV}) d_{jip}\right]
 + g^{(P)}_1\delta_{ji}\delta_{p8}\right\}\cdot\nonumber\\ && \times
 \left\{g^{(P)}_8\left[-i\alpha_{PV} f_{nmp}+(1-\alpha_{PV}) d_{nmp}\right]
 + g^{(P)}_1\delta_{nm}\delta_{p8}\right\}\ , \nonumber\\   
\label{eq:2.19}\end{eqnarray}
where the $2\times 2$-matrix $Z$ is defined as
\begin{equation}
 Z = \left(\begin{array}{cc} 1 & -i \\ i & 1 \end{array}\right)\ .
\label{eq:2.20}\end{equation}

\begin{figure}   
% {\includegraphics[3.2in,8.5in][11in,10in]{ynfig/su3.tmepar.ps}}
  {\includegraphics[3.2in,8.5in][11in,10in]{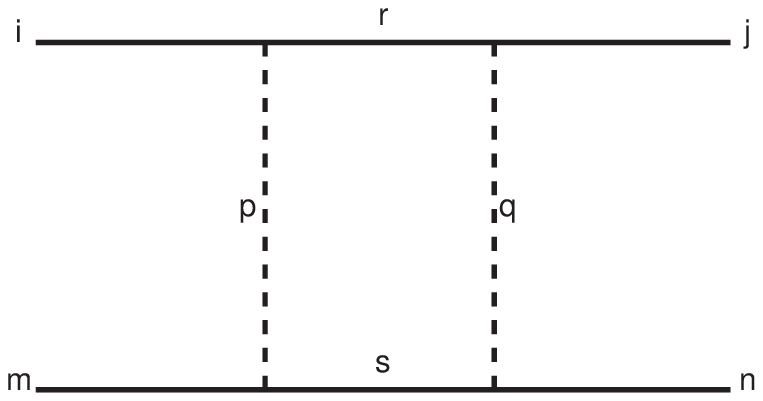}}
 \caption{Octet representation indices TME-parallel-graphs. 
         The solid lines denote baryons with labels $i,m,j,n,r,s$.
         The dashed lines with labels $p,q$ refer to the
         pseudoscalar mesons.}
 \label{fig.su3.tmepar}
  \end{figure}

\begin{figure}   
% {\includegraphics[3.2in,8.5in][11in,10in]{ynfig/su3.tmecrs.ps}}
  {\includegraphics[3.2in,8.5in][11in,10in]{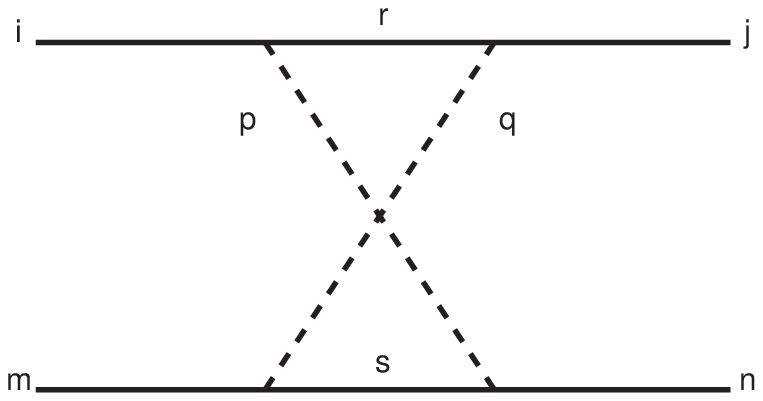}}
 \caption{Octet representation indices TME-crossed-graphs.
         The solid lines denote baryons with labels $i,m,j,n,r,s$.
         The dashed lines with labels $p,q$ refer to the
         pseudoscalar mesons.}
 \label{fig.su3.tmecrs}
  \end{figure}

$\bullet$ {\bf Two-Meson-Exchange}: 
The SU(3) matrix elements for the parallel (//) and crossed (X)
TME-graphs Fig.~\ref{fig.su3.tmepar} and Fig.~\ref{fig.su3.tmecrs} are given by 
\begin{eqnarray}
      M^{(//)}_{tme}(j,n;i,m)&=& \sum_{p,q,r,s}^{\prime}
      H_{2}(j,r,q)\ H_{1}(r,i,p) \nonumber\\                    
                      &\times&H_{2}(n,s,q)\ H_{1}(s,m,p) \\ && \nonumber\\              
      M^{(X)}_{tme}(j,n;i,m)&=& \sum_{p,q,r,s}^{\prime}
      H_{2}(j,r,q)\ H_{1}(r,i,p) \nonumber\\                    
                      &\times&H_{1}(n,s,q)\ H_{2}(s,m,p)               
\label{eq:2.21}\end{eqnarray}

%$\bullet$ \underline{Pair-Meson-Exchange}:
%The SU(3) matrix elements for the graphs with meson-pair vertices, 
%the so-called MPE-graphs \ref{diagtme} are given by 
%\begin{eqnarray}
%      M_{(1-pair)}(j,n;i,m)&=&H_{pair}(j,i,s)\ O(q,p,s) \nonumber\\                    
%                      &\times& H_{2}(m,r,q)\ H_{1}(r,m,p)\\ && \nonumber\\       
%      M_{(2-pair)}(j,n;i,m)&=&H_{pair}(j,i,s)\ O(q,p,s) \nonumber\\                    
%                      &\times& O(q,p,r)\ H_{pair}(n,m,p)       
%\label{eq:2.14}\end{eqnarray}
Again, like in the OBE-case, the numerical values of the SU(3) matrix elements 
for TME can be computed easily making a computer program.

%$\bullet$ \underline{Potentials on Isospin-basis}:

%---------------------------------------------------------------------------------

%---------------------------------------------------------------------------------

\section{ MPE interactions and SU(3) }
\label{sec:3}       

%% \begin{multicols}{2}
\subsection{Pair Couplings and SU(3)-symmetry}
\label{sec:3a}       
Below, $\sigma, {\bf a}_0, {\bf A}_1, \ldots $ are short-hands for 
respectively  the nucleon densities $\bar{\psi} \psi$,
$\bar{\psi}\mbox{\boldmath $\tau$}\psi$,
$\bar{\psi}\gamma_5\gamma_\mu\mbox{\boldmath $\tau$}\psi, \ldots $.

The SU(3)-octet and -singlet mesons, denoted by the subscript $8$
respectively $1$, are in terms of the physical ones defined as follows:
\begin{enumerate}
\item[(i)] \underline{Pseudo-scalar-mesons}:
\begin{eqnarray*}
   \eta_1 &=& \cos\theta_{PV} \eta' - \sin\theta_{PV} \eta \\       
   \eta_8 &=& \sin\theta_{PV} \eta' + \cos\theta_{PV} \eta              
\end{eqnarray*}
Here, $\eta'$ and $\eta$ are the physical pseudoscalar mesons 
 $\eta(957)$ respectively $\eta(548)$.
\item[(ii)] \underline{Vector-mesons}:         
\begin{eqnarray*}
   \phi_1 &=& \cos\theta_{V} \omega  - \sin\theta_{V} \phi \\       
   \phi_8 &=& \sin\theta_{V} \omega + \cos\theta_{V} \phi              
\end{eqnarray*}
Here, $\phi$ and $\omega$ are the physical vector mesons 
 $\phi(1019)$ respectively $\omega(783)$.
\end{enumerate}
Then, one has the following SU(3)-invariant pair-interaction 
Hamiltonians:\\
 1.\ SU(3)-singlet couplings $S^\alpha_\beta = 
\delta^\alpha_\beta \sigma/\sqrt{3}$:
\begin{eqnarray*}
 {\cal H}_{S_1PP} &=& \frac{g_{S_1PP}}{\sqrt{3}}\left\{
\mbox{\boldmath $\pi$}\cdot\mbox{\boldmath $\pi$} + 
  2 K^\dagger K + \eta_8\eta_8\right\}\cdot \sigma
\end{eqnarray*}
 2.\ SU(3)-octet symmetric couplings I, 
 $S^\alpha_\beta = (S_8)^\alpha_\beta \Rightarrow (1/4) Tr\{ S[P,P]_+\}$:
\begin{eqnarray*}
 {\cal H}_{S_8PP} &=& 
\frac{g_{S_8PP}}{\sqrt{6}}\left\{\vphantom{\frac{A}{A}}\right.
 ({\bf a}_0\cdot\mbox{\boldmath $\pi$})\eta_8 + 
 \frac{\sqrt{3}}{2}{\bf a}_0\cdot(K^\dagger \mbox{\boldmath $\tau$}K) 
 \nonumber \\
 &+& \frac{\sqrt{3}}{2}
 \left\{(K_0^\dagger\mbox{\boldmath $\tau$}K)\cdot\mbox{\boldmath $\pi$}+ 
 h.c.\right\} \nonumber\\ &-&
  \frac{1}{2}\left\{(K_0^\dagger K)\eta_8 + h.c. \right\} 
 \nonumber \\ &+& 
  \frac{1}{2}f_0\left(\mbox{\boldmath $\pi$}\cdot\mbox{\boldmath $\pi$} 
 - K^\dagger K -\eta_8\eta_8\right) 
 \left.\vphantom{\frac{A}{A}}\right\}
\end{eqnarray*}
 3.\ SU(3)-octet symmetric couplings II, 
 $S^\alpha_\beta = (B_8)^\alpha_\beta \Rightarrow (1/4) Tr\{ B^\mu [V_\mu, P]_+\}$:
\begin{eqnarray*}
 {\cal H}_{B_8VP} &=&
 \frac{g_{B_8VP}}{\sqrt{6}}\left\{\vphantom{\frac{A}{A}}\right. 
 \frac{1}{2}\left[\left({\bf B}_1^\mu\cdot\mbox{\boldmath $\rho$}_\mu\right) \eta_8 +
 \left({\bf B}_1^\mu\cdot\mbox{\boldmath $\pi$}_\mu\right) \phi_8 \right] 
  \nonumber \\ &+&
 \frac{\sqrt{3}}{4}\left[{\bf B}_1\cdot(K^{*\dagger}\mbox{\boldmath $\tau$}K)
 + h.c. \right] 
  \nonumber \\ &+&
 \frac{\sqrt{3}}{4}\left[(K_1^\dagger\mbox{\boldmath $\tau$} K^*)\cdot
 \mbox{\boldmath $\pi$}
 +(K_1^\dagger\mbox{\boldmath $\tau$} K)\cdot\mbox{\boldmath $\rho$} + h.c. \right] 
  \nonumber \\ &-&
 \frac{1}{4}\left[(K_1^\dagger\cdot K^*) \eta_8 + (K_1^\dagger\cdot K) \phi_8 
 + h.c. \right] 
  \nonumber \\ &+&
  \frac{1}{2}H^0\left[\mbox{\boldmath $\rho$}\cdot\mbox{\boldmath $\pi$} 
% -\frac{1}{2}\left(K^{*\dagger}\cdot K+ K^\dagger\cdot K^* \right)
 -\frac{1}{2}\left(K^{*\dagger}\cdot K+ h.c. \right)
 -\phi_8\eta_8 \right]
 \left.\vphantom{\frac{A}{A}}\right\}                            
\end{eqnarray*}
 4.\ SU(3)-octet a-symmetric couplings I, 
 $A^\alpha_\beta = (V_8)^\alpha_\beta \Rightarrow 
 (-i/\sqrt{2}) Tr\{ V^\mu [P,\partial_\mu P]_-\}$:
\begin{eqnarray*}
 {\cal H}_{V_8PP} &=& g_{A_8PP}\left\{\vphantom{\frac{A}{A}}\right.
 \frac{1}{2}\mbox{\boldmath $\rho$}_\mu\cdot\mbox{\boldmath $\pi$}\times
 \stackrel{\leftrightarrow}{\partial^\mu}\!\!
 \mbox{\boldmath $\pi$}+\frac{i}{2}\mbox{\boldmath $\rho$}_\mu\cdot(K^\dagger
 \mbox{\boldmath $\tau$}\!\!\stackrel{\leftrightarrow}{\partial^\mu}\!\! K) 
 \nonumber \\
 &+& \frac{i}{2}\left(\vphantom{\frac{A}{A}} K^{* \dagger}_\mu \mbox{\boldmath $\tau$}
(K\!\!\stackrel{\leftrightarrow}{\partial^\mu}\!\!\mbox{\boldmath $\pi$}) 
 - h.c. \right)
 +i\frac{\sqrt{3}}{2}\left(\vphantom{\frac{A}{A}} K^{* \dagger}_\mu\cdot
 \right.\nonumber\\  && \left. (K\cdot\stackrel{\leftrightarrow}
{\partial^\mu}\!\! \eta_8) - h.c. \vphantom{\frac{A}{A}}\right) 
 +\frac{i}{2}\sqrt{3} \phi_\mu (K^\dagger\stackrel{\leftrightarrow}{\partial^\mu}\!\! K)
 \left.\vphantom{\frac{A}{A}}\right\}
%\mbox{\Large /}2
\end{eqnarray*}
 5.\ SU(3)-octet a-symmetric couplings II, 
 $A^\alpha_\beta = (A_8)^\alpha_\beta \Rightarrow 
 (-i/\sqrt{2}) Tr\{ A^\mu [P,V_\mu]_-\}$:
\begin{eqnarray*}
 {\cal H}_{A_8VP} &=& g_{A_8VP}\left\{\vphantom{\frac{A}{A}}\right.
 {\bf A}_1\cdot\mbox{\boldmath $\pi$}\times\mbox{\boldmath $\rho$}
 \nonumber\\ &+&
  \frac{i}{2}{\bf A}_1\cdot\left[(K^\dagger\mbox{\boldmath $\tau$} K^*)  
  -(K^{*\dagger}\mbox{\boldmath $\tau$} K)\right] \nonumber \\
 &-&
  \frac{i}{2}\left(\left[(K^\dagger\mbox{\boldmath $\tau$}K_A)\cdot
 \mbox{\boldmath $\rho$}
 + (K_{A}^\dagger\mbox{\boldmath $\tau$}K^*)\cdot\mbox{\boldmath $\pi$}
 \right] - h.c.\right) \nonumber \\
 &-&
  i\frac{\sqrt{3}}{2}\left(\left[(K^\dagger\cdot K_A)\phi_8
 +(K_A^\dagger\cdot K^*)\eta_8\right] - h.c. \right) \nonumber \\
 &+&
  \frac{i}{2}\sqrt{3} f_1 \left[K^\dagger\cdot K^*-K^{*\dagger}\cdot K\right]  
 \left.\vphantom{\frac{A}{A}}\right\}
\end{eqnarray*}
The relation with the pair-couplings of \cite{RS96ab} and paper I is 
 $g_{S_1PP}/\sqrt{3}= g_{(\pi\pi)_0}/m_\pi$, 
$g_{A_8VP}= g_{(\pi\rho)_1}/m_\pi$ etc.

\subsection{Computations MPE-graphs SU(3)-factors}
\label{sec:3b}
\begin{figure}   
% {\includegraphics[3.2in,8.5in][11in,10in]{ynfig/su3.onepair.ps}}
  {\includegraphics[3.2in,8.5in][11in,10in]{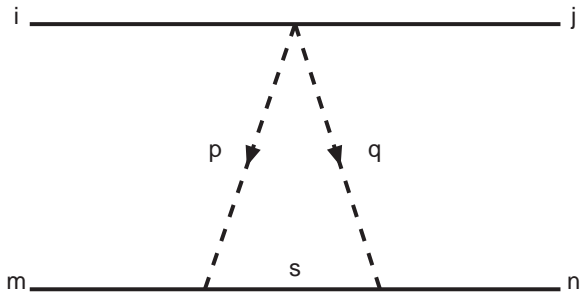}}
 \caption{Octet representation indices MPE one-pair-graphs.
         The solid lines denote baryons with labels $i,m,j,n,s$.
         The dashed lines with labels $p,q$ refer to the
         pseudoscalar etc. mesons.}
 \label{onepair.su3}   
  \end{figure}

\begin{figure}   
% {\includegraphics[3.2in,8.5in][11in,10in]{ynfig/su3.twopair.ps}}
  {\includegraphics[3.2in,8.5in][11in,10in]{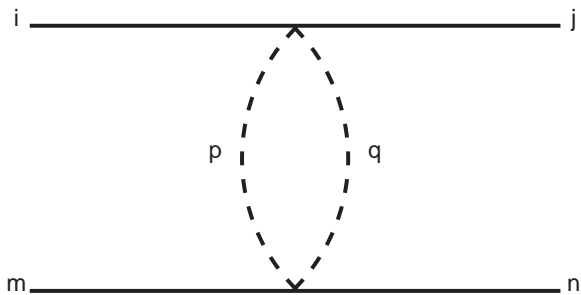}}
 \caption{Octet representation indices MPE two-pair-graphs.
         The solid lines denote baryons with labels $i,m,j,n$.
         The dashed lines with labels $p,q$ refer to the
         pseudoscalar etc. mesons.}
 \label{twopair.su3}   
  \end{figure}

The SU(3) matrix elements for the graphs with meson-pair vertices, 
the so-called MPE-graphs Fig.~\ref{onepair.su3} and Fig.~\ref{twopair.su3} are, 
using the cartesian-octet representation  
in section~\ref{sec:2c}, given by 
\begin{eqnarray}
      M_{(1-pair)}(j,n;i,m)&=& \sum_{p,q,r,s}^\prime
          H_{pair}(j,i,s)\ O(q,p,s) \nonumber\\                    
        &\times& H_{2}(m,r,q)\ H_{1}(r,m,p)\\ && \nonumber\\       
      M_{(2-pair)}(j,n;i,m)&=& \sum_{p,q,r,s=1}^\prime
          H_{pair}(j,i,s)\ O(q,p,s) \nonumber\\                    
        &\times& O(q,p,r)\ H_{pair}(n,m,p)       
\label{eq:3.14}\end{eqnarray}
Again, like in the OBE-case, the numerical values of the SU(3) matrix elements 
for MPE can be computed straightforwardly making a computer program.

%---------------------------------------------------------------------------------

%---------------------------------------------------------------------------------

\section{Broken SU(3)-Couplings and Form Factors }
\label{sec:4}

\subsection{Broken SU(3) BBM-couplings}             
\label{sec:4a}

In our models, breaking of the SU(3) symmetry is introduced in several
places. First of all, we use the physical masses for the baryons and
mesons. Second, we allow for the fact that the $\Lambda$ and $\Sigma^0$
have the same quark content, and so there is an appreciable mixing
between the isospin-pure $\Lambda$ and $\Sigma^0$ states~\cite{Dal64}.
Although exact ${\rm SU}(2) \subset {\rm SU}(3)$ symmetry requires that 
$f_{\Lambda\Lambda\pi^0}=0$,
$\Lambda$-$\Sigma^0$ mixing and the interaction $\Sigma^0\rightarrow
\Lambda+\pi^0$ result in a non-zero coupling constant for the physical
$\Lambda$-hyperon, derived by Dalitz and von Hippel \cite{Dal64}. 
This $\Lambda-\Sigma^0$-mixing leads also to a non-zero coupling of the $\Lambda$
to the other $I=1$ mesons: $\rho(760), a_0(980), a_1(1270)$, as well as to the 
$I=1$-pairs. For the details of these OBE-couplings see e.g. \cite{RSY99}, 
equations (2.15)-(2.17). The corresponding so-called CSB-potentials are 
included in the ESC-model for OBE, TME, and MPE.

In paper I of this series we have shown that the NNM-coupling constants are    
described pretty well by the $^{3}P_{0}$ mechanism \cite{Mic69,LeY73,Cha80}. 
In this paper we use the predictions of the $^3P_0$-model for the 
$F/(F+D)$-ratios as well. Therefore, it is most natural to use for 
the description of a possible flavor symmetry breaking of the coupling constants  
the $^{3}P_{0}$ mechanism as well, like in \cite{RSY99}.
In \cite{RSY99} it is argued that a symmetric treatment 
of the `moving'-quarks and the pair-quarks in the $^3P_0$-coupling process is
appropriate, since this leads to a covariant vertex.
Therefore, in \cite{RSY99}
the $^{3}P_{0}$-Hamiltonian for the $BBM$-couplings is taken as follows
\begin{eqnarray}
&&   H_{I} = \int d^{3}x \int d^{3}y\  F(x-y)\cdot \nonumber\\
&& \times \left[\bar{q}(x) O_{\bar{q} q} q(x) \right]^{(1)}
 \otimes \left[\bar{q}(y) O_{\bar{q} q} q(y) \right]^{(2)}
\label{eq:3.1}\end{eqnarray}
where the quark-field operators are vectors in flavor-space, with
components $q_{i}=(u,d,s)$ and $\bar{q}_{i}=(\bar{u},\bar{d},\bar{s})$.
(In the following we will refer to the non-strange quarks $u$ and $d$ as $n$-quarks.)
It is understood in (\ref{eq:3.1}) that the first factor creates or
annihilates a quark-pair, whereas the second factor `moves' a quark
from the baryon into the meson or vice versa. 
O is a matrix in quark-flavor space, which, supposing no quark-mixing,
is diagonal. However since it will
in general break SU(3)- and SU(2)-symmetry by using the form
\begin{equation}
    (O_{\bar{q} q})_{i,j}= \left(\begin{array}{ccc}
                         \gamma_{u} &   0        &     0     \\
                            0       & \gamma_{d} &     0     \\
                            0       &    0       & \gamma_{s}
                          \end{array}\right)\ ,  \\
\label{eq:3.2}
\end{equation}
where the pair-creation constants                           
$\gamma_{u},\gamma_{d}$, and $\gamma_{s}$ in principle could be unequal.
The CSB described above is on the quark-level due to $\gamma_u \neq \gamma_d$.
For a more detailed description of some properties of the Hamiltonian in
(\ref{eq:3.1}) and $(O_{\bar{q} q})_{i,j}$ 
we refer to \cite{RSY99}. 

Here, we assume that there is, also 'flavor-symmetry-breaking' (FSB) of the 
'medium strong' kind, i.e. $\gamma_n = \gamma_u = \gamma_d \neq \gamma_s$.
We introduce this medium strong SU(3)-breaking according to the 
$^3P_0$-model by a modification of the $s\bar{s}$-coupling, using the ratio
$\Delta_{FSB}= \gamma_{s}/\gamma_{n}-1$. 
For $\Delta_{FSB} =0$ there is no SU(3)-breaking, while 
for $\Delta_{FSB} \neq 0$ there is. 

%C----------------------------------------------------------------------
%C NEW SCHEME:
%C----------------------------------------------------------------------
The $I=1$-meson couplings for {\it NN} are determined in the process of {\it NN}-fitting.
This fixes $\gamma_n$. In the coupling of the $I=1/2$-mesons, $K, K^*,\kappa$, 
, only one $s\bar{s}$-operator is active and the SU(3)-breaking of the 
coupling is given by 
\begin{eqnarray}
   \Delta g_{\Lambda N K} &=& \Delta_{FSB}\ \hat{g}_{\Lambda N K}\ , \ 
%  \Delta g_{\Lambda N K} &=& -(1-\Delta_{FSB})\ g_{\Lambda N K}\ , \ 
%  \Delta g_{\Sigma  N K} = -(1-\Delta_{FSB})\ g_{\Sigma  N K} , \nonumber\\    
%  \Delta g_{\Xi\Lambda K} &=& -(1-\Delta_{FSB})\ g_{\Xi\Lambda K}\ , \
%  \Delta g_{\Xi\Sigma  K}= -(1-\Delta_{FSB})\ g_{\Xi\Sigma  K} , \nonumber\\    
\label{eq:3.3} \end{eqnarray}
and completely similar expressions for 
$  \Delta g_{\Sigma  N K} $, $\Delta g_{\Xi\Lambda K} $, and                               
$  \Delta g_{\Xi\Sigma K} $.                                              
Here, $\hat{g}_{\Lambda N K}$ etc. are calculated as usual in terms of
$g_{NN\pi}$ and the SU$(3)$-scheme with $\alpha_{PV}$. 
Similar formulas we used for the SU(3)-breaking in the case of
the vector, axial-vector, and scalar mesons.

In the case of the $I=0$-mesons two $s\bar{s}$-operators are active in the 
baryon-baryon coupling. Now, the $I=0$-mesons have a $n\bar{n}$- and a $s\bar{s}$-
component. In our scheme only the coupling of the $s\bar{s}$ 
is affected by the SU(3)-breaking.
Therefore, it is natural to transform first to the so-called 'ideal' $q\bar{q}$-basis,
applying the SU(3)-breaking, and transform back to the physical basis. This scheme
is as follows: 
\begin{enumerate}
\item[a.] The $I=0$ $|n\bar{n}\rangle = |u\bar{u}+d\bar{d}\rangle/\sqrt{2}$- 
and the $-|s\bar{s}\rangle$-states
are in $SU(3)$ linear combinations of the 
$\{1\}$ and $\{8\}$-octet states. Likewise, the coupling of the $n\bar{n}$- and
$s\bar{s}$-quark pairs to baryons are the same linear combinations, i.e.
\begin{eqnarray}
 \hat{g}_{n\bar{n}} &=& \cos\theta_I\ \hat{g}_{1} + \sin\theta_I\ \hat{g}_8\ , \nonumber\\
 \hat{g}_{s\bar{s}} &=&-\sin\theta_I\ \hat{g}_{1} + \cos\theta_I\ \hat{g}_8\ ,              
\label{eq:3.4} \end{eqnarray}
where $\cos\theta_I= \sqrt{2/3}$, and $\sin\theta_I=\sqrt{1/3}$.
\item[b.] 
Since for the $s\bar{s}$-coupling process two strange quark pairs are involved, 
and none in the $n\bar{n}$-coupling, the FSB is given on the level of the quark-pair
coupling by:
\begin{equation}
 g_{n\bar{n}} \rightarrow \hat{g}_{n\bar{n}}\ \ ,\ \ 
 g_{s\bar{s}} \rightarrow \left(1+\Delta_{FSB}\right)^2\hat{g}_{s\bar{s}}\ .
\label{eq:3.5} \end{equation}
\item[c.] The translation of this breaking to the level of the 
$\{1\}$ and $\{8\}$-octet couplings is the inverse transformation of (\ref{eq:3.4}), 
and from there to the physical mesons. For example in the case of the vector mesons we
have
\begin{eqnarray}
   g_{\omega} &=& \cos\Delta\theta_V\ g^V_{1} + \sin\Delta\theta_V\ g^V_8\ , \nonumber\\
   g_{\phi } &=&-\sin\Delta\theta_V\ g^V_{1} + \cos\Delta\theta_V\ g^V_8\ ,              
\label{eq:3.6} \end{eqnarray}
where $\Delta\theta_V = \theta_V-\theta_I$.
The similar procedure is used for the pseudoscalar, scalar, and axial-vector mesons.
\end{enumerate}

This breaking applies to the NNM-, YNM-, and YYM-couplings, containing 
the free parameter $\Delta_{PS}$ for the pseudoscalar mesons, and one parameter
$\Delta_V$ which is used for the vector, axial-vector, and scalar mesons.

We note that this breaking somewhat differs from that used in NSC97 \cite{RSY99}, 
which was based on an SU(6)$_W$-scheme. The problem with the latter, from our 
viewpoint is, that the states of for example the vector nonet are a mixture
of $W=0$ and $W=1$, making the implementation of SU(3)-breaking less 
straightforward as in the scheme described above.

The implementation of this scheme in practice is done as follows. 
We start, for example in the 
case of the vector mesons for the g-couplings, with the parameter set 
$(\hat{g}_\rho,\hat{g}_\omega,\theta_V,\alpha_V)$ and compute all couplings in the 
usual SU(3)-scheme, giving $\hat{g}_{NN\rho},\hat{g}_{\Sigma\Sigma\rho}$, etc.
%\hat{g}_{\Lambda\Sigma\rho},\hat{g}_{\Lambda N K},\hat{g}_{\Sigma N K},
%\hat{g}_{\Lambda\Sigma\rho},\hat{g}_{\Lambda N K},\hat{g}_{\Sigma N K},
This defines the singlet $\{1\}$ couplings
\begin{eqnarray*}
 \hat{g}_1 &=& \left[\hat{g}_\omega - \sin\theta_V \hat{g}_8\right]/\cos\theta_V\ ,
\end{eqnarray*}
where the octet $\{8\}$ coupling for nucleons is given by 
$\hat{g}_8 = (4\alpha_V-1) \hat{g}_{NN\rho}/\sqrt{3}$, and similarly for $\Lambda$, 
$\Sigma$, and $\Xi$. Then, we compute $\hat{g}_{n\bar{n}}$ and $\hat{g}_{s\bar{s}}$ 
using (\ref{eq:3.4}). Subsequently we compute the symmetry breaking by the transformation
etc. as described above, 
and finally we compute the coupling constants $g_\omega, g_\phi$, etc.

\noindent We finish this discussion by noticing that for the $I=1$-mesons 
$\pi, \rho, a_0, a_1$ for all baryon
couplings $g = \hat{g}$, because then only n-quarks are 'active'.

\subsection{Form Factors}                                                      
\label{sec:4b}
Also in this work, like in the NSC97-models \cite{RSY99}, the form 
factors depend on the SU(3) assignment of the mesons,  
In principle, we introduce form factor masses 
$\Lambda_{8}$ and $\Lambda_{1}$ for the $\{8\}$ and $\{1\}$ members of each 
meson nonet, respectively. In the application to ${\it YN}$ and ${\it YY}$, we allow
for SU(3)-breaking, by using different cut-offs for the strange mesons
$K$, $K^{*}$, and $\kappa$. Moreover, for the $I=0$-mesons we assign the 
cut-offs as if there were no meson-mixing. For example we assign $\Lambda_1$ 
for $\eta', \omega, \epsilon$, and $\Lambda_8$ for $\eta, \phi, S^*$, etc.
% update -----------------------------------------------------------
For the axial-mesons we use a single cut-off $\Lambda^A$.
% update -----------------------------------------------------------
%-endtheory------------------------------------------------------------------

\section{ ESC04-model: Fitting ${\it NN}\oplus {\it YN}$-data}                               
\label{sec:5} 
Like in the {\it NN}-fit, described in I, also in the 
simultaneous $\chi^2$-fit of the {\it NN}- and {\it YN}-data,
it appeared again that the OBE-couplings could be constraint successfully
by the 'naive' predictions of the QPC-model \cite{LeY73}. Although these 
predictions, see I, section \ref{sec:4}, are 'bare' ones, we tried to keep during the 
searches many OBE-couplings rather closely in the neighborhood of the predicted    
values.
Also, it appeared that we could either fix the $F/(F+D)$-ratios 
to those as suggested by the QPC-model, 
or apply the same restraining strategy as for the OBE-couplings.                      

In the simultaneous $\chi^2$-fit of the {\it NN}- and {\it YN}-data a {\it single set
of parameters} was used. Of course, it is to be expected that the accurate and 
very numerous {\it NN}-data essentially fix most of the parameters. Only some of 
the parameters, for example certain $F/(F+D)$-ratios, are influenced by the 
{\it YN}-data. 

\subsection{ Parameters and Nucleon-nucleon Fit}                                  
\label{sec:5a} 
% update --------------------------------------------------------------
For the cut-off masses $\Lambda$ we used as free parameters 
$\Lambda_8^P, \Lambda_8^V,\Lambda_1^V$, and $\Lambda^A$. 
The $I=0$ cut-off masses for the pseudoscalar and scalar mesons were
fixed to $\Lambda_1^P=900$ MeV, and $\Lambda_1^S \approx 1100$ MeV. 
% update --------------------------------------------------------------

The treatment of the broad mesons $\rho$ and $\epsilon$ is similar to that in the 
OBE-models \cite{NRS78,MRS89}. For the $\rho$-meson the same parameters are used 
as in these references. However, for the $\epsilon=f_0(760)$ assuming 
$m_\epsilon=760$ MeV and $\Gamma_\epsilon = 640$ MeV the Bryan-Gersten parameters      
\cite{Bry72} are used. For the chosen mass and width they are: 
$ m_1=496.39796$ MeV, $m_2=1365.59411$ MeV, and $\beta_1=0.21781, \beta_2=0.78219$.
The 'mass' of the diffractive exchanges were all fixed to $m_P=309.1$ MeV.

Summarizing the parameters we have for {\it NN}:
\begin{enumerate}
\item QPC-constrained: $ g_{NN\rho}, g_{NN\omega}$, \\
 $f_{NN\rho},f_{NN\omega}, f_{NNa_1}, g_{a_0},g_{NN\epsilon}, g_{NNA_2},g_{NNP}$, 
\item Pair couplings: $g_{NN(\pi\pi)_1},f_{NN(\pi\pi)_1}, g_{NN(\pi\rho)_1}$,\\
 $g_{NN\pi\omega}, g_{NN\pi\eta}, g_{NN\pi\epsilon}$, 
\item Cut-off masses: $\Lambda_8^P, \Lambda_8^V, \Lambda_8^S, \Lambda_1^V, \Lambda^A$.
\end{enumerate}
Of course, also the couplings for the pseud-scalar mesons $f_{NN\pi},f_{NN\eta'}$ were 
fitted. 
The pair coupling $g_{NN(\pi\pi)_0}$ was kept fixed at a small, 
but otherwise arbitrary value.

The {\it NN}-data used are the same as in I, and we refer the reader to this paper
for a description of the employed phase shift analysis \cite{Sto93,Klo93}. Differences 
with I are that here we did not fit the {\it NN}-low energy parameters and the 
deuteron binding energy explicitly.

\subsection{ Parameters and Hyperon-nucleon Fit}                                  
\label{sec:5b} 
All 'best' low-energy {\it YN}-data are included in the fitting, This is a selected set of   
35 low-energy {\it Y\!N}-data, the same set has been used in \cite{MRS89} and \cite{RSY99}.
We added 3 (preliminary) total $\Sigma^+p$ X-sections 
from the  recent KEK-experiment E289 \cite{Kanda05}.
In section~\ref{sec:8} these are given together with the results.
Next to these we added 'pseudo-data'
for the $\Lambda p$ and $\Lambda\Lambda$ scattering length's and effective ranges, 
in fm:
\begin{eqnarray}
 \hat{a}_{\Lambda p}(^1S_0) &=& -1.95 \pm 0.10\ \ ,\ \  \hat{r}_{\Lambda p}(^1S_0) =  2.90\ , 
\nonumber\\
 \hat{a}_{\Lambda p}(^3S_1) &=& -1.86 \pm 0.10\ \ ,\ \  \hat{r}_{\Lambda p}(^1S_0) =  2.70\ , 
\nonumber\\
 \hat{a}_{\Lambda\Lambda}(^1S_0) &=& -3.00 \pm 0.10\ , 
\label{eq:5.1}\end{eqnarray}
The $\Lambda p$-values which are suggested by the experience in several hyper-nuclear 
applications of the NSC97-models. Also, during the fitting checks were done to 
prevent the occurrence of bound states. 
Parameters, typically strongly influenced by the {\it YN}-data, are 
\begin{enumerate}
\item $F/(F+D)$-parameters: $\alpha_{PV}$, $\alpha_V^m$, $\alpha_S$,
\item SU(3)-symmetry breaking: $\Delta_{FSB}$.                               
\end{enumerate}
Notice that the strange octet-mesons $K$ etc. were given the same form factors as their
non-strange companions. 
So, because of {\it YN} we have introduced 4 extra free parameters.
We notice that the need to avoid bound states in the {\it YN} and {\it YY} systems has 
in particularly some influence on the trio $g_\epsilon, g_\omega$, and $g_P$.
Of particular importance of this was the introduction of the zero in the scalar-meson
form factors, see paper I for a detailed description. Like in I, also here we used 
a fixed zero by taking $U = 750$ MeV.

%\newpage
\section{ Coupling Constants, $F/(F+D)$ Ratios, and Mixing Angles}             
\label{sec:6} 

%\subsection{ BBM Coupling Constants, $F/(F+D)$ Ratios, and Mixing Angles}             
%\label{sec:6a} 
%----------------------------------------------------------------------
% 3P0 material:
Like in paper I, we constrained the OBE-couplings
by the 'naive' predictions of the QPC-model \cite{Mic69}.  
We kept during the 
searches all OBE-couplings in the neighborhood of these predictions, but a little
less so than in paper I.
The same has been done for all $\alpha=F/(F+D)$-ratios, i.e. for BBM- and the 
{\it BB}-Pair-couplings. In fact, all $F/(F+D)$-ratios were fixed, except the 
ratio $\alpha_V^m$ for vector mesons and $\alpha_S$ for the scalar mesons.                   

The mixing for the pseudoscalar, vector, and scalar mesons, as well as the 
handling of the diffractive potentials, has been described elsewhere, see
e.g. \cite{MRS89,RSY99}. The mixing etc. of the axial-vector mesons is completely
the same as for the vector etc. mesons, and also need not be discussed here.

In Table~\ref{table4} we give the fitted ESC04 meson couplings and parameters.
% corrected 28 nov. 2005: 
\begin{table}
\caption{Meson couplings and parameters employed in the ESC04-potentials.
%        shown in Figs.~\protect\ref{pap1fig3} to \protect\ref{pap1fig9}.
         Coupling constants are at ${\bf k}^{2}=0$.
         An asterisk denotes that the coupling constant is not searched,
         but constrained via SU(3) are simply put to some value used in     
         previous work.
         The used widths of the $\rho$ and $\varepsilon$ are 146 MeV and 640 MeV
         respectively.}
%\begin{tabular}{cdrrd}
\begin{ruledtabular}
\begin{tabular}{crrrr}
meson & mass (MeV) & $g/\sqrt{4\pi}$ & $f/\sqrt{4\pi}$ & $\Lambda$ (MeV) \\
\hline
 $\pi$         &  138.04 \hspace{3.5mm} &           &   0.2631   &    833.63    \\
 $\eta$        &  548.80 \hspace{3.5mm} &           &   0.1933$^\ast$   &     ,, \hspace{6mm}  \\
 $\eta'$       &  957.50 \hspace{3.5mm} &           &   0.1191   &    900.00    \\
 $\rho$        &  770.00 \hspace{3.5mm} &  0.7800   &   3.4711   &    839.53    \\
 $\phi$        & 1019.50 \hspace{3.5mm} &--0.3788   & --0.0494$^\ast$   &      ,, \hspace{6mm} \\
 $\omega$      &  783.90 \hspace{3.5mm} &  3.0138   &   0.4467   &    869.84    \\
 $a_1 $        & 1270.00 \hspace{3.5mm} &  2.5426   &            &    945.66    \\
 $f_1 $        & 1420.00 \hspace{3.5mm} &  0.8896$^\ast$   &     &     ,,  \hspace{6mm} \\
 $f_1'$        & 1285.00 \hspace{3.5mm} &  1.2544   &            &      ,, \hspace{6mm}  \\
 $a_{0}$       &  962.00 \hspace{3.5mm} &  0.9251   &            &   1159.88    \\
 $f_{0}$       &  993.00 \hspace{3.5mm} &--0.8162   &            &      ,, \hspace{6mm}  \\
 $\varepsilon$ &  760.00 \hspace{3.5mm} &  3.4635   &            &   1101.61    \\
 $a_{2}$       &  309.10 \hspace{3.5mm} &  0.0000   &            &              \\
 $f_2  $       &  309.10 \hspace{3.5mm} &  0.0000   &            &              \\
 $f_2' $       &  309.10 \hspace{3.5mm} &  0.0000   &            &              \\
 Pomeron       &  309.10 \hspace{3.5mm} &  1.9651   &            &              \\
%\hline\hline
\end{tabular}
\end{ruledtabular}
\label{table4}
\end{table}

In Table \ref{tab.3p0} we compare the fitted meson coupling constants with 
the 'naive' predictions of the QPC-model.
For the QPC-predictions in Table~\ref{tab.3p0}, see paper I. One sees that 
the fitted parameters are rather close to those of the QPC-model, and even
more so than in paper I. 
Notice that we omitted here the pion coupling, which requires a different $\gamma_M$
factor in the QPC-model, see remarks in I. Also, we see that the deviation between the
scalar and vector couplings from the QPC-model relations, 
$g_\epsilon-g_\omega \approx 3(g_{a_0}- g_\rho)$, which seems a purely isospin factor.
\begin{table}[hbt]
 \caption{{\it NN}+{\it YN}: ESC04 Couplings and $^3P_0$-Model Relations.}       
\begin{ruledtabular}
\begin{tabular}{lccccc} & & & & &  \\
  Meson          & $r_M[{\rm fm]}$ & $X_M$ & $\gamma_M$ & $^3P_0$ & ESC04 \\
                 &       &       &          &         &      \\
\hline
                 &       &       &          &         &      \\
%%\pi(140)$      & 0.56  & $5/6$ &  4.19    & $f=0.27$& $f=0.26$ \\
%%               &       &       &          &         &      \\
%%               &       &       &          & $g=3.67$& $g=3.58$ \\
%%               &       &       &          &         &      \\
 $\rho(770)$     & 0.56  & $1/2$ &  1.53    & $g=0.78$& $g=0.78$ \\
                 &       &       &          &         &      \\
 $\omega(783)$   & 0.56  & $3/2$ &  1.53    & $g=2.40$& $g=3.01$ \\
                 &       &       &          &         &      \\
 $a_0(962)$      & 0.56  &$\sqrt{3}/2$ & 1.53    & $g=0.79$& $g=0.92$ \\
                 &       &       &          &         &      \\
 $\epsilon(760)$ & 0.56  &$3\sqrt{3}/2$&  1.53    & $g=2.11$& $g=3.46$ \\
                 &       &       &          &         &      \\
 $a_1(1270)$     & 0.56  &$3\sqrt{3}/2$&  1.53    & $g=2.73$& $g=2.54$ \\
                 &       &       &          &         &      \\
%\hline\hline
\end{tabular}
\end{ruledtabular}
\label{tab.3p0}         
\end{table}
%----------------------------------------------------------------------
In Table~\ref{tab.su3par} the SU(3) singlet and octet couplings are 
listed i.e. $\hat{g}$ etc., and also $F/(F+D)$-ratios and mixing angles.
\begin{table}[hbt]
\caption{Coupling constants, $F/(F+D)$-ratio's, mixing angles etc.
 The values with $\star )$ have
 been determined in the fit to the {\it YN}-data. The other parameters
 are theoretical input or determined by the fitted parameters and
 the constraint from the {\it NN}-analysis. }
\begin{ruledtabular}
\begin{tabular}{ccccll} & & & & &  \\
  mesons    &   & $\{1\}$  & $\{8\}$  &  $F/(F+D)$   &    angles   \\
            &   &          &          &           &             \\
\hline
            &   &          &          &           &             \\
  ps-scalar  & f & 0.1852   &  0.2631 & $\alpha_{PV}=0.4668
  ^{\star)}$ & $\theta_{P} = -23.00^{0}$ \\
            &   &          &          &           &             \\
  vector    & g & 2.6218   &  0.7800  & $\alpha^{e}_{V}=1.0$ &
 $\theta_{V} = 37.50^{0} $   \\
            &   &          &          &           &             \\
            & f &  0.3845  &  3.4711  & $\alpha^{m}_{V}=0.2760
  ^{\star)}$ &  \\
            &   &          &          &           &             \\
  axial     & g &  1.5023  &  2.5426  & $\alpha_{A}=0.2340$ &
 $\theta_{A} =-23.00^{0~\star)}$ \\
            &   &          &          &           &             \\
  scalar    & g &  3.1688  &  0.9251  & $\alpha_{S}=0.8410$ &
 $\theta_{S} = 40.32^{0~\star)}$ \\
            &   &          &          &           &             \\
  diffractive   & g &  1.9651  & 0.0000   &  $\alpha_{D}=1.000$ &
 $\psi_{D}= 0.0^{0~\star)}$\\
            &   &          &          &           &             \\
%\hline
\end{tabular}
\end{ruledtabular}
\label{tab.su3par}  \end{table}
 \vspace*{\fill}

In Table \ref{tab.partwo} and in Table~\ref{tab.parthree}
we list the couplings of the physical mesons to the 
nucleons $(Y=1)$, and the hyperons with $Y=0$. 
These were computed using the FSB-scheme, described above. We found (ESC04a)
$\Delta_{FSB}(PV)=-0.258$, and $\Delta_{FSB}(V,S,A)=-0.267$.     

%\begin{widetext}
\begin{table}[hbt]
\caption{Coupling constants for pseudoscalar and vector meson
$Y=0$ and $Y= \pm 1$ exchanges.}
\begin{ruledtabular}
\begin{tabular}{lcrrrrrr}  & & & & & & & \\
 M  & & \multicolumn{1}{c}{ NNM } & \multicolumn{1}{c}{$\Lambda\Lambda M$} &
     \multicolumn{1}{c}{$\Lambda\Sigma M$} & \multicolumn{1}{c}{$\Sigma\Sigma M$}
     & \multicolumn{1}{c}{$\Lambda NM$} & \multicolumn{1}{c}{$\Sigma N M$}  \\
        &     &            &              &             &   &   &    \\
\hline
        &     &            &              &             &   &   &    \\
$\pi$ & g & 3.57602 & \multicolumn{1}{c}{CSB} & 2.47895 & 4.23203
              &   \multicolumn{1}{c}{---}      &    \multicolumn{1}{c}{---}       \\
      & f & 0.26306 & \multicolumn{1}{c}{CSB} & 0.16196 & 0.24559
              &   \multicolumn{1}{c}{---}      &    \multicolumn{1}{c}{---}       \\
        &     &            &              &             &   &    &   \\
$\eta$  &  f  &  0.19333   & --0.02028    &    \multicolumn{1}{c}{---}      &   0.21534
              &   \multicolumn{1}{c}{---}      &    \multicolumn{1}{c}{---}       \\
        &     &            &              &             &   &    &   \\
$\eta^{'}$&f  &  0.11908   &   0.14213    &    \multicolumn{1}{c}{---}      &   0.11671
              &   \multicolumn{1}{c}{---}      &    \multicolumn{1}{c}{---}       \\
        &     &            &              &             &   &    &   \\
$ K  $  &  g  &   \multicolumn{1}{c}{---}      &    \multicolumn{1}{c}{---}       &    \multicolumn{1}{c}{---}      &   \multicolumn{1}{c}{---}
              & --3.22933   &   0.19837    \\
        &  f  &   \multicolumn{1}{c}{---}      &    \multicolumn{1}{c}{---}       &    \multicolumn{1}{c}{---}      &   \multicolumn{1}{c}{---}
              & --0.21786   &   0.01296    \\
        &     &            &              &             &   &  &     \\
$\rho$  &  g  & 0.78000 & \multicolumn{1}{c}{CSB} & \multicolumn{1}{c}{---} & 1.56000
              &   \multicolumn{1}{c}{---}      &    \multicolumn{1}{c}{---}       \\
        &  f  & 3.47113 & \multicolumn{1}{c}{CSB} & 2.90094 & 1.91768
              &   \multicolumn{1}{c}{---}      &    \multicolumn{1}{c}{---}       \\
        &     &            &              &             &   &  &     \\
$\phi$  &  g  &--0.37884   & --0.94125    &    \multicolumn{1}{c}{---}      & --0.94125
              &   \multicolumn{1}{c}{---}      &    \multicolumn{1}{c}{---}       \\
        &  f  &--0.04944   & --1.34461    &    \multicolumn{1}{c}{---}      &   1.07066
              &   \multicolumn{1}{c}{---}      &    \multicolumn{1}{c}{---}       \\
        &     &            &              &            &   &  &     \\
$\omega$&  g  &  3.01376   &   2.21121    &    \multicolumn{1}{c}{---}      &   2.21121
              &   \multicolumn{1}{c}{---}      &    \multicolumn{1}{c}{---}       \\
        &  f  &  0.44671   & --1.40149    &    \multicolumn{1}{c}{---}      &   2.04507
              &   \multicolumn{1}{c}{---}      &    \multicolumn{1}{c}{---}       \\
        &     &            &              &             &    &  &    \\
$K^{\star}$&  g  &   \multicolumn{1}{c}{---}      &    \multicolumn{1}{c}{---}       &    \multicolumn{1}{c}{---}      &   \multicolumn{1}{c}{---}
              &--0.99079   & --0.57203    \\
        &  f  &   \multicolumn{1}{c}{---}      &    \multicolumn{1}{c}{---}       &    \multicolumn{1}{c}{---}      &   \multicolumn{1}{c}{---}
              &--2.28171   &   1.13927    \\
        &     &            &              &             &    &  &    \\
$a_1 $  &  g  & 2.54264 & \multicolumn{1}{c}{CSB} & 2.24770 & 1.19215
              &   \multicolumn{1}{c}{---}      &    \multicolumn{1}{c}{---}       \\
        &     &            &              &             &   &  &     \\
$f_1 $  &  g  &  0.88961   & --0.66726    &    \multicolumn{1}{c}{---}      &   2.57849
              &   \multicolumn{1}{c}{---}      &    \multicolumn{1}{c}{---}       \\
        &     &            &              &            &   &  &     \\
$f'_1  $&  g  &  1.25438   &   1.40489    &    \multicolumn{1}{c}{---}      &   1.09111
              &   \multicolumn{1}{c}{---}      &    \multicolumn{1}{c}{---}       \\
        &     &            &              &             &    &  &    \\
$K_1$   &  g  &   \multicolumn{1}{c}{---}      &    \multicolumn{1}{c}{---}       &    \multicolumn{1}{c}{---}      &   \multicolumn{1}{c}{---}
              &--1.58137   &   0.99042    \\
        &     &            &              &             &    &  &    \\
%\hline
\end{tabular}
\end{ruledtabular}
  \label{tab.partwo} \end{table}
 
\begin{table}[hbt]
\caption{Coupling constants for scalar meson and `diffractive'
$Y=0$ and $Y= \pm 1$ exchanges. Nomenclature scalar mesons: 
$\delta= a_0(962)$, $\epsilon=f_0(760)$, $S^*=f_0(993)$, $\kappa=K_0^*(900)$.}
\begin{ruledtabular}
\begin{tabular}{lcrrrrrr}  & & & & & & & \\
 M  & & \multicolumn{1}{c}{ NNM } & \multicolumn{1}{c}{$\Lambda\Lambda M$} &
  \multicolumn{1}{c}{$\Lambda\Sigma M$} & \multicolumn{1}{c}{$\Sigma\Sigma M$}
   & \multicolumn{1}{c}{$\Lambda NM$} & \multicolumn{1}{c}{$\Sigma N M$} \\
        &     &            &              &             &   &   &    \\
\hline
        &     &            &              &             &   &   &    \\
$\delta$& g & 0.92511 & \multicolumn{1}{c}{CSB} &  0.16975 & 1.55621
              &   \multicolumn{1}{c}{---}      &    \multicolumn{1}{c}{---}       \\
        &     &            &              &             &    &  &    \\
$S^*$   &  g  &--0.81620   & --1.36993    &    \multicolumn{1}{c}{---}      & --1.23870
              &   \multicolumn{1}{c}{---}      &    \multicolumn{1}{c}{---}       \\
        &     &            &              &             &    &  &    \\
$\epsilon$   & g  &  3.46354   &   2.58418    &    \multicolumn{1}{c}{---}      &   2.79258
              &   \multicolumn{1}{c}{---}      &    \multicolumn{1}{c}{---}       \\
        &     &            &              &             &    &  &    \\
$\kappa$&  g  &   \multicolumn{1}{c}{---}      &    \multicolumn{1}{c}{---}       &    \multicolumn{1}{c}{---}      &   \multicolumn{1}{c}{---}
              &--1.05063   & --0.46283    \\
        &     &            &              &             &    &  &    \\
$A_{2}$ & g &  0.00000& \multicolumn{1}{c}{CSB} &  0.00000 &  0.00000
              &   \multicolumn{1}{c}{---}      &    \multicolumn{1}{c}{---}       \\
        &     &            &              &             &    &  &    \\
%$f'_2$  &  g  &--0.03455   & --0.01337    &    \multicolumn{1}{c}{---}      & --0.01337
%              &   \multicolumn{1}{c}{---}      &    \multicolumn{1}{c}{---}       \\
%        &     &            &              &             &    &  &    \\
%$f_2$   &  g  &  0.00116   &   0.03147    &    \multicolumn{1}{c}{---}      &   0.03147
%              &   \multicolumn{1}{c}{---}      &    \multicolumn{1}{c}{---}       \\
%        &     &            &              &             &    &  &    \\
$P$ &g  &  1.96510   &  1.96510     &    \multicolumn{1}{c}{---}
&   1.96510  & \multicolumn{1}{c}{---} & \multicolumn{1}{c}{---} \\
        &     &            &              &             &    &  &    \\
$K^{\star\star}_2$&  g  &   \multicolumn{1}{c}{---}      &    \multicolumn{1}{c}{---}       &    \multicolumn{1}{c}{---}      &   \multicolumn{1}{c}{---}
              &  0.00000   &   0.00000    \\
        &     &            &              &             &    &  &    \\
%\hline
\end{tabular}
\end{ruledtabular}
  \label{tab.parthree} \end{table}
 
%\end{widetext}
%--------------------------------------------------------------------------

%\subsection{ MPE Coupling Constants and $F/(F+D)$ Ratios}             
%\label{sec:6b} 
In Table~\ref{tab.gpair} we listed the fitted Pair-couplings for the MPE-potentials.
We recall that only One-pair graphs are included, in order to avoid double
counting, see paper I. The $F/(F+D)$-ratios are all fixed, assuming heavy-boson 
domination of the pair-vertices. The ratios are taken from the QPC-model for 
$Q\bar{Q}$-systems with the same quantum numbers as the dominating boson.
The {\it BB}-Pair couplings are computed, assuming unbroken SU(3)-symmetry, from the 
{\it NN}-Pair coupling and the $F/(F+D)$-ratio using SU(3).
 
\begin{table}[hbt]
\caption{Pair-meson coupling constants employed in the ESC04 MPE-potentials.     
         Coupling constants are at ${\bf k}^{2}=0$.}
%        An asterisk denotes that the coupling constant is fixed at its
%        theoretical value as given in \protect\cite{RS96b}.}                 
\begin{ruledtabular}
\begin{tabular}{cclcc}
 $J^{PC}$ & SU(3)-irrep & $(\alpha\beta)$  & $g/4\pi$  & $F/(F+D)$ \\
\colrule
 $0^{++}$ & $\{1\}$  & $g(\pi\pi)_{0}$   &  ---    &  ---    \\
 $0^{++}$ & ,,       & $g(\sigma\sigma)$ &  ---    &  ---    \\
 $0^{++}$ &$\{8\}_s$ & $g(\pi\eta)$      &--0.1860 &  1.000  \\
%$0^{++}$ &          & $g(\pi\eta')$     &  ---    &  ---    \\
 $1^{--}$ &$\{8\}_a$ & $g(\pi\pi)_{1}$   &--0.0024 &  1.000  \\
          &          & $f(\pi\pi)_{1}$   &  0.1310 &  0.400  \\
 $1^{++}$ & ,,       & $g(\pi\rho)_{1}$  &  0.8864 &  0.643  \\
 $1^{++}$ & ,,       & $g(\pi\sigma)$    &--0.0241 &  0.643  \\
 $1^{++}$ & ,,       & $g(\pi P)$        &  0.0    &  ---    \\
 $1^{+-}$ &$\{8\}_s$ & $g(\pi\omega)$    &--0.1722 &  0.467  \\
\end{tabular}
\end{ruledtabular}
\label{tab.gpair}
\end{table}

Unlike in \cite{RS96ab}, we did not fix pair couplings using
a theoretical model, based on heavy-meson saturation and chiral-symmetry.
So, in addition to the 14 parameters used in \cite{RS96ab} we now have
6 pair-coupling fit parameters. 
In Table~\ref{tab.gpair} the fitted pair-couplings are given.
Note that the $(\pi\pi)_0$-pair coupling gets contributions from the $\{1\}$ and
the $\{8_s\}$ pairs as well, giving in total $g_{(\pi\pi)}=0.10$, which has the
same sign as in \cite{RS96ab}. The $f_{(\pi\pi)_1}$-pair coupling has opposite
sign as compared to \cite{RS96ab}. In a model with a more complex and realistic
meson-dynamics \cite{SR97} this coupling is predicted as found in the present 
ESC-fit. The $(\pi\rho)_1$-coupling agrees nicely with $A_1$-saturation, see 
\cite{RS96ab}. We conclude that the pair-couplings are in general not well
understood, and deserve more study.

In the ESC-model described here, is fully consistent with SU(3)-symmetry  
using a straightforward extension of the {\it NN}-model to {\it YN} and {\it YY}.
For example $g_{(\pi\rho)_1} = g_{A_8VP}$, and
besides $(\pi\rho)$-pairs one sees also that $K K^*(I=1)$- and 
$K K^*(I=0)$-pairs contribute to the {\it NN} potentials.
All $F/(F+D)$-ratio's are taken fixed with heavy-meson saturation in mind.
The approximation we have made in this paper is to neglect the baryon mass
differences, i.e. we put $m_\Lambda = m_\Sigma = m_N$. This because we
have not yet worked out the formulas for the inclusion of these mass 
differences, which is straightforward in principle.

%\clearpage 
\section{ ESC04-model , $N\!N$-Results}                                  
\label{sec:7} 
\subsection{ Parameters and Nucleon-nucleon Fit}                                  
\label{sec:7a} 
For a more detailed discussion on the {\it NN}-fitting we refer to I. Here, we fit only to the
1993 Nijmegen representation of the $\chi^2$-hypersurface of the 
{\it NN} scattering data below $T_{\rm lab}=350$ MeV \cite{Sto93,Klo93}. 
This in contrast to I where
also low-energy parameters are fitted for {\it np} and {\it nn}. 
In this simultaneous fit of {\it NN} and {\it YN}, we obtained 
for the phase shifts a $\chi^2/Ndata =1.22$. 
In Table~\ref{table4} the meson parameters are given for the ESC04a-model.
In Table~\ref{tab.chidistr} the distribution of the $\chi^2$ is 
shown for the ten energy bins, which can be compared with a similar table in paper I.
Also, for a comparison with paper I, and for use of this model for the description
of {\it NN}, we give in Tables \ref{tab.nnphas1} and \ref{tab.nnphas2} the nuclear-bar
phases for {\it pp} in case $I=1$, and for {\it np} in case $I=0$.
The deuteron was not fitted, and we have for the binding energy $E_B= 2.224797$MeV, which
is very close to the $E_B(experiment)=2.224644$.

\begin{table}
\caption{$\chi^2$ and $\chi^2$ per datum at the ten energy bins for the    
 Nijmegen93 Partial-Wave-Analysis. $N_{\rm data}$ lists the number of data
 within each energy bin. The bottom line gives the results for the 
 total $0-350$ MeV interval.
 The $\chi^{2}$-access for the ESC04-model in the ${\it NN}+{\it YN}$-fit is denoted    
 by  $\Delta\chi^{2}$ and $\Delta\hat{\chi}^{2}$, respectively.}  
\begin{ruledtabular}
\begin{tabular}{crrrrrr} & & & & & \\
 $T_{\rm lab}$ & $\sharp$ data & $\chi_{0}^{2}$\hspace*{5.5mm}&
 $\Delta\chi^{2}$&$\hat{\chi}_{0}^{2}$\hspace*{3mm}&
 $\Delta\hat{\chi}^{2}$ \\ &&&&& \\ \hline
0.383 & 144 & 137.5549 & 22.9 & 0.960 & 0.159  \\
  1   &  68 &  38.0187 & 53.2 & 0.560 & 0.783  \\
  5   & 103 &  82.2257 &  7.1 & 0.800 & 0.068  \\
  10  & 209 & 257.9946 & 53.1 & 1.234 & 0.183  \\
  25  & 352 & 272.1971 & 62.5 & 0.773 & 0.177  \\
  50  & 572 & 547.6727 &240.3 & 0.957 & 0.420  \\
  100 & 399 & 382.4493 & 73.6 & 0.959 & 0.184  \\
  150 & 676 & 673.0548 &104.4 & 0.996 & 0.154  \\
  215 & 756 & 754.5248 &214.4 & 0.998 & 0.284  \\
  320 & 954 & 945.3772 &333.1 & 0.991 & 0.349  \\ \hline
      &    &     &     &     &    \\
Total &4233&4091.122&1164.6 &0.948 &0.268  \\
      &    &     &     &     &     \\
\end{tabular}
\end{ruledtabular}
\label{tab.chidistr} 
\end{table}

\begin{table}[h]
\caption{ ESC04 {\it pp} and {\it np} nuclear-bar phase shifts in degrees.}
\begin{ruledtabular}
\begin{tabular}{crrrrr}  & & & & & \\
 $T_{\rm lab}$ & 0.38& 1 & 5  & 10 & 25 \\ \hline
     &    &     &     &     &     \\
 $\sharp$ data &144  & 68  & 103 & 290& 352 \\
     &    &     &     &     &   \\
$\Delta \chi^{2}$& 24  & 53  & 7   & 53  & 62  \\
     &    &     &     &     &    \\ \hline
     &    &     &     &     &    \\
%$^{1}S_{0}(np)$ & 54.26  & 61.68 & 62.99 & 59.15& 49.77  \\
 $^{1}S_{0}$ & 14.62  & 32.62 & 54.71 & 55.07& 48.39  \\
 $^{3}S_{1}$ & 159.39 & 147.77& 118.23& 102.70& 80.78\\
 $\epsilon_{1}$ & 0.03  & 0.11 & 0.66 & 1.13 & 1.71 \\
 $^{3}P_{0}$ & 0.02   &  0.13 & 1.56 & 3.69 & 8.58  \\
 $^{3}P_{1}$ & -0.01  &-0.08  &-0.87  & -2.01  & -4.85  \\
 $^{1}P_{1}$ & -0.05  &-0.19  &-1.52  & -3.12  & -6.46  \\
 $^{3}P_{2}$ &  0.00  & 0.01  & 0.21  &  0.64  &  2.44  \\
 $\epsilon_{2}$ &-0.00  &-0.00 &-0.05 &-0.19 &-0.79 \\
 $^{3}D_{1}$ & 0.00  &-0.01  &-0.19  & -0.69  & -2.85  \\
 $^{3}D_{2}$ & 0.00  & 0.01  & 0.22  &  0.85  &  3.72  \\
 $^{1}D_{2}$ & 0.00  & 0.00  & 0.04  &  0.16  &  0.67  \\
 $^{3}D_{3}$ & 0.00  & 0.00  & 0.00  &  0.00  &  0.01  \\
 $\epsilon_{3}$ & 0.00  & 0.00 & 0.01 & 0.08 & 0.56 \\
 $^{3}F_{2}$ & 0.00  & 0.00  & 0.00  &  0.01  &  0.10  \\
 $^{3}F_{3}$ & 0.00  & 0.00  &-0.00  & -0.03  & -0.22  \\
 $^{1}F_{3}$ & 0.00  & 0.00  &-0.01  & -0.07  & -0.42  \\
 $^{3}F_{4}$ & 0.00  & 0.00  & 0.00  &  0.00  &  0.02  \\
 $\epsilon_{4}$ & 0.00  & 0.00 & 0.00 &-0.00 &-0.05 \\
%$^{3}G_{3}$ &-0.00 &-0.00  &-0.00  &-0.00  & -0.00  & -0.05  \\
%$^{3}G_{4}$ & 0.00 & 0.00  & 0.00  & 0.00  &  0.01  &  0.17  \\
%$^{1}G_{4}$ & 0.00 & 0.00  & 0.00  & 0.00  &  0.00  &  0.04  \\
%$^{3}G_{5}$ &-0.00 &-0.00  &-0.00  &-0.00  & -0.00  & -0.01  \\
%$\epsilon_{5}$ & 0.00 & 0.00  & 0.00 & 0.00 & 0.00 & 0.04 \\
%$^{3}H_{4}$ & 0.00 & 0.03  & 0.11  & 0.21  &  0.36  &  0.57  \\
%$^{3}H_{5}$ &-0.01 &-0.08  &-0.29  &-0.51  & -0.75  & -1.07  \\
%$^{1}H_{5}$ &-0.03 &-0.16  &-0.50  &-0.82  & -1.13  & -1.48  \\
%$^{3}H_{6}$ & 0.00 & 0.01  & 0.04  & 0.11  &  0.22  &  0.44  \\
%$\epsilon_{6}$ &-0.00 &-0.03  &-0.11 &-0.22 &-0.35 &-0.53 \\
     &    &     &     &     &    \\
%\hline\hline
\end{tabular}
\end{ruledtabular}
\label{tab.nnphas1}   
\end{table}

\begin{table}[h]
\caption{ ESC04 {\it pp} and {\it np} nuclear-bar phase shifts in degrees.}
\begin{ruledtabular}
\begin{tabular}{crrrrr}  & & & & & \\
 $T_{\rm lab}$ &  50 &100 & 150 & 215 & 320 \\ \hline
     &    &     &     &     &     \\
 $\sharp$ data &572  &399  &676  & 756& 954 \\
     &    &     &     &     &   \\
$\Delta \chi^{2}$& 240 & 74  & 104 & 214 & 333 \\
     &    &     &     &     &    \\ \hline
     &    &     &     &     &    \\
%$^{1}S_{0}(np)$ & 39.00  & 24.54 & 14.19 & 3.76 & -9.20  \\
 $^{1}S_{0}$ & 38.40  & 24.05 & 13.51 & 2.79 &--10.60 \\
 $^{3}S_{1}$ & 63.01  & 43.67 & 31.44 & 19.93 & 6.42 \\
 $\epsilon_{1}$ & 1.94  & 2.18 & 2.56 & 3.22 & 4.57 \\
 $^{3}P_{0}$ & 11.65  &   9.84& 5.19 &-1.31 &-10.86 \\
 $^{3}P_{1}$ & -8.23  &-13.19 &-17.25 & -21.85 & -28.09 \\
 $^{1}P_{1}$ & -9.75  &-14.12 &-17.78 & -22.09 & -28.12 \\
 $^{3}P_{2}$ &  5.75  & 11.02 & 14.21 &  16.29 &  16.96 \\
 $\epsilon_{2}$ &-1.68  &-2.68 &-2.97 &-2.83 &-2.17 \\
 $^{3}D_{1}$ &-6.58  &-12.61 &-17.06 & -21.36 & -26.14 \\
 $^{3}D_{2}$ & 8.92  & 17.08 & 21.90 &  24.70 &  24.83 \\
 $^{1}D_{2}$ & 1.66  & 3.80  & 5.88  &  8.09  &  10.16 \\
 $^{3}D_{3}$ & 0.18  & 1.05  & 2.15  &  3.42  &  4.60  \\
 $\epsilon_{3}$ & 1.62  & 3.52 & 4.88 & 6.02 & 6.97 \\
 $^{3}F_{2}$ & 0.32  & 0.74  & 0.98  &  0.97  &  0.13  \\
 $^{3}F_{3}$ &-0.66  &-1.46  &-2.09  & -2.74  & -3.73  \\
 $^{1}F_{3}$ &-1.13  &-2.20  &-2.90  & -3.58  & -4.65  \\
 $^{3}F_{4}$ & 0.10  & 0.42  & 0.88  &  1.57  &  2.71  \\
 $\epsilon_{4}$ &-0.19  &-0.52 &-0.81 &-1.12 &-1.46 \\
 $^{3}G_{3}$ &-0.27 &-0.99  &-1.89  &-3.10  & -4.88  \\
 $^{3}G_{4}$ & 0.72  & 2.13  & 3.54  &  5.16  &  7.23  \\
 $^{1}G_{4}$ & 0.15  & 0.40  & 0.65  &  1.00  &  1.60  \\
 $^{3}G_{5}$ &-0.06  &-0.21  &-0.36  & -0.49  & -0.58  \\
 $\epsilon_{5}$ & 0.21  & 0.72 & 1.26 & 1.91 & 2.75 \\
%$^{3}H_{4}$ & 0.03  & 0.11  & 0.21  &  0.36  &  0.57  \\
%$^{3}H_{5}$ &-0.08  &-0.29  &-0.51  & -0.75  & -1.07  \\
%$^{1}H_{5}$ &-0.16  &-0.50  &-0.82  & -1.13  & -1.48  \\
%$^{3}H_{6}$ & 0.01  & 0.04  & 0.11  &  0.22  &  0.44  \\
%$\epsilon_{6}$ &-0.03  &-0.11 &-0.22 &-0.35 &-0.53 \\
     &    &     &     &     &    \\
%\hline\hline
\end{tabular}
\end{ruledtabular}
\label{tab.nnphas2}   
\end{table}

\section{ ESC04-model , {\it Y\!N}-Results}                                  
\label{sec:8} 
%\section{Fit to {\it Y\!N} total cross sections}

In combined {\it N\!N} and {\it Y\!N} fit, the used {\it Y\!N} scattering data 
from Refs.~\cite{Ale68}-\cite{Ste70}, are shown in Table~\ref{tab.reslts1}.
Since we know from the experience with the NSC97 models rather 
well the favored s-wave scattering lengthes for $\Lambda N$, we 
added values for these as pseudo-data, see section~\ref{sec:5b}.
The {\it N\!N} interaction puts very strong constraints on most
of the parameters, and so we are left with only a limited set of
parameters which have some freedom to steer the {\it YN} channels.
Like in the NSC97 models we exploit here (i) the magnetic vector-meson 
$F/(F+D)$ ratio $\alpha_V^m$, (ii) the scalarmeson $F/(F+D)$ ratio 
$\alpha_S$, and the flavor symmetry breaking parameter$\Delta_{FSB}$.
We did not break SU(3) by introducing independent cut-off parameters
for the strange mesons $K, K^*$ etc., but $\Lambda_K= \Lambda_\pi$ and
similar for the other meson-nonets.
%The fitted parameters are given in Table~\ref{tabparfit}.           
The fitted parameters are given in Table~\ref{table4} and Table~\ref{tab.gpair}.

The aim of the present study was to construct a realistic potential model 
for baryon-baryon with parameters that are optimal theoretically, but at
the sametime describes the baryon-baryon scattering data very satisfactory.

This model can then be used with a great deal of confidence  
in calculations of hypernuclei and in their predictions for the $S=-2$,
$-3$, and $-4$ sectors. Especially for the latter application, these
models will be the first models for the $S<-1$ sector to have
their theoretical foundation in the {\it N\!N} and {\it Y\!N} sectors.
 
%\clearpage

\begin{table}[h]
\caption{Comparison of the calculated and experimental values for the
         38 {\it YN}-data that were included in the fit. The superscipts
         $RH$ and $M$ denote, respectively,
         the Rehovoth-Heidelberg Ref.~\protect\cite{Ale68}
         and Maryland data Ref.~\protect\cite{Sec68}.
         Also included are 3 $\Sigma^+p$ X-sections at $p_{lab}=400, 500, 650$ MeV
         from Ref.~\protect\cite{Kanda05}.
         The laboratory momenta
         are in MeV/c, and the total cross sections in mb.}
\begin{center}
\begin{ruledtabular}
\begin{tabular}{cccccc} \\
\multicolumn{2}{c}{$\Lambda p\rightarrow \Lambda p$} & $\chi^{2}=0.8$    &
\multicolumn{2}{c}{$\Lambda p\rightarrow \Lambda p$} & $\chi^{2}=3.3$   \\
$p_{\Lambda}$ & $\sigma^{RH}_{exp}$ & $\sigma_{th}$ &
$p_{\Lambda}$ & $\sigma^{M }_{exp}$ & $\sigma_{th}$ \\
\hline & & & & &\\
145 &   180$\pm$22 &182.1    & 135 &   209.0$\pm$58 &195.6\\
185 &   130$\pm$17 &135.7    & 165 &   177.0$\pm$38 &157.4\\
210 &   118$\pm$16 &112.6    & 195 &   153.0$\pm$27 &125.9\\
230 &   101$\pm$12 & 97.1    & 225 &   111.0$\pm$18 &100.8\\
250 &    83$\pm$ 9 & 84.4    & 255 &    87.0$\pm$13 & 81.0\\
290 &    57$\pm$ 9 & 63.2    & 300 &    46.0$\pm$11 & 59.0\\
       & & & & &\\
\multicolumn{2}{c}{$\Sigma^+ p\rightarrow \Sigma^+ p$} & $\chi^{2}=4.7$ &
\multicolumn{2}{c}{$\Sigma^- p\rightarrow \Sigma^- p$} & $\chi^{2}=4.3$ \\
$p_{\Sigma^+}$ & $\sigma_{exp}$ & $\sigma_{th}$ &
$p_{\Sigma^-}$ & $\sigma_{exp}$ & $\sigma_{th}$ \\
\hline & & & & &\\
 145 & 123$\pm$62 & 104.3 & 142.5 & 152$\pm$38 & 133.7 \\
 155 & 104$\pm$30 &  94.3 & 147.5 & 146$\pm$30 & 128.9 \\
 165 &  92$\pm$18 & 85.4  & 152.5 & 142$\pm$25 & 124.4 \\
 175 &  81$\pm$12 & 77.4  & 157.5 & 164$\pm$32 & 120.0 \\
 400 &  75$\pm$25 & 26.6  & 162.5 & 138$\pm$19 & 115.9 \\
 500 &  26$\pm$20 & 24.9  & 167.5 & 113$\pm$16 & 111.9 \\
 650 &  52$\pm$40 & 21.9  &       &            &       \\
     & & & & &\\
\multicolumn{2}{c}{$\Sigma^- p\rightarrow \Sigma^0 n$} & $\chi^{2}=6.4$ &
\multicolumn{2}{c}{$\Sigma^- p\rightarrow \Lambda n$} & $\chi^{2}=4.4$ \\
$p_{\Sigma^+}$ & $\sigma_{exp}$ & $\sigma_{th}$ &
$p_{\Sigma^-}$ & $\sigma_{exp}$ & $\sigma_{th}$ \\
\hline & & & & &\\
110 & 396$\pm$91 & 183.2  &110& 174$\pm$47& 219.9 \\
120 & 159$\pm$43 & 160.0  &120& 178$\pm$39& 188.7 \\
130 & 157$\pm$34 & 141.1  &130& 140$\pm$28& 163.6 \\
140 & 125$\pm$25 & 125.5  &140& 164$\pm$25& 143.0 \\
150 & 111$\pm$19 & 112.3  &150& 147$\pm$19& 126.0 \\
160 & 115$\pm$16 & 101.1  &160& 124$\pm$14& 111.7 \\
       & & & & &\\
\multicolumn{3}{l}{$r_{R}^{exp}=0.468\pm 0.010$}  &
\multicolumn{2}{l}{$r_{R}^{th }=0.473$} & $\chi^{2}=0.2$ \\
%\hline\hline \\
\end{tabular}
\end{ruledtabular}
\end{center}
\label{tab.reslts1}
\end{table}
 
The $\chi^2$ on the 38 {\it Y\!N} scattering data for the ESC04     
model is given in Table~\ref{tab.reslts1}.                            
The capture ratio at rest, given in the last
column of the table, for its definition see e.g. \cite{RSY99}.
This capture ratio turns out to be rather constant in the momentum
range from 100 to 170 MeV/$c$. 
% update -------------------------------------------------------------
Obviously, for very low momenta the cross sections are almost completely dominated 
by $S$ waves. 
%For $p_\Sigma \rightarrow 0$ one has $r_F \rightarrow 0$. 
For a discussion of the capture ratio at rest $r_R$ see \cite{Swa62,Swa71,Fuj98}. 
We obtained $r_R = 0.473$, which is close to the experimental value 
$r_R^{exp} = 0.468 \pm 0.010$.
%The value of $r_C$ is rather sensitive to the $\Sigma^0-\Sigma^-$ mass difference.
% update -------------------------------------------------------------

The $\Sigma^+ p$ nuclear-bar phase shifts as a function of energy are given in
Table~\ref{spphas}. Notice that the $^3S_1$-phase shows repulsion, except for 
very low energies. This means that the the potential has a weak long range 
attractive tail.

The $\Lambda N$ nuclear-bar phase shifts as a function of energy are given in
Table~\ref{lpphas}. The $^3S_1$-phase shows that there is a resonance below
the $\Sigma N$-threshold, the so-called analogue of the deuteron. This signals
the fact that  in the $\Sigma N (^3S_1,I=1/2)$-state there is a strong attraction.

\begin{table}
\caption{ESC04 nuclear-bar $\Sigma^{+} p$ phases in degrees.}
\begin{ruledtabular}
\begin{tabular}{crrrrr} & & & & & \\ 
 $p_{\Sigma^{+}}$ & 200 & 400 & 600 & 800 & 1000 \\ \hline
     &    &     &     & &  \\
 $T_{\rm lab}$ & 16.7 & 65.5 & 142.8 & 244.0 & 364.5 \\ \hline
     &    &     &     & &  \\
 $^{1}S_{0}$ & 39.05 & 26.07  & 10.11&--4.50 &--17.46\\
 $^{3}S_{1}$ & 1.26  &--0.21  &--3.75 &--6.80 &--10.11\\
 $\epsilon_{1}$ &--3.38 &--4.54 &--2.89 &  0.57 & 3.82 \\
 $^{3}P_{0}$ & 5.91 & 10.76 & 3.87 &  -7.62& -19.84 \\
 $^{1}P_{1}$ & 4.62 & 21.50  & 35.55& 38.36& 35.06 \\
 $^{3}P_{1}$ &--3.28&--9.20 &--13.96&--17.52&--19.59\\
 $^{3}P_{2}$ & 1.29 &  7.61  & 14.89 &19.30 &20.58 \\
 $\epsilon_{2}$ & -0.44&-1.26  &-2.72 &-2.61 & -0.20 \\
 $^{3}D_{1}$ & 0.34 & 1.54  &  1.73 &-0.63  &  -5.70 \\
 $^{1}D_{2}$ & 0.36&  2.22 & 5.276 &  8.20 &  9.51\\
 $^{3}D_{2}$ & -0.53& -2.81 & -5.48 &-8.49  & -11.95 \\
 $^{3}D_{3}$ & 0.06 & 0.97  &  3.18 & 5.91  & 8.40 \\
     &    &     &     &  &   \\
%\hline\hline
\end{tabular}
\end{ruledtabular}
\label{spphas} 
\end{table}
 
\begin{table}
\caption{ESC04 nuclear-bar $\Lambda p$ phases in degrees.}
\begin{ruledtabular}
\begin{tabular}{crrrrrrr} & & & & & & & \\ 
%\hline\hline
 $p_{\Lambda}$ & 100 & 200 & 300 & 400 & 500 & 600 &633.4 \\
 \hline    &    &     &     & & & & \\
 $T_{\rm lab}$ & 4.5 & 17.8 & 39.6 & 69.5 & 106.9& 151.1& 167.3 \\ 
 \hline    &    &     &     & & & & \\
 $^{1}S_{0}$ & 22.30 & 29.20 & 27.49 & 22.59 & 16.68&10.62 & 8.65\\
 $^{3}S_{1}$ & 17.72 & 26.17 & 28.37 & 28.95 & 32.25&55.52&102.55 \\
 $\epsilon_{1}$ & 0.07& 0.30 & 0.48 & 0.25 & -1.18 & -8.43& 17.32 \\
 $^{3}P_{0}$ & 0.03 & 0.15  &0.06  &-0.79 &-2.74&-5.66 &-6.76 \\
 $^{1}P_{1}$ &-0.02 &-0.12 &-0.52&-1.45&-3.01&-5.09&-5.86\\
 $^{3}P_{1}$ & 0.03 & 0.13 & 0.14 &-0.17& -0.90&-1.93&-2.23 \\
 $^{3}P_{2}$ & 0.13&  0.89& 2.41 & 4.32 & 6.10 & 7.47 &  7.84\\
 $\epsilon_{2}$ & 0.00& 0.00 &-0.04 &-0.14 & -0.30& -0.52&-0.64\\
 $^{3}D_{1}$ & 0.00 & 0.02  & 0.12  & 0.40 & 1.04 & 3.11  & 2.19 \\
 $^{1}D_{2}$ & 0.00& 0.07 &0.24 & 0.74 & 1.54 &2.51 &2.84  \\
 $^{3}D_{2}$ & 0.00& 0.06 & 0.30 & 0.82 & 1.62& 2.53 & 2.82  \\
     &    &     &     &  & & & \\
%\hline\hline
\end{tabular}
\end{ruledtabular}
\label{lpphas} 
\end{table}

\begin{figure}[hbt]
% \resizebox{\textwidth}{!}    
  \resizebox{3.5cm}{!}       
  {\includegraphics[200,000][400,850]{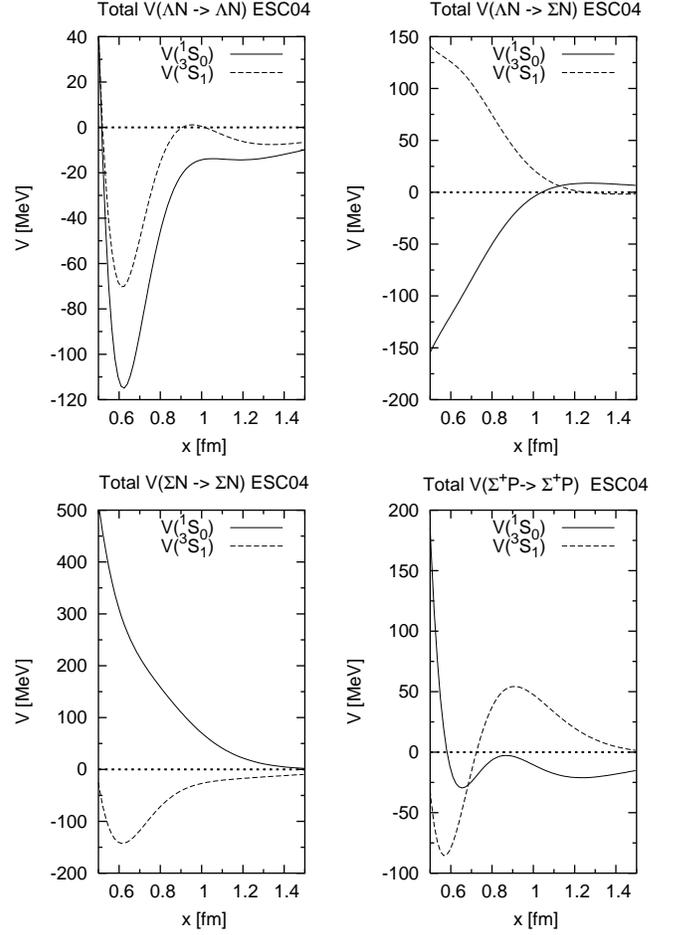}}
 \caption{Total potentials in the partial waves $^1S_0$ and $^3S_1$, for 
 $I=1/2$- and $I=3/2$-states.}
 \label{totfig1}
  \end{figure}

\begin{figure}[hbt]
% \resizebox{\textwidth}{!}    
  \resizebox{3.5cm}{!}       
  {\includegraphics[200,000][400,850]{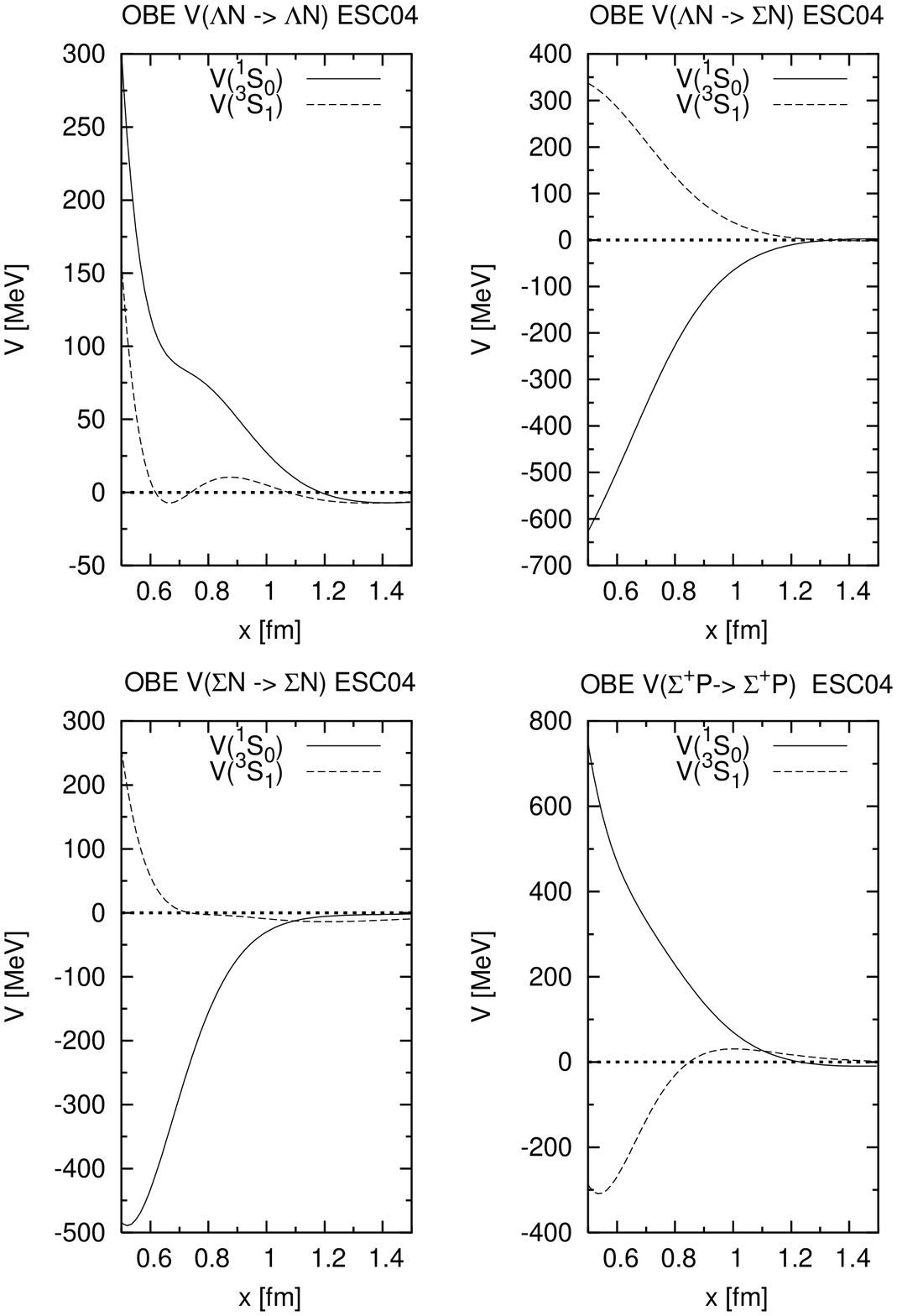}}
 \caption{OBE potentials in the partial waves $^1S_0$ and $^3S_1$, for 
 $I=1/2$- and $I=3/2$-states.}
 \label{obefig1}
  \end{figure}

\begin{figure}[hbt]
% \resizebox{\textwidth}{!}    
  \resizebox{3.5cm}{!}       
  {\includegraphics[200,000][400,850]{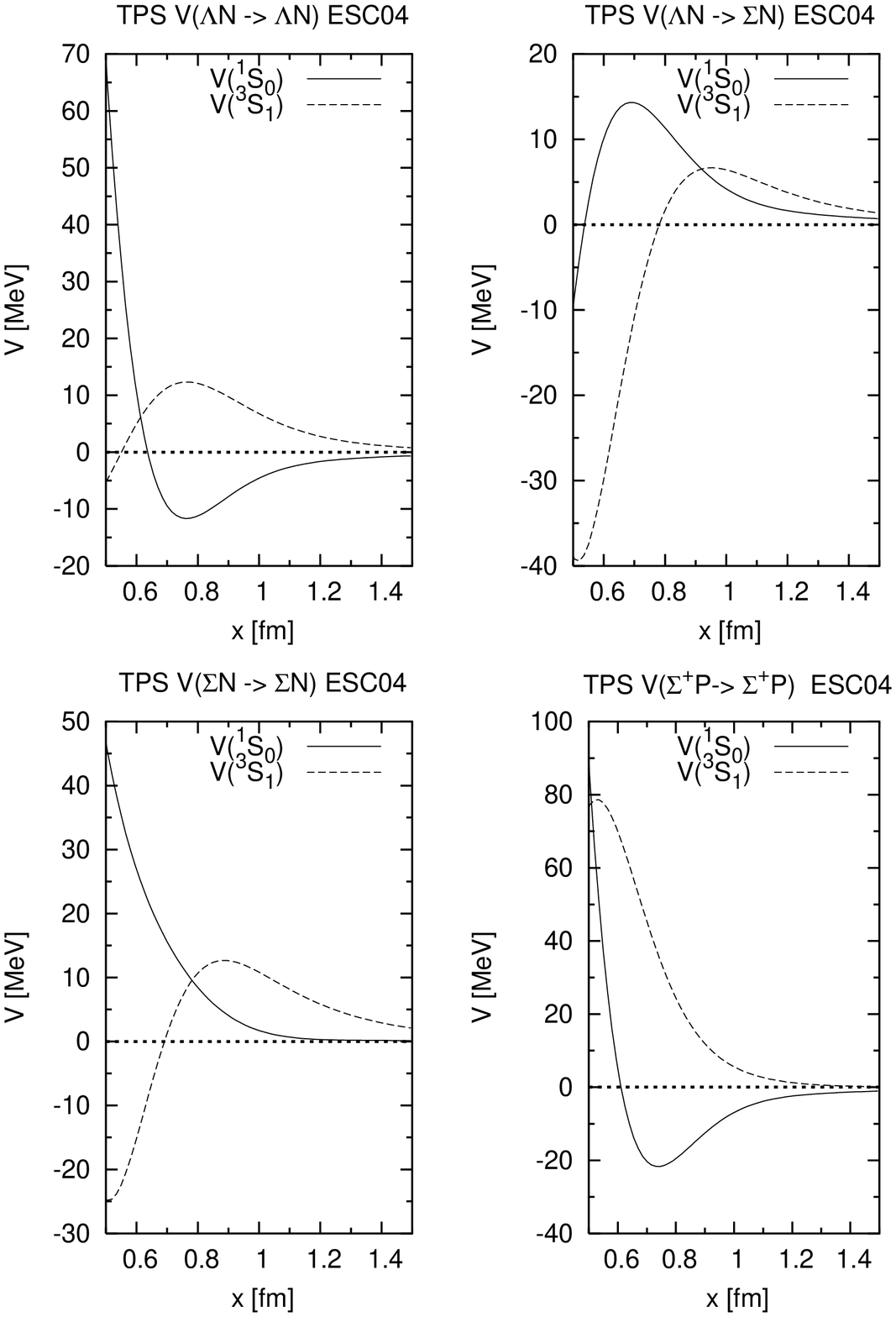}}
 \caption{TPS potentials in the partial waves $^1S_0$ and $^3S_1$, for 
 $I=1/2$- and $I=3/2$-states.}
 \label{tmefig1}
  \end{figure}

\begin{figure}[hbt]
% \resizebox{\textwidth}{!}    
  \resizebox{3.5cm}{!}       
  {\includegraphics[200,000][400,850]{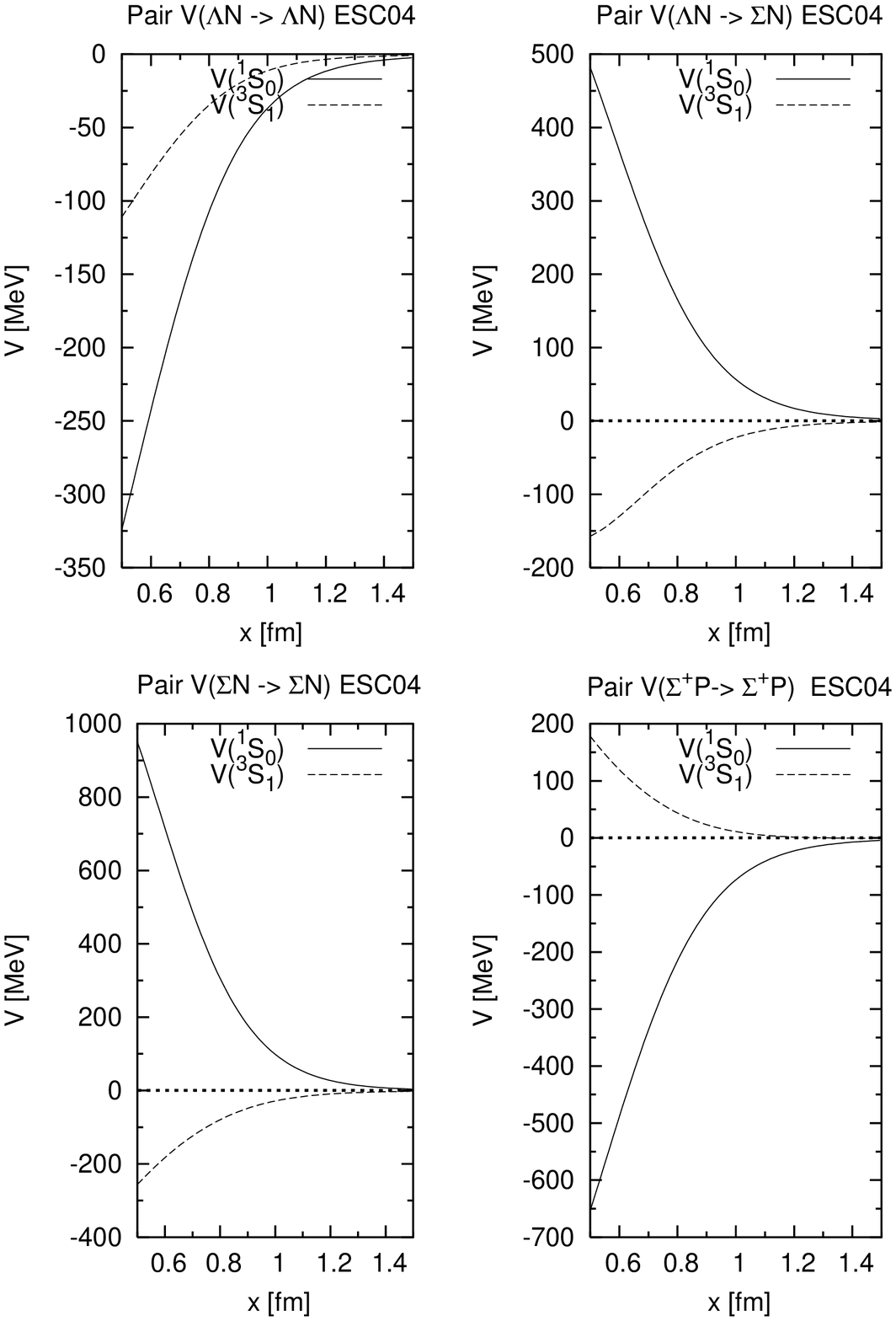}}
 \caption{Pair potentials in the partial waves $^1S_0$ and $^3S_1$, for 
 $I=1/2$- and $I=3/2$-states.}
 \label{pairfig1}
  \end{figure}

\begin{figure}[hbt]
% \resizebox{\textwidth}{!}    
  \resizebox{3.5cm}{!}       
  {\includegraphics[200,000][400,850]{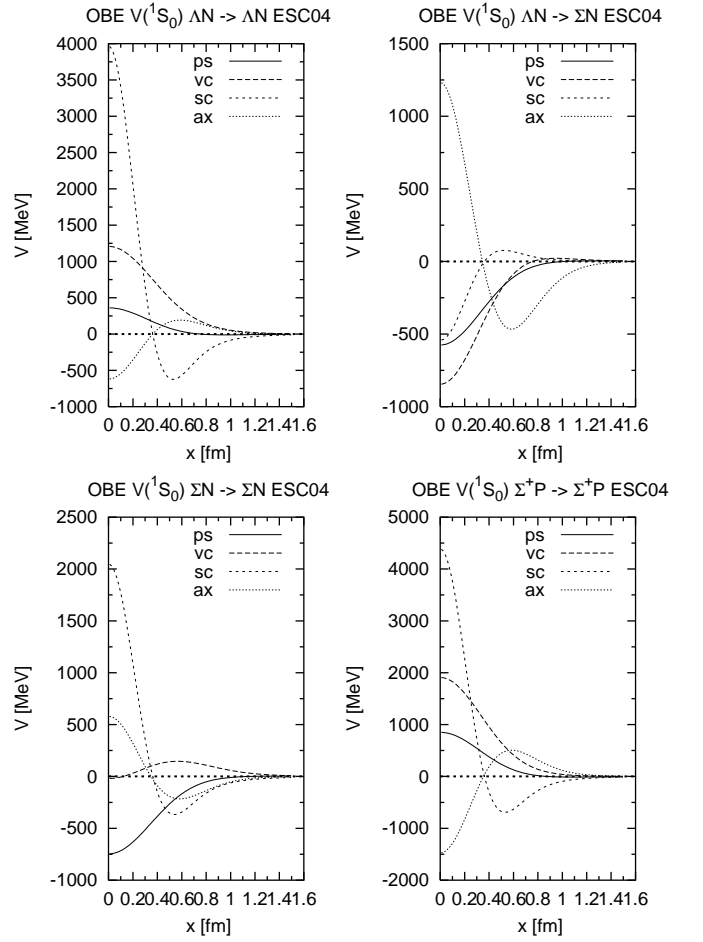}}
 \caption{OBE-potentials in the $^1S_0$ partial waves, for pseudoscalar (PS),
 vector (VC), scalar (SC), and axial-vector (AX) exchange, in the 
 $I=1/2$- and $I=3/2$-states.}
 \label{mesfig1}
  \end{figure}

\begin{figure}[hbt]
% \resizebox{\textwidth}{!}    
  \resizebox{3.5cm}{!}       
  {\includegraphics[200,000][400,850]{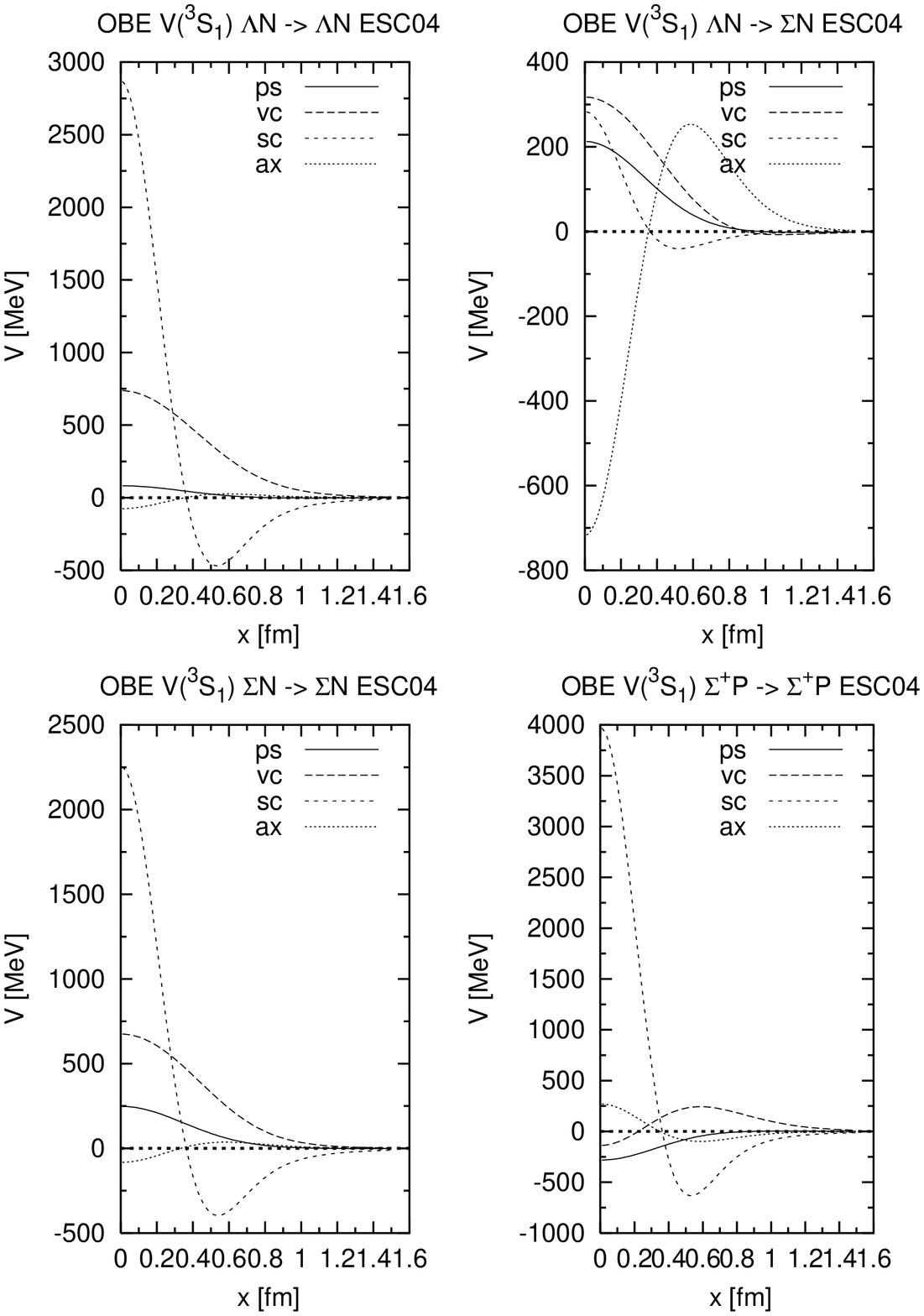}}
 \caption{OBE-potentials in the $^3S_1$ partial waves, for pseudoscalar (PS),
 vector (VC), scalar (SC), and axial-vector (AX) exchange, in the 
 $I=1/2$- and $I=3/2$-states.}
 \label{mesfig2}
  \end{figure}

%-----------------------------------------------------------------
 In Fig.~\ref{totfig1} we plot the total potentials for the S-wave channels
$\Lambda N \rightarrow \Lambda N$, $\Lambda N \rightarrow \Sigma N$,
and $\Sigma N  \rightarrow \Sigma N$. The same is done in Fig.~\ref{obefig1},
Fig.~\ref{tmefig1}, and Fig.~\ref{pairfig1} for respectively the OBE-, TME-,
and MPE-contributions.
In Fig.~\ref{mesfig1} and Fig.~\ref{mesfig2} we show for the same channels the
OBE-contributions from the different types of mesons: the pseudoscalar, the
vector, the scalar, and the axial-vector mesons.
% update ---------------------------------------------------------
From these figures one can notice e.g. (i) the total potentials are dominated
by the OBE- and MPE-contributions, (ii) the OBE- and MPE-potentials are often opposite 
to each other. For example, the $\Lambda N$ elastic potentials are attractive 
due to the sizeable attractive contributions from the MPE-potentials
overcoming the OBE ones.

Finally, all ESC-potentials described in this paper are available on 
the Internet \cite{ynonline}.
% update ---------------------------------------------------------

%-----------------------------------------------------------------
% potential figures plot04:

%-----------------------------------------------------------------
% \newpage
%\begin{widetext}
%\end{widetext}

%\begin{widetext}
%\end{widetext}

%\begin{widetext}
%\end{widetext}

%\begin{widetext}
%\end{widetext}

%\begin{widetext}
%\end{widetext}

%\begin{widetext}
\section{ Brief Comparison ESC04 Models}                                 
\label{sec:11} 
In this section we display some global comparison between the different
ESC04a-d models emerging from the different options, mentioned above.

In Table~\ref{tab.ESC04a-d.1} we give the FSB- and $a_{PV}$-parameters and 
the $\chi^2$ obtained in the simultaneous ${\it NN} \oplus {\it YN}$-fitting. 
\begin{table}[hbt]
\caption{$\Delta_{FSB}$-, $a_{PV}$-parameters, and the $\chi^2$'s for {\it NN} and {\it YN}.}        
\label{tab.ESC04a-d.1}
\begin{center}
\begin{ruledtabular}
 \begin{tabular}{cc|cccc|cc}
%\hline
       & & $a_{PV}$ & $\Delta_{FSB}(PV)$ & $\Delta_{FSB}(V)$ & &
 $\chi^2_{p.d.p.}(NN)$ & $\chi^2({\it YN})$ \\
 \hline
ESC04a & & 0.5 & -0.258 & -0.267 & & 1.22 & 24.2 \\
ESC04b & & 1.0 & -0.214 & -0.280 & & 1.20 & 49.5 \\
ESC04c & & 0.5 &  0.000 &  0.000 & & 1.28 & 23.0 \\
ESC04d & & 1.0 &  0.000 &  0.000 & & 1.33 & 26.0 \\
%\hline
 \end{tabular}
\end{ruledtabular}
\end{center}
\end{table}

%\noindent Here, a special remark on the treatment of the pseudoscalar couplings in the 
%$TME$- and $MPE$-potentials is to the point. 
%We have chosen the $a_{PV}$-parameter, 
%which regulates the mixture of pseudosclar and pseudo-vector coupling to be 1/2.
%This seems to fit the {\it NN}+{\it YN}-data better than purely pseudo-vector coupling ($a_{PV}=1$). 

%For the latter, we have typically $\chi^2_{{\it YN}} \approx 40-60$, compared to 
%$\chi^2_{{\it YN}}  \approx 24$ in the case $a_{PV}=1/2$. 
%On the other hand, the {\it NN}-fit gives $\chi^2_{\rm NN data} \approx 1.17$ for $a_{PV}=1$, which
%somewhat better as for $a_{PV}=1/2$.

In Table~\ref{tab.ESC04a-d.2} we give the $F/(F+D)$-ratio's,and $\theta_S$.                 
The latter because in these models there is no imposed constraint on the
parameters $(\alpha_S, \theta_S)$. The vector mixing angle $\theta_V$ is for all
models the same. This is also the case for the axial mixing angle where we fixed
$\theta_A = \theta_{PS} = -23.0^o$. In this table it is remarkable that 
whereas $\alpha_S$ is constant, there is a big difference w.r.t. $\theta_S$.
Furthermore, one notices that for most models $\alpha^m_V$ is close to 
the estimates from static and non-static SU(6) \cite{Sak65}.
As a final point we mention that the $F/(F+D)$-ratio's for the pair-couplings are very
similar to the values of ESC04b, given above.
\begin{table}[hbt]
\caption{$F/(F+D)$-ratio's for OBE-couplings, and the scalar-meson mixong 
 angle $\theta_S$ in degrees.}        
\label{tab.ESC04a-d.2}
\begin{center}
\begin{ruledtabular}
 \begin{tabular}{cc|cccccc}
%\hline
       & & $\alpha_{PV}$ & $\alpha^e_V$ & $\alpha^m_V$ & $\alpha_A$ & 
 $\alpha_S$ & $\theta_S$  \\  
 \hline
ESC04a & & 0.467 & 1.0 & 0.276 & 0.234 & 0.841 & 40.32  \\
ESC04b & & 0.403 & 1.0 & 0.316 & 0.246 & 0.841 & 40.31  \\
ESC04c & & 0.510 & 1.0 & 0.306 & 0.234 & 0.841 & 22.09  \\
ESC04d & & 0.499 & 1.0 & 0.430 & 0.234 & 0.841 & 11.45  \\
%\hline
 \end{tabular}
\end{ruledtabular}
\end{center}
\end{table}

In Table~\ref{tab.ESC04a-d.3} we list the $\Lambda N$ scattering lengths and          
effective ranges. Here, $(a_s,r_s)$ are these quantities for $\Lambda N(^1S_0)$ and
$(a_t,r_t)$ for $\Lambda N(^3S_1)$. Here we repeat the different options used to
distinguish the different models.
\begin{table}[hbt]
\caption{$\Lambda N $ scattering lengths and effective ranges in fm.}     
\label{tab.ESC04a-d.3}
\begin{center}
\begin{ruledtabular}
 \begin{tabular}{cc|ccc|cccc}
%\hline
       & & SFB & $a_{PV}$ & & $a_s$ & $a_t$ & $r_s$ & $r_t$ \\  
 \hline
ESC04a & & yes & 0.5 & & -2.073& -1.537 & 2.998  & 2.773 \\
ESC04b & & yes & 1.0 & & -1.957& -1.689 & 3.156  & 2.823 \\
ESC04c & & no  & 0.5 & & -1.946& -1.850 & 3.473  & 2.900 \\
ESC04d & & no  & 1.0 & & -1.941& -1.858 & 3.570  & 3.133 \\
%\hline
 \end{tabular}
\end{ruledtabular}
\end{center}
\end{table}
In Table~\ref{tab.ESC04a-d.4} we list the scattering lengths and          
effective ranges for $\Sigma^+p$ and $\Lambda \Lambda$.
\begin{table}[hbt]
\caption{$\Sigma^+p$ and $\Lambda\Lambda $ scattering lengths and effective ranges in fm.}
\label{tab.ESC04a-d.4}
\begin{center}
\begin{ruledtabular}
 \begin{tabular}{lc|ccccc|cc}
%\hline
       & & $a_s$ & $a_t$ & $r_s$ & $r_t$ & & $a_s(\Lambda\Lambda)$ & $r_s(\Lambda\Lambda)$ \\  
 \hline
ESC04a & &-4.09 & -0.020 & 3.49 & -3356  & & -1.149 & 4.482 \\
ESC04b & &-2.87 & +0.179 & 4.10 & -34.20 & & -1.245 & 4.453 \\
ESC04c & &-3.87 & +0.077 & 3.72 & -253.5 & & -1.081 & 4.463 \\
ESC04d & &-3.43 & +0.217 & 3.98 & -28.94 & & -1.323 & 4.401 \\
%\hline
 \end{tabular}
\end{ruledtabular}
\end{center}
\end{table}

%----------------------------------------------------------------------------------

%In Table~\ref{tab.ESC04a-d} we display some global comparison between the different
%ESC04-models emerging from the different options.
%\begin{table}[hbt]
%\caption{Comparison ESC04-models. $U_{\Lambda,\Sigma}$ in MeV at normal density.     
%}
%\label{tab.ESC04a-d}
%\begin{center}
% \begin{tabular}{|l|cc|cc|ccc|}
% \hline
%       & FSB & $a_{PV}$ & $\chi^2_{p.d.p.}({\it NN})$ & $\chi^2({\it YN})$ & $U_\Lambda$ & $U_{\Sigma}$ 
% & $U_\Xi$ \\
% \hline
%ESC04a & yes & 0.5 & 1.22  &  24.2  & --40   & --36 & +17.5 \\
%ESC04b & yes & 1.0 & 1.20  &  49.5  & --40   & --26 & $ >> 0 $\\
%ESC04c & no  & 0.5 & 1.28  &  23.0  & --46   & --36 &  --7.3 \\
%ESC04d & no  & 1.0 & 1.33  &  26.0  & --46   & --27 &  --21.4 \\
% \hline
% \end{tabular}
%\end{center}
%\end{table}

%-----------------------------------------------------------------
% NEW G-MATRIX BEGIN:
%-----------------------------------------------------------------

\section{G-matrix interactions and hypernuclei}
\label{sec:10} 

\subsection{Properties of $\Lambda N$ and $\Sigma N$ G-matrices}

The free-space {\it Y\!N} scattering data are too sparse
to discriminate clearly the {\it Y\!N} interaction models. 
Then, it is very helpful to test the interaction models
in analyses of various hypernuclear phenomena. 
Effective {\it Y\!N} interactions used in models of hypernuclei
can be derived from the free-space {\it Y\!N} interactions 
most conveniently using the G-matrix theory.
In the previous work~\cite{RSY99}, the G-matrix results were
used as an important guidance to discriminate especially 
the spin-dependent parts in the interaction models. 
Here, the versions a $\sim$ f of the NSC97 model were 
designed so as to specify their different strengths of 
$\Lambda N$ spin-spin interactions, and among them those of 
NSC97e and NSC97f were demonstrated to be consistent with 
the hypernuclear data. Afterwards, the plausibility of our
approach has been confirmed by successful calculations for 
$s$-shell hypernuclei~\cite{Hiya01}~\cite{Nogga02}~\cite{Nemu02} 
using NSC97e/f or their simulated versions.

Let us perform the G-matrix analyses for ESC04a-d in the same way.
The G-matrix equations for {\it Y\!N} pairs in nuclear matter are 
solved with the simple QTQ prescription (the gap choice) for 
the intermediate-state spectra, which means that no potential 
term is taken into account in the off-shell propagation.
As discussed for NSC97~\cite{RSY99},
the QTQ prescription is accurate enough to investigate 
properties of {\it Y\!N} G-matrix interactions. 
The nucleon energy spectra in the {\it Y\!N} G-matrix equation 
are obtained from the {\it N\!N} G-matrices for ESC04(N\!N), where 
the phenomenological three-nucleon interaction (TNI) is taken 
into account so as to assure nuclear saturation.
The details for TNI are explained in the next subsection.

In this work, the properties of the G-matrix interactions derived 
from ESC04a-d models are compared often with those of NSC97e/f.
The calculated values for NSC97e/f in this work are slightly 
different from those in \cite{RSY99} because of different choice 
of the nucleon spectra. 
Hereafter, a two-particle state with isospin ($T$), spin ($S$), 
orbital and total angular momenta ($L$ and $J$) is represented
as $^{2T+1,2S+1}L_J$. An isospin quantum number is often omitted,
when it is evident.

\begin{table}[ht]
\caption{Values of $U_\Lambda$ at normal density and 
partial wave contributions for ESC04a-d and NSC97e/f obtained 
from the G-matrix calculations with the QTQ intermediate
spectra. All entries are in MeV.}
\label{Gmat1}
\begin{center}
\begin{ruledtabular}
 \begin{tabular}{lc|cccccccc|c}
%\hline
& & $^1S_0$ & $^3S_1$ & $^1P_1$ & $^3P_0$ & $^3P_1$ & $^3P_2$ & $D$ & & 
$U_\Lambda$ \hspace{0mm} \\
 \hline
ESC04a & & --13.7 & --20.5 & 0.6 & 0.2 & 0.5 & --4.5 & --1.0 & & --38.5 \hspace{0mm}  \\
ESC04b & & --13.3 & --22.6 & 0.5 &--0.0& 0.6 & --4.3 & --1.1 & & --40.2 \hspace{0mm}  \\
ESC04c & & --13.9 & --28.5 & 2.9 &  0.0& 1.3 & --6.5 & --1.3 & & --46.0 \hspace{0mm}  \\
ESC04d & & --13.6 & --26.6 & 3.2 &--0.2& 0.9 & --6.4 & --1.4 & & --44.1 \hspace{0mm}  \\
\hline
NSC97e & & --12.7 & --25.5 & 2.1 & 0.5 & 3.2 & --1.3 & --1.2 & & --34.8 \hspace{0mm}  \\
NSC97f & & --14.3 & --22.4 & 2.4 & 0.5 & 4.0 & --0.7 & --1.2 & & --31.8 \hspace{0mm}  \\
%\hline
 \end{tabular}
\end{ruledtabular}
\end{center}
\end{table}

In Table~\ref{Gmat1} we show the potential energies $U_\Lambda$
for a zero-momentum $\Lambda$ and their partial-wave contributions
$U_\Lambda(^{2S+1}L_J)$ at normal density $\rho_0$ ($k_F$=1.35 fm$^{-1}$).
A statistical factor $(2J+1)$ is included in $U_\Lambda(^{2S+1}L_J)$.
The total contributions $U_\Lambda$ should be compared 
to the experimental value of about $-30$ MeV.
In an appearance, the values for ESC04a-d seem to be rather 
worse than those for NSC97e/f. It should be noted, however, 
that the shallower values of $U_\Lambda$ for NSC97e/f are 
owing to the strongly repulsive contributions of their 
$P$-state interactions. The sums of even-state contributions 
for ESC04a/b (ESC04c/d) are similar to (slightly larger than) 
those for NSC97e/f. One should notice here that the even-state 
strengths of NSC97e/f are proved to be attractive enough to 
reproduce appropriately $\Lambda$ binding energies in 
$s$-shell hypernuclei~\cite{Hiya01}~\cite{Nogga02}~\cite{Nemu02}.
Thus, we can say that the remarkable difference between ESC04a-d
and NSC97e/f appears in the $P$-state interactions: Those of 
ESC04a-d and NSC97e/f are attractive and repulsive, respectively.
If the attractive $P$-state interactions of ESC04a-d are considered
to be reasonable, one should take into account another repulsive 
contribution in order to reproduce the value of 
$U_\Lambda \sim -30$ MeV, as discussed later.
%%%%%%%%%%%%%%%%%%%%%%%%%%%%%%%%%%%%
Though there are no clear-cut data for $\Lambda N$ $P$-state 
interactions, an important consideration was given by Millener,
supporting attractive $P$-state interactions~\cite{Mil01}.
He claims that the attractive $P$-state interaction 
is consistent with the 6.0 MeV separation observed in the 
$(K^-,\pi^-)$ reaction for the two $({1/2}^-)$ states of 
$^{13}_\Lambda$C composed of the 
$^{12}$C$(0^+,2^+)\otimes p_\Lambda$ configurations.  

%As well known, the choice of a continuous intermediate-energy 
%spectrum (CIES) in the G-matrix equation leads to the considerable
%gain of $U_\Lambda$. In Table~\ref{Gmat1}, the values 
%in parentheses give the results obtained with the CIES choice.
%Comparing them with the corresponding values with the QTQ choice,
%the CIES treatment turns out to give the energy gains of
%2.9 MeV, 2.5 MeV, 3.2 MeV and 3.5 MeV, respectively, 
%for ESC04a, ESC04b, NSC97e and NSC97f.
%These values are more or less similar to each other, being
%far smaller than the value for NSC89 with the extremely strong 
%$\Lambda N$-$\Sigma N$ coupling part~\cite{Yam94}. 
%Namely, the $\Lambda N$-$\Sigma N$ coupling parts of ESC04a/b are
%moderate as well as those of NSC97e/f. This feature is also 
%reflected in the moderate value of the $\Sigma$ conversion width 
%$\Gamma_\Sigma$ in nuclear matter as shown later.

\begin{table}[ht]
\caption{Contributions to $U_\Lambda$ at normal density from 
spin-independent, spin-spin, $LS$ and tensor parts of the 
G-matrix interactions derived from ESC04a-d and NSC97e/f.
All entries are in MeV.}
\label{Gmat2}
\begin{center}
\begin{ruledtabular}
 \begin{tabular}{lc|cccccc}
%\hline
 & & $U_0(S)$ & $U_{\sigma\sigma}(S)$ & $U_0(P)$ & $U_{\sigma\sigma}(P)$ 
 & $U_{LS}(P)$ & $U_{T}(P)$  \\
 \hline
ESC04a & & --8.55 & 1.73  & --0.27 & --0.25 & --0.45 & 0.08  \\
ESC04b & & --8.96 & 1.44  & --0.27 & --0.22 & --0.41 & 0.10  \\
ESC04c & & --10.6 & 1.09  & --0.19 & --0.86 & --0.65 & 0.18  \\
ESC04d & & --10.1 & 1.19  & --0.20 & --0.96 & --0.58 & 0.17  \\
 \hline
NSC97e & & --9.55 & 1.06  &   0.38 & --0.44 & --0.46 & 0.17  \\
NSC97f & & --9.18 & 1.71  &   0.52 & --0.50 & --0.48 & 0.23  \\
%\hline
 \end{tabular}
\end{ruledtabular}
\end{center}
\end{table}

In order to see the spin-dependent features of the $\Lambda N$ G-matrix 
interactions more clearly, it is convenient to derive contributions
to $U_\Lambda$ from the spin-independent, spin-spin, $LS$, and 
tensor components of the G matrices, which are denoted as
$U_0$, $U_{\sigma\sigma}$, $U_{LS}$, $U_T$, respectively.
These quantities in $S$ and $P$ states can be transformed from values
of $U_\Lambda(^{2S+1}L_J)$ using Eq.(7.1) in Ref.~\cite{RSY99}.
The obtained values are shown in Table~\ref{Gmat2}.
The $S$-state contributions $U_0(S)$ for ESC04a-d are found to
be not remarkably different from those for NSC97e/f.
The relative ratio of $U_\Lambda(^1S_0)$ and $U_\Lambda(^3S_1)$
is related to the contribution $U_{\sigma\sigma}(S)$ from the
spin-spin interaction. 
Various analyses suggest that the reasonable strength of the 
$S$-state spin-spin interaction is between those of NSC97e/f.
Then, the spin-spin parts of ESC04a-d are found to be in this 
region, though they are slightly different from each other.

The features of the $P$-state interactions are indicated by
the values of $U_0(P)$, $U_{\sigma\sigma}(P)$, $U_{LS}(P)$
and $U_T(P)$ in Table~\ref{Gmat2}.
The negative (positive) values of $U_0(P)$ for ESC04a-d (NSC97e/f)
are due to the attractive (repulsive) interactions. 
The spin-spin, $LS$ and tensor strengths of 
ESC04a/b are slightly weaker than those of NSC97e/f.
On the other hand, the spin-spin and $LS$ strengths of
ESC04c/d are rather stronger than the others.
Let us discuss here the $LS$ parts more in detail,
because the clear data of the spin-orbit splittings have been
obtained in the $\gamma$-ray experiments. The values of 
$U_{LS}(P)$ are composed of the contributions of two-body 
$SLS$ interaction (attractive) and $ALS$ interaction (repulsive).
In order to compare clearly the $SLS$ and $ALS$ components,
it is convenient to derive the strengths of the 
$\Lambda$ $l$-$s$ potentials in hypernuclei. 
In the same way as in~\cite{RSY99},
we use the following expression derived with 
the Scheerbaum approximation~\cite{Sch76},
\begin{eqnarray}
   && U^{ls}_\Lambda(r) = K_\Lambda\, \frac{1}{r} \, \frac{d\rho}{dr}\,
      {\bf l}\cdot{\bf s} , \nonumber\\
   && K_\Lambda= -\frac{\pi}{3} (S_{SLS}+S_{ALS}),  \nonumber\\
   && S_{SLS,ALS} = \frac{3}{\bar q} \int^\infty_0\!\!\!\! r^3
                   j_1(\bar q r)\, G_{SLS,ALS}(r)\, dr
             \label{eq:so}
\end{eqnarray}
where $G_{SLS}(r)$ and $G_{ALS}(r)$ are $SLS$ and $ALS$ parts of
G-matrix interactions in configuration space, respectively, 
and $\rho(r)$ is a nuclear density distribution.
We take here $\bar q=0.7$ fm$^{-1}$.
Table~\ref{Gmat3} shows the values of $K_\Lambda$ and $S_{SLS,ALS}$
obtained from the $SLS$ and $ALS$ parts of the $\Lambda N$ G-matrix
interactions calculated at $k_F=1.0$ fm$^{-1}$ in the cases of
ESC04a-d and NSC97e/f.
It is found here that the obtained values for ESC04a/b 
are smaller than those for NSC97e/f,
because the $SLS$ ($ALS$) parts of the former are
less attractive (more repulsive) than those of the latter.
On the other hand, the spin-orbit strengths of
ESC04c/d are rather stronger than the others.
In comparison of the experimental data, 
even the smallest $K_\Lambda$ value in the case of 
% update --------------------------------------------------------
ESC04b is too large~\cite{Hiya00,Fuj04}. 
% update --------------------------------------------------------
\begin{table}[ht]
\begin{center}
\caption{Strengths of $\Lambda$ spin-orbit splittings for 
ESC04a-d and NSC97e/f.
See the text for the definitions of $K_\Lambda$ and $S_{SLS,ALS}$.}
\vskip 0.5cm
\begin{ruledtabular}
\begin{tabular}{lc|ccc}
%\hline
       & & $S_{SLS}$ & $S_{ALS}$ & $K_\Lambda$  \\
\hline 
 ESC04a & &  --24.9 &12.1 & 13.4\\
 ESC04b & &  --22.3 &13.2 &  9.5\\
 ESC04c & &  --36.6 &10.2 & 27.6\\
 ESC04d & &  --32.7 &10.1 & 23.6\\
 \hline
 NSC97e & &  --26.0 & 9.8 & 16.9\\
 NSC97f & &  --26.9 & 9.5 & 18.1\\
%\hline
\end{tabular}
\end{ruledtabular}
\label{Gmat3}
\end{center}
\end{table}

%%%%%%%%%%%%%% Fig.1 %%%%%%%%%%%%%%%
%-----------------------------------------------------------------
%\begin{widetext}
%\onecolumngrid
%\begin{center}
 \begin{figure}    
 \vspace{-5cm}
     \resizebox{10cm}{15cm}
% \resizebox{\textwidth}{!}
 {\includegraphics*[0.0in,0in][8.0in,10in]{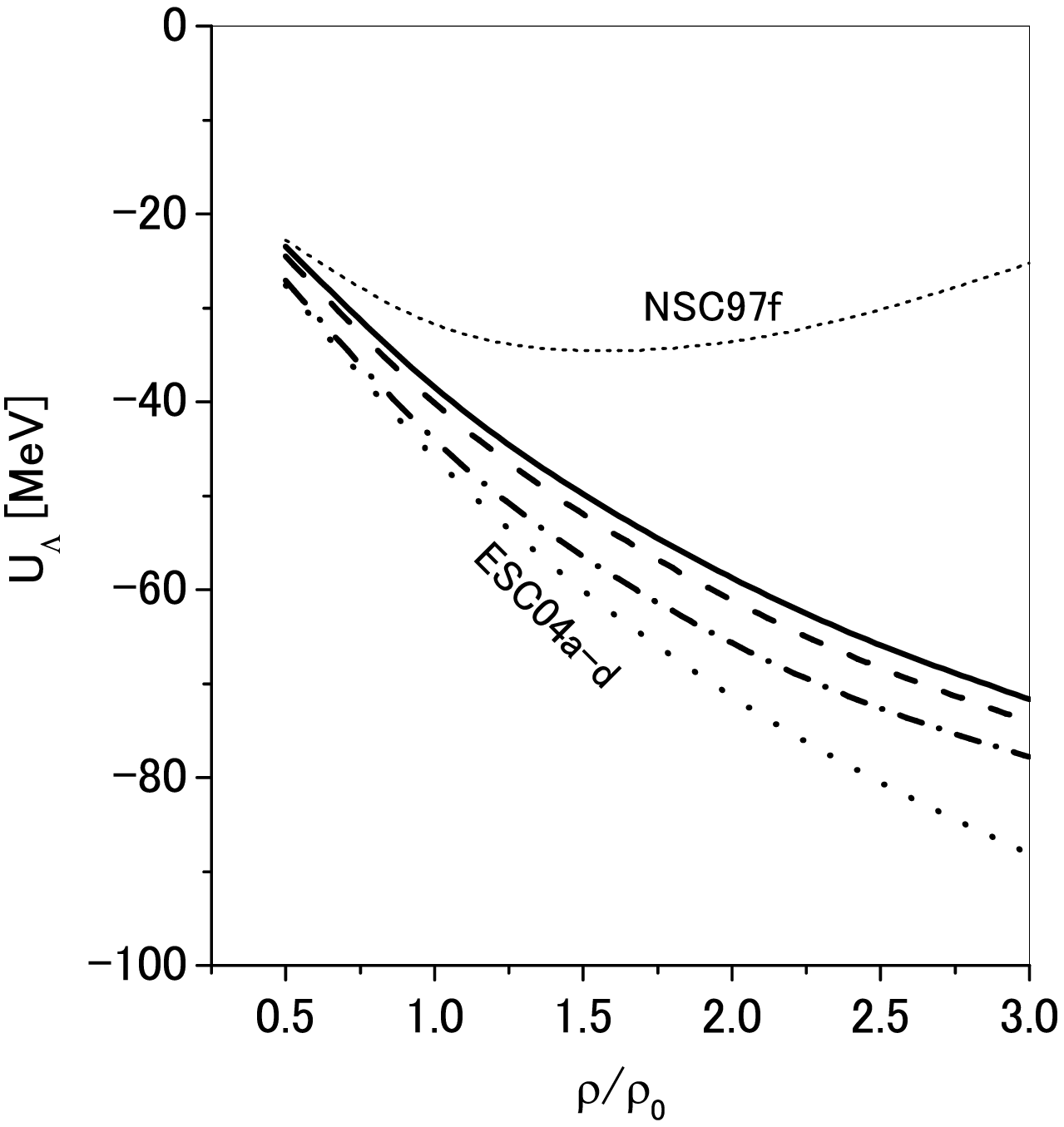}}
\caption{
Calculated values of $U_\Lambda$ as a function of $\rho/\rho_0$
for ESC04a (solid curve), ESC04b (dashed curve), ESC04c (dotted curve)
and ESC04d (dot-dashed curve). The thin dashed curve is for NSC97f.
}
\label{fig.Gfig1}
 \end{figure}
%\end{center}

%%%%%%%%%%%%%% Fig.2 %%%%%%%%%%%%%%%
%-----------------------------------------------------------------
%\onecolumngrid
 \begin{figure}   
 \vspace{-7.5cm}
 \resizebox{10cm}{15.25cm}
% \resizebox{!}{!}
 {\includegraphics*[0,0in][8in,10in]{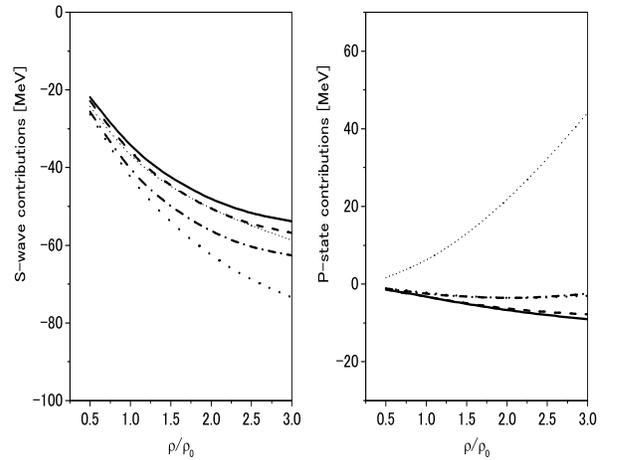}}
\caption{
$S$- and $P$-state contributions to $U_\Lambda$ are shown
in the left and right panels, respectively, as a function of 
$\rho/\rho_0$. The solid, dashed, dotted and dot-dashed curves
are for ESC04a-d, respectively. The thin dashed curve is for NSC97f.
}
\label{fig.Gfig2}
\end{figure}

%-----------------------------------------------------------------

In Fig.~\ref{fig.Gfig1}, 
the calculated values of $U_\Lambda$ are drawn as a function
of $\rho/\rho_0$ up to the high-density region. 
Their $S$- and $P$-contributions are shown in the left and
right sides of Fig.~\ref{fig.Gfig2}, respectively.
In these figures, solid, dashed, dotted and dot-dashed curves 
are for ESC04a-d, respectively. For comparison, 
the result for NSC97f is drawn by the thin dashed curve.
The $U_\Lambda$ values for ESC04a-d are found to become far more 
attractive with increase of density than those of NSC97f,
Comparing the partial-wave contributions for ESC04a-d
with those for NSC97f, we find that the $S$-state 
contributions are more or less similar to each other and 
the distinct difference comes from the $P$-state contributions. 
The difference between the $P$-state interactions in
ESC04 and NSC97 models turn out to be magnified dramatically
in the high-density region.

%%%%%%%%%%%%%% Fig.3 %%%%%%%%%%%%%%%
%-----------------------------------------------------------------
%\begin{widetext}
% \onecolumngrid        ! probeersel
 \begin{figure}   
 \vspace{-5cm}
 \resizebox{10cm}{15.25cm}
% \resizebox{!}{!}
%{\includegraphics*[0,0][8in,10in]{ynfig/Gfig3.EPS}}
 {\includegraphics*[0,0in][8in,10in]{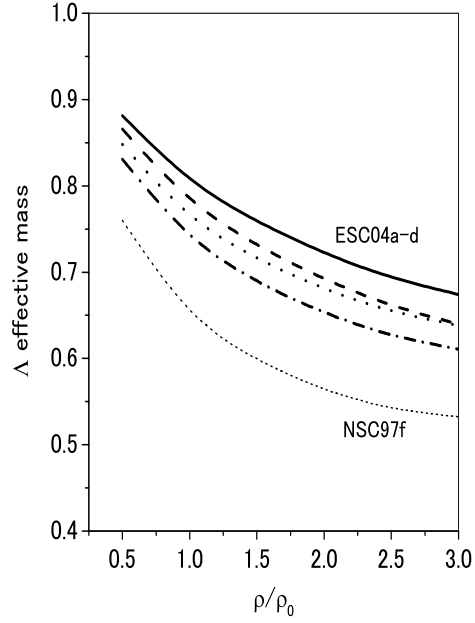}}
\caption{
Calculated values of $m^*_\Lambda$ as a function of $\rho/\rho_0$
for ESC04a (solid curve), ESC04b (dashed curve), ESC04c (dotted curve)
and ESC04d (dot-dashed curve). The thin dashed curve is for NSC97f.
}
\label{fig.Gfig3}
 \end{figure}
%-----------------------------------------------------------------

The $\Lambda$ effective mass $M^*_\Lambda$ in nuclear matter
is an important quantity which is related to the property
of the underlying $\Lambda N$ interaction. 
Here, we calculate a global effective mass defined by
\begin{eqnarray}
\frac{M^*_\Lambda}{M_\Lambda} = 
\left(1+\frac{d U_\Lambda}{d T_\Lambda}\right)^{-1} \ ,
\end{eqnarray}
where $T_\Lambda$ denotes $\Lambda$ kinetic energy. 
The calculated values of $m^*_\Lambda=M^*_\Lambda/M_\Lambda$
at normal density are 
0.81 (ESC04a), 0.79 (ESC04b), 0.77 (ESC04c), 0.74 (ESC04d),
0.67 (NSC97e) and 0.66 (NSC97f).
In Fig.~\ref{fig.Gfig3} the calculated values of 
$m^*_\Lambda$ are drawn as a function of $\rho/\rho_0$ by
solid (ESC04a), dashed (ESC04b), dotted (ESC04c) and 
dot-dashed (ESC04d) curves. The thin dashed curve is for NSC97f.
We find here that the calculated values for NSC97f are 
distinctively smaller than the values for ESC04a-d.
The reason why the $m^*_\Lambda$ values for NSC97e/f are small
is because their repulsive $P$-state interactions contribute
to the derivatives $d U_\Lambda/d T_\Lambda$
as large positive quantities.
%Namely, the difference of $m^*_\Lambda$ between ESC04 and NSC97
%models is originated mainly from their different $P$-state interactions.
In Ref.\cite{Yam00}, one of authors (Y.Y.) and collaborators
analyzed the measured $\Lambda$ energy spectra in heavy hypernuclei
with special attention to $\Lambda$ effective masses. They concluded 
that the small value of $m^*_\Lambda$ obtained from NSC97f leads to 
too broad level distances, and the adequate value of $m^*_\Lambda$ 
is around 0.8 at normal density. Thus, the $m^*_\Lambda$ values for 
ESC04a-d turn out to be more reasonable than those for NSC97 models.

%\begin{widetext}
% \onecolumngrid
\begin{table}[ht]
\caption{Values of $U_\Sigma$ at normal density and partial 
wave contributions for ESC04a-d and NSC97f (in MeV).}
\label{Gmat4}
\begin{center}
\begin{ruledtabular}
 \begin{tabular}{cc|cc|rrrrrrrc|r}
%\hline
      & & T  & & $^1S_0$ & $^3S_1$ & $^1P_1$ & $^3P_0$ & $^3P_1$ & $^3P_2$ 
& $D$ && $U_\Sigma$ \hspace{0mm} \\
 \hline
ESC04a& & $1/2$ & & 11.6 & --26.9 & 2.4 & 2.7 & --6.4 & --2.0 & --0.8 && \\
&&$3/2$ & &  --11.3& 2.6 & --6.8 & --2.3 & 5.9 & --5.1 & --0.2 && --36.5  \\
 \hline
ESC04b& & $1/2$ & &  9.6 & --25.3 & 1.8 & 1.6 & --5.4 & --2.1 & --0.7 && \\
&&$3/2$ & & --9.6 & 9.9 & --5.5 & --1.9 & 5.4 & --4.6 & --0.2 && --27.1  \\
 \hline
ESC04c& & $1/2$ &&  6.4 & --20.6 & 2.4 & 2.9 & --6.7 & --1.6 & --0.9 && \\
&&$3/2$ & & --10.7& 6.9 & --8.8 & --2.6 & 6.0 & --5.8 & --0.2 && --33.2  \\
 \hline
ESC04d& & $1/2$ &&  6.5 & --21.0 & 2.6 & 2.4 & --6.7 & --1.7 & --0.9 && \\
&&$3/2$ & & --10.1&14.0 & --8.5 & --2.6 & 5.9 & --5.7 & --0.2 && --26.0 \\
 \hline
NSC97f& & $1/2$ &&  14.9& --8.3  & 2.1 & 2.5 &--4.6 & 0.5 & --0.5 && \\
&&$3/2$ & & --12.4 &--4.1 &--4.1& --2.1 & 6.0 &--2.8 &--0.1 && --12.9 \\
%\hline
 \end{tabular}
\end{ruledtabular}
\end{center}
\end{table}

Next, let us show the properties of $\Sigma N$ G-matrix 
interactions. We solve here the $\Sigma N$ starting channel 
G-matrix equation in the QTQ prescription. 
%In this treatment, 
%there appears no imaginary part due to an energy-conserving 
%$\Sigma N$-$\Lambda N$ transition: The $\Sigma$ conversion width 
%is out of our consideration in this work.
% update ------------------------------------------------------
In this treatment, there appears no imaginary part, due to an energy-conserving 
$\Sigma N$-$\Lambda N$ transition. Although it is possible to derive the 
$\Sigma$ conversion width $\Gamma_\Sigma$ by taking into account the $\Lambda$
and $N$ potentials in the intermediate states, we choose not to discuss this
rather complex issue in this paper.
% update ------------------------------------------------------
In Table~\ref{Gmat4} we show the calculated values of $U_\Sigma$
at normal density for ESC04a-d and NSC97f.
Here, the $U_\Sigma$ values for ESC04a-d are found to be 
far more attractive than that for NSC97f, because the 
$^3S_1(\Sigma N,I=1/2)$ ($^1S_0(\Sigma N,I=1/2)$) 
%%$^{1/2,3}S_1$ ($^{1/2,1}S_0$) 
contributions for the former are remarkably more 
attractive (less repulsive) than that for the latter.

It has been pointed out that the $\Sigma$-wells in nuclei
might actually be repulsive, based on the $\Sigma$-atomic 
data~\cite{BFG94} and the quasi-free spectrum 
of $(K^-,\pi\-)$ reaction~\cite{Dab99}.
Recently, the $(\pi^-,K^-)$ experiment has been performed
in order to study the $\Sigma$-nucleus potentials~\cite{Noumi}.
They demonstrated that the observed spectrum can be reproduced
with a strongly repulsive potential. The theoretical analyses
for this data also indicate that the $\Sigma$-nucleus 
potential most likely is repulsive~\cite{Kohno04}.
If we consider these analyses seriously,
it is rather problematic how to understand repulsive 
$\Sigma$-nucleus potentials on the basis of the ESC model.
It should be noted, however, that there is no decisive evidence
for the repulsive $\Sigma$-nucleus potential experimentally 
in the present.

%%%%%%%%%%%%%%%%%%%%%%%%%%%%%%%%%%%%%%%%%%%%%%%%%%%%%%%%%%%%
\subsection{Three-body and nuclear medium effects}

\begin{figure}   
  {\includegraphics[1.7in,6.0in][ 9.5in, 8.5in]{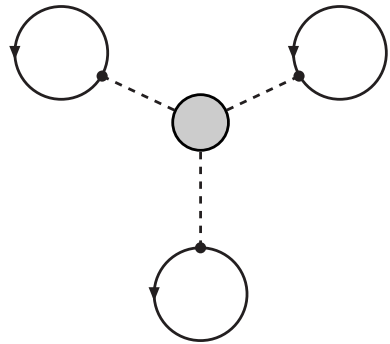}}
 \caption{Triple-meson three-body interaction.}
  \label{fig.3bf1} 
  \end{figure}

A natural possibility is the presence of three-body forces (3BF) 
in hypernuclei generating effective two-body forces, which could 
(partially) solve this well-depth issue. 
Since a thorough investigation is outside the scope of this paper, 
we discuss three-body and nuclear medium effects here in a simple 
phenomenological way. As for example discussed in \cite{Jack83}, 
three-body effects in a nuclear medium could be described roughly 
by using effective triple-meson vertices, like in Fig.~\ref{fig.3bf1}. 
Here, the meson lines could be e.g. scalar-, vector-, 
pomeron-exchanges, etc. In view of the big cancellations in the 
two-body case for $\omega + P + \epsilon$-potentials, 
one expects also similar cancellations to take place in Fig.~\ref{fig.3bf1}. 
One also expects that the density dependent corrections in the nuclear 
medium give intermediate range (weak) attraction, and short range repulsion. 
In this short and simple discussion of the possible implications, 
we only consider the repulsive component. Fig.~\ref{fig.3bf1} could be 
viewed upon as the exchange of a meson between two-nucleons, 
while it is scattered intermediately by a third one. Then, 
it is natural to describe such an effect by a change in the propagator,
i.e. by a change of the mass. Here, we analyze the effects by taking 
into account the change of the vector-meson masses using the form 
\begin{equation}
% m_V(\rho) = m_V\left( 1 -\alpha \rho + \beta \rho^{5/3}\right)\ .
m_V(\rho) = m_V\ \exp(-\alpha_V \rho)\ ,
\label{eq:10.1} \end{equation}
where $\alpha_V$ is treated as an adjustable parameter.

%%%%%%%%%%%%%% Fig.3bf %%%%%%%%%%%%%%%
%---------------------------------------------------------------------
%---------------------------------------------------------------------

On the basis of the SU(3) properties of the ESC model, the changes of
vector-meson masses in a nuclear medium induce the density-dependent 
effective repulsions in a rather universal manner in {\it N\!N}, {\it Y\!N} and 
{\it Y\!Y} channels. Then, our first step is to investigate this effect in 
usual nuclear matter. Since for the scalar exchanges $\epsilon +P$ 
we expect big cancellations, also in the many-body case, we here for 
simplicity only change the vector-meson masses for an analysis of 
the sensitivity of e.g. the well-depths w.r.t. medium effects.

For convenience, our medium-induced effects are handled in comparison
with the three-nucleon interaction (TNI) introduced by 
Lagaris-Pandharipande~\cite{LaPa81}, which is represented in simple 
forms of density-dependent two-body interactions. Here, we refer 
their parameter sets TNI2 and TNI3, reproducing nuclear 
incompressibility 250 MeV and 300 MeV, respectively. Their TNI is 
composed of the attractive part (TNA) and the repulsive part (TNR).
Our modeling for the repulsive component through the change of 
vector-meson masses corresponds only to their TNR. Hereafter,
TNR (TNA) parts of TNI2 and TNI3 are denoted as TNR2 and TNR3 
(TNA2 and TNA3), respectively.

%%%%%%%%%%%%%% Fig.4 %%%%%%%%%%%%%%%
%-----------------------------------------------------------------
 \begin{figure}   
\vspace{-7.5cm}
\resizebox{10cm}{15.25cm}
% \resizebox{!}{!}
 {\includegraphics*[0.0in,0in][8.0in,10in]{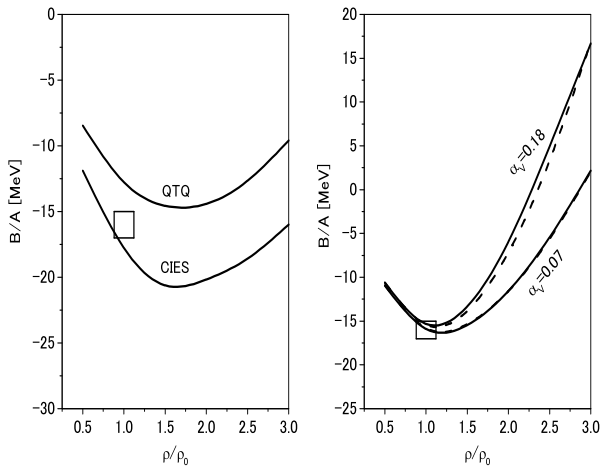}}
\caption{
Binding energy per nucleon $B/A$ in symmetric nuclear matter 
as a function of $\rho/\rho_0$ (saturation curve).
The box represents the area in which saturation occurs empirically.
In the left panel, the two solid curves show the G-matrix 
results for ESC04(N\!N) with the QTQ and CIES prescriptions.
In the right panel, the dashed curves are obtained by adding 
the TNI contributions on the QTQ result, and the solid curves 
are for the medium-corrected versions ESC04$^*$(N\!N).
See the text for the detail.
}
\label{fig.Gfig4}
 \end{figure}
%-----------------------------------------------------------------

In the left panel of Fig.~\ref{fig.Gfig4}, we show the saturation
curves of symmetric nuclear matter, namely binding energy per 
nucleon as a function of $\rho$, which are obtained from the 
G-matrix calculations with ESC04(N\!N). 
Here, the upper curve denoted as ``QTQ'' is calculated with the
QTQ prescription. The lower one denoted as ``CIES'' is obtained 
with the choice of a continuous intermediate-energy spectrum in 
the G-matrix equation. The CIES result is known to simulate well
the result including the three-hole line contributions~\cite{Baldo}.
In the following procedure, however, we use the QTQ result
because our G-matrix analyses for hypernuclear systems are
based on the QTQ prescription in this paper.
The box in the figure show the area where nuclear saturation is 
expected to occur empirically, and the energy minimums of both 
curves of ``QTQ'' and ``CIES'' are found to deviate from this area.  
In order to realize the nuclear saturation, three-body effects
should be added on the contributions of ESC04(N\!N) in the same
way as the cases of using the other {\it N\!N} interaction models:
Here, we use the above mentioned TNI. The dashed curves in the 
right panel of Fig.~\ref{fig.Gfig4} are obtained by adding the TNI
contributions on the (QTQ) G-matrix results, where the reduction 
factor 0.8 is multiplied on the TNA part so as to give the energy 
minimum at an adequate value of $-15\ \sim\ -16$ MeV. The two curves
in the figure correspond to the cases of adopting TNI2 and TNI3. 
Then, the saturation condition is found to be satisfied nicely.
Hereafter, when we use the TNI together with ESC04(N\!N),
the factor 0.8 is always multiplied on the TNA part.
In addition, the nucleon energy spectra obtained in the case of 
adopting TNI2 are adopted in the {\it Y\!N} G-matrix equations 
in this work.

Next, we perform the G-matrix calculations with ESC04(N\!N) 
in which the vector-meson masses are changed according to 
(\ref{eq:10.1}). Hereafter, the medium-corrected versions
of ESC04 are denoted as ESC04$^*(\alpha_V)$ including the 
parameter $\alpha_V$. In the right side of Fig.~\ref{fig.Gfig4},
the two solid curves are obtained by adding the contributions of 
TNA2 and TNA3 multiplied by 0.8 to the G-matrix results.
It should be noted here that the TNR parts are switched off
because they are substituted by our medium-induced repulsions.
Namely, the TNA parts are used here as phenomenological 
substitutes for the three-body attractive effects 
which are out of our present scope.
The parameter $\alpha_V$ in (\ref{eq:10.1}) is chosen 
so as to simulate the TNR contributions:
The two solid curves in the figure are obtained by choosing 
$\alpha_V=0.07$ fm$^3$ and $\alpha_V=0.18$ fm$^3$,
which turn out to be quite similar to the dashed curves
obtained by adding the TNI2 and TNI3 contributions, 
respectively, on the original G-matrix result.
Thus, it turns out that the density dependence of our 
medium-induced repulsion is very similar to that of TNR.
Although this similarity is of no fundamental meaning,
it is nicely demonstrated that our medium-induced repulsion 
plays the same role as TNR for nuclear saturation.

\begin{table*}[ht]
\caption{Calculated values of $U_\Lambda$ and $U_\Sigma$ 
at normal density for ESC04a-d$^*(\alpha_V=0.18)$. 
Their $S$-state contributions are also given.
All entries are in MeV. }
\label{Gmat5}
\begin{center}
\begin{ruledtabular}
 \begin{tabular}{lc||ccc|cc||ccccc|c}
%\hline
& & $^{1/2,1}S_0$ & $^{1/2,3}S_1$ & & $U_\Lambda$ &  
& $^{1/2,1}S_0$ & $^{1/2,3}S_1$ 
& $^{3/2,1}S_0$ & $^{3/2,3}S_1$ & & $U_\Sigma$  \\
 \hline
ESC04a$^*$   & & --12.0 & --15.8 & & --30.6
             & &   12.2 & --26.4 & --11.0&  8.6 & & --27.9  \\
ESC04b$^*$   & & --11.6 & --18.5 & & --33.0
             & &   10.1 & --25.5 & --9.0 & 15.2 & & --19.7  \\
ESC04c$^*$   & & --12.3 & --25.1 & & --39.3
             & &    7.9 & --20.0 & --10.3& 12.7 & & --23.6  \\
ESC04d$^*$   & & --12.0 & --23.0 & & --37.2
             & &    8.3 & --20.3 & --9.6 & 19.1 & & --16.6  \\
%\hline
 \end{tabular}
\end{ruledtabular}
\end{center}
\end{table*}

Let us study the effects of the medium corrections in the {\it Y\!N}
sectors of the ESC models. 
Then, a prospective way is to perform calculations for the values 
of $\alpha_V=0.07$ fm$^3$ and $\alpha_V=0.18$ fm$^3$ which
induce repulsions similar to TNR2 and TNR3, respectively, 
in nucleon matter. In the following analysis, we investigate
mainly the case of $\alpha_V=0.18$ fm$^3$.
% 
%The medium-corrected versions of ESC04a-d are denoted
%as ESC04a-d$^*(\alpha_V)$.
In Table~\ref{Gmat5}, the calculated values of $U_\Lambda$ 
and $U_\Sigma$ at normal density and their $S$-state contributions 
are shown in the case of taking $\alpha_V=0.18$ fm$^3$.
Comparing these values with those in Table~\ref{Gmat1}
and Table~\ref{Gmat4}, we find that the repulsive contributions
are substantial both for $U_\Lambda$ and $U_\Sigma$.
In the case of $U_\Lambda$, the $U_\Lambda$ values for
ESC04a-d are too attractive in comparison with the empirical 
indication of $U_\Lambda \sim -30$ MeV. These overbinding values
turn out to be improved substantially by our medium-induced repulsion.
Especially, the values for ESC04a/b$^*(\alpha_V)$ are noted to
agree well with the above empirical value.
Similar repulsive contributions are seen also in the case
of $U_\Sigma$, though the resulting values are still negative. 
However, it is important that the repulsive contribution is 
large in the $^{3}S_1(I=3/2)$ state, as discussed later.

It should be emphasized here that the spin-dependent features 
of the $\Lambda N$ G-matrix interaction are not really affected by 
our medium-induced repulsion. For instance, the values 
of $U_{\sigma \sigma}(S)$ become small only by 0.05 MeV (ESC04a),
0.09 MeV (ESC04b), 0.10 MeV (ESC04c) and 0.11 MeV (ESC04d)
in the case of taking $\alpha_V=0.18$ fm$^3$.
%%%%%%%%%%%%%%%%%%%%%%%%%%%%
%(This is related to the fact that the spin-spin potentials for the
%vector mesons have volume integral zero. This as also the case 
%for the pseudoscalar mesons, which is a reason for neglecting  
%the medium effects for those mesons here.)
%%%%%%%%%%%%%%%> confusing for readers???
In the cases of the $P$-state contributions such as 
$U_{\sigma\sigma}(P)$,  $U_{LS}(P)$ and $U_{T}(P)$,
the changes are negligibly small.
The change of the effective mass $m^*_\Lambda$ is also small:
The $m^*_\Lambda$ values for ESC04a-d$^*(\alpha=0.18)$ are 
smaller by only $0.01 \sim 0.02$ than those for ESC04a-d. 
These facts suggest interesting possibilities of using 
ESC04a-d$^*(\alpha_V)$ in various spectroscopic studies of 
$\Lambda$ hypernuclei, where the parameter $\alpha_V$ can be 
adjusted so as to reproduce experimental values of $B_\Lambda$
with almost no influence on spin-dependent structures of 
$\Lambda$ hypernuclei.
Then, we stress that the meaning of the above choice 
$\alpha_V=0.18$ fm$^3$ is only for its similarity to TNR3.

%%%%%%%%%%%%%% Fig.5 %%%%%%%%%%%%%%%
%-----------------------------------------------------------------
 \begin{figure}   
 \vspace{-5cm}
 \resizebox{10cm}{15.25cm}
% \resizebox{!}{!}
 {\includegraphics*[0,0in][8in,10in]{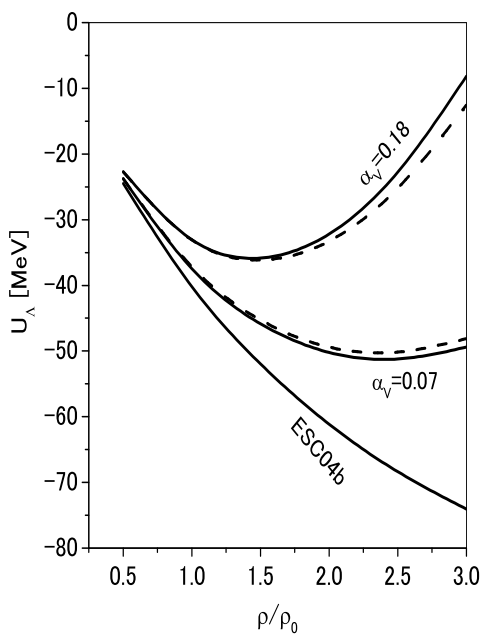}}
\caption{
Calculated values of $U_\Lambda$ as a function of $\rho/\rho_0$.
The three solid curves are for ESC04b, ESC04b$^*(\alpha_V=0.07)$
and ESC04b$^*(\alpha_V=0.18)$. The two dashed curves are 
obtained by adding the TNR2 and TNR3 contributions on
the result for ESC04b. See the text for the detail.
}
\label{fig.Gfig5}
 \end{figure}
%-----------------------------------------------------------------

Our medium-induced repulsions are related intimately to the 
problem of maximum masses of neutron stars. As well known, 
the repulsive three-body force in high-density neutron matter, 
embodied in TNR, plays an essential role for a stiffening of 
the EOS of neutron-star matter, assuring the observed maximum 
mass of neutron stars. However, the hyperon mixing in neutron-star
matter brings about the remarkable softening of the EOS, 
which cancels the effect of the repulsive three-body force
for the maximum mass.
In order to avoid this serious problem, Nishizaki, Takatsuka 
and one of the authors (Y.Y.) \cite{TNY02,NYT01} introduced the conjecture that 
the TNR-type repulsions work universally for ${\it YN}$ and ${\it YY}$ 
as well as for $N\!N$. 
They showed that the role of the TNR for stiffening the EOS
can be recovered clearly by this assumption. 
Our model of the medium-induced repulsion explains their 
assumption quite naturally. In Fig.~\ref{fig.Gfig5}, 
we draw the values of $U_\Lambda$ as a function of $\rho/\rho_0$
in some cases:
The three solid curve are for ESC04b and ESC04b$^*(\alpha_V=0.07)$
and ESC04b$^*(\alpha_V=0.18)$, and the two dashed curves are 
obtained by adding the TNR2 and TNR3 contributions on
the result for ESC04b. 
It is found, here, our medium-induced repulsions for
$\alpha_V=0.07$ and $\alpha_V=0.18$ are very similar to
the TNR2 and TNR3 contributions, respectively,
as well as the case of nuclear saturation curves.
Thus, it is clearly demonstrated that our medium-induced
repulsions, which works universally among octet baryons,
will assure the stiffening of the EOS.

In neutron-star matter, the chemical equilibrium condition for 
$\Sigma^-$ given by $\mu_{\Sigma^-}=\mu_n+\mu_{e^-}$ makes 
$\Sigma^-$ mixing more favorable than $\Lambda$ mixing controlled
by $\mu_\Lambda=\mu_n$ in cases of neglecting strong interactions.
Then, it is an important problem whether the well depth of 
$\Sigma^-$ is attractive or repulsive in neutron matter.
As shown in Table~\ref{Gmat5}, our medium-induced repulsion for 
$\Sigma N$ contributes dominantly in the $^{3}S_1(I=3/2)$-state
with the largest statistical weight. Thus, this repulsive effect 
appears most strongly in the $\Sigma^-$ well depth in neutron 
matter given by the $I=3/2$ $\Sigma N$ interaction.

In our analysis for hypernuclear systems, we do not consider 
the three-body attraction, such as TNA, which plays an important
role for nuclear saturation as well as the three-body repulsion 
such as TNR and our medium-induced effect.
The origin of such a part is considered to be in meson-exchange 
three-body correlations, being initiated by Fujita-Miyazawa~\cite{FM}. 
Possible counterparts in our hyperonic matter will be 
studied in future.
%%%%%%%%%%> Tom!  Is this description is OK???   
%%%%%%%%%%>       in CONCLUSION AND DISCUSSION ???

\subsection{Double-$\Lambda$ states}

Here, we study the $\Lambda \Lambda$ $(^{11}S_0)$ interactions,
for which the experimental information can be obtained
from the data of double-$\Lambda$ hypernuclei.
In the past, NHC-D~\cite{NRS77} has been used popularly as 
a standard meson-theoretical model for $S=-2$ interactions. 
The reason was because this interaction is compatible with strong 
$\Lambda \Lambda$ attraction ($\Delta B_{\Lambda \Lambda}=4 \sim 5$
MeV) supported by earlier data on double-$\Lambda$ hypernuclei. 
This strong $\Lambda \Lambda$ attraction of NHC-D is due to its 
specific feature that only the scalar singlet meson is taken 
into account. Since the discovery of NAGARA event identified 
uniquely as $^{\ 6}_{\Lambda\!\Lambda}$He~\cite{Tak01} in 2001,
the $\Lambda \Lambda$ interaction is established to be rather
less attractive ($\Delta B_{\Lambda\!\Lambda } \approx 1 $ MeV).
Then, it is quite important to investigate what values of
$\Delta B_{\Lambda\!\Lambda}$ are obtained for ESC04 models.

Let us here evaluate the values of 
$\Delta B_{\Lambda\!\Lambda}(^{\ 6}_{\Lambda\!\Lambda}$He),
taking account of the $\Lambda \Lambda$-$\Xi N$ coupling effect
explicitly. For this purpose, we adopt the three-body model 
composed of the $\alpha\!+\!\Lambda\!+\!\Lambda$ and 
$\alpha\!+\!\Xi\!+\!N$ configurations.
The effective $\Lambda \Lambda$-$\Lambda \Lambda$ and
$\Lambda \Lambda$-$\Xi N$ interactions~\cite{Lans04} 
are derived in the G-matrix framework as follows:
We solve the $\Lambda \Lambda$-$\Xi N$-$\Sigma \Sigma$
coupled-channel G-matrix equation for a $\Lambda \Lambda$
pair in nuclear matter, and represent the resultant
$\Lambda \Lambda$-$\Lambda \Lambda$ and $\Lambda \Lambda$-$\Xi N$ 
G-matrices as local potentials in coordinate space.
These G-matrix interactions depend on the nucleon 
Fermi momentum $k_F$ of nuclear matter.  
Then, it is a problem what value of $k_F$ should be chosen 
in our calculation for $^{\ 6}_{\Lambda\!\Lambda}$He.
In similar calculations for $^5_\Lambda$He, the value of 
$k_F$ parameter included in the $\Lambda N$ G-matrix interaction
was chosen around 1.0 fm$^{-1}$~\cite{Yam85}. This value of 
$k_F \sim 1.0$ fm$^{-1}$ agree qualitatively with the value 
derived from the average nuclear density felt by the $\Lambda$
particle in $^5_\Lambda$He. Because a sophisticated estimation 
of the $k_F$ value is not necessary for our purpose of 
demonstrating features of the interaction models, we choose 
this plausible value of $k_F$= 1.0 fm$^{-1}$ in the present
calculations for $^{\ 6}_{\Lambda\!\Lambda}$He.

Using our $\Lambda \Lambda$-$\Lambda \Lambda$ and 
$\Lambda \Lambda$-$\Xi N$ G-matrix interactions, three-body 
variational calculations are performed in the Gaussian basis 
functions~\cite{Yama89}, where the $\Xi N$-$\Xi N$ interaction
is not taken into account for simplicity.
It should be noted that in our approach
high-lying $\Lambda \Lambda$-$\Xi N$-$\Sigma \Sigma$ correlations
are renormalized into the 
$\Lambda \Lambda$-$\Lambda \Lambda$ and $\Lambda \Lambda$-$\Xi N$
G-matrices, and low-lying $\Lambda \Lambda$-$\Xi N$ correlations
are treated in the model space composed of 
$\alpha\!+\!\Lambda\!+\!\Lambda$ and $\alpha\!+\!\Xi\!+\!N$ 
configurations. In order to avoid the double counting of the 
$\Lambda \Lambda$-$\Xi N$ coupling interaction, it is necessary 
that the high-lying $\Lambda \Lambda$-$\Xi N$ correlations are 
not included in our three-body model space. A practical way for
this problem is to take the two $\Lambda$ ($\Xi$ and $N$) 
coordinates from the center of mass of $\alpha$ core, not the 
relative coordinate between them explicitly, because the 
short-range correlations are taken into account unfavorably 
in this model space.

As for the interactions between the $\alpha$ cluster and valence 
particles ($\Lambda$, $\Xi$, $N$), we adopt the phenomenological
potentials: For $\alpha$-$\Lambda$ and $\alpha$-$\Xi$ interactions,
we use the two-range Gaussian potentials given in Ref.\cite{Lans04}. 
Here, the former is fitted so as to reproduce the $\Lambda$ binding
energies of $^4_\Lambda$H and $^4_\Lambda$He. The strength of the 
latter (named as Xa1~\cite{Lans04}) is determined in consideration
of the experimental indication that the $\Xi$ well depth is roughly
half of the $\Lambda$ one. On the other hand, we use the 
Kanada-Kaneko potential~\cite{Kanada} for the $\alpha$-$N$ 
interaction, designed so as to reproduce scattering phase shifts.
In the $\alpha\!+\!\Xi\!+\!N$ channel, we take into account
the orthogonality condition between $\alpha$ and $N$.

\begin{table}[ht]
\caption{
$\Delta B_{\Lambda\!\Lambda}(^{\ 6}_{\Lambda\!\Lambda}$He)
values (in MeV) are calculated with G-matrix interactions 
derived from ESC04a-d, NSC97 and NHC-D. 
(The hard-core radius in NHC-D is taken as 0.53 fm.)
}
\label{Gmat6}
\begin{center}
\begin{ruledtabular}
\begin{tabular}{lc|cc}
%\hline
       & & $\Delta B_{\Lambda\!\Lambda}$ (MeV) 
& $P_{\Xi N}$ (\%) \\
\hline
ESC04a  & & 1.36 & 0.44  \\
ESC04b  & & 1.37 & 0.45  \\
ESC04c  & & 0.97 & 1.15  \\
ESC04d  & & 0.98 & 1.18  \\
\hline
NSC97f  & & 0.34 & 0.19  \\
NHC-D   & & 1.05 & 0.14  \\
%\hline
\end{tabular}
\end{ruledtabular}
\end{center}
\end{table}

In Table~\ref{Gmat6} we show the calculated values of
$\Delta B_{\Lambda\!\Lambda}(^{\ 6}_{\Lambda\!\Lambda}$He)
and $\Xi N$ admixture probabilities $P_{\Xi N}$ in the cases 
of using ESC04a-d, NSC97f and NHC-D. 
(In the calculation for NHC-D, the hard-core radius 
in the $^{11}S_0$ state is taken as 0.53 fm, and
the $\Sigma \Sigma$ channel is not taken into account.)
The effect of the medium-induced repulsion is not so remarkable
in this case, because the $\Lambda \Lambda$ G-matrix is calculated 
at low density ($k_F=$ 1.0 fm$^{-1}$).
For instance, the calculated values for ESC04a$^*(\alpha=0.18)$
are $\Delta B_{\Lambda\!\Lambda}$=1.24 MeV and $P_{\Xi N}$=0.44 \%.
The calculated $\Delta B_{\Lambda\!\Lambda}$ values should be 
compared with the experimental value 
$1.01 \pm 0.20^{+0.18}_{-0.11}$ MeV~\cite{Tak01}.
Then, the calculated values for ESC04a-d are considered to
be more or less reasonable in the present scope of our
simple three-body model.

On the other hand, the value of $\Delta B_{\Lambda\!\Lambda}$ for 
NSC97f turns out to be rather too small compared with the 
experimental value: The $\Lambda \Lambda$ interaction of NSC97f 
is concluded to be too weak.
It is interesting that our result for NSC97f is quite similar to 
the Yamada's result~\cite{Yamada04}, obtained from the 
sophisticated variational calculation with direct use of NSC97f.
This means that our model-space approach with G-matrix effective
interactions simulates nicely the real space approach with 
free-space interactions. It was pointed out by Yamada that the 
$\Lambda \Lambda$-$\Xi N$-$\Sigma \Sigma$ coupling treatment
leads to the less $\Lambda \Lambda$ binding than the
$\Lambda \Lambda$-$\Xi N$ one due to the existence of
a pseudo bound state in the case of NSC97f. It should be noted 
that such a pseudo bound state does not appear in the case of 
ESC04a-d. In \cite{Vidana} the similar result was obtained for 
NSC97f by the G-matrix calculation. 
% update ------------------------------------------------------------
On the other hand, the importance of the rearrangement effect of the 
$\alpha$-core for $\Delta B_{\Lambda\Lambda}(^6_{\Lambda\Lambda}$He) 
has been pointed out by \cite{Kohno03,Usmani04,Nemu05}. It is an 
open problem to study core-rearrangement effects on the basis of the
ESC04 models.
% update ------------------------------------------------------------

The most striking feature of ESC04a-d is the far stronger 
$\Lambda \Lambda$-$\Xi N$ coupling than NSC97f and NHC-D. 
In Table~\ref{Gmat6}, this feature is seen in larger value of 
$P_{\Xi N}$ in the case of ESC04a-d. In particular, it is 
very curious that the $\Lambda \Lambda$-$\Xi N$ couplings of 
ESC04c/d are extremely strong. As shown in Ref.~\cite{Lans04}, 
such a coupling effect appears dramatically in
$^{\ 5}_{\Lambda\!\Lambda}$H and $^{\ 5}_{\Lambda\!\Lambda}$He
because of the small energy differences between ground 
$\Lambda$-$\Lambda$ states and $\Xi$-$\alpha$ states. 
A comprehensive study on the $\Lambda \Lambda$-$\Xi N$ 
coupling is now in progress on the basis of ESC04a-d.
%%%%%%%%%%%%%%%%%%%%%%%%%%%

\subsection{Properties of $\Xi N$ G-matrix}

There is no $\Xi N$ scattering data at present.
We have only uncertain information on $\Xi$-nucleus interactions
experimentally. We consider that the most reliable data
in the present stage was given by the BNL-E885 experiment~\cite{E885},
in which they measured the missing mass spectra for 
the $^{12}$C$(K^-, K^+)$X reaction.  Reasonable agreement between 
this data and theory is realized by assuming a $\Xi$-nucleus potential
$U_\Xi (\rho)=-V_0 f(r)$ with well depth $V_0 \sim 14$ MeV within 
the Wood-Saxon (WS) prescription.
%Another information was obtained from events of simultaneous emissions
%of two $\Lambda$ hypernuclei (twin $\Lambda$ hypernuclei) in the 
%KEK-E176 experiment.~\cite{Aoki93}  The obtained values of binding 
%energies $B_{\Xi^-}$ suggest the attractive $\Xi$-nucleus potential.

Let us here derive the potential energies $U_\Xi$ 
using the G-matrix theory in the same way as the cases of
$U_\Lambda$ and $U_\Sigma$. In the past, NHC-D gave rise to 
attractive values of $U_\Xi$, while strongly repulsive values were 
obtained for the other Nijmegen models. Then,
it is very curious what values of $U_\Xi$ are obtained for ESC04a-d.

%In the same way as the $\Sigma N$ case,
%we solve the $\Xi N$ starting channel G-matrix equation
%in the QTQ prescription, for simplicity, in which
%there appears no imaginary part due to an energy-conserving 
%$\Xi N$-$\Lambda \Lambda$ transition.
% update ----------------------------------------------------------
In the same way as in the $\Sigma N$ case, we solve the $\Xi N$ starting 
channel G-matrix equation in the QTQ prescription. Likewise, 
the $\Xi$ conversion width $\Gamma_\Xi$, due to an energy-conserving 
$\Xi N$-$\Lambda \Lambda$ transition, is not calculated here.
% update ----------------------------------------------------------
In this case, the channel-coupling treatments are performed
for $\Lambda \Lambda$-$\Xi N$-$\Sigma \Sigma$ and
$\Xi N$-$\Lambda \Sigma$-$\Sigma \Sigma$ channels.

\begin{table}[ht]
\caption{Values of $U_\Xi$ at normal density and partial 
wave contributions for ESC04a-d and ESC04d$^*(\alpha_V=0.18)$.
For comparison, the result for NHC-D is also shown
(hard-core radii are taken as 0.50 fm in all channels.). 
All entries are in MeV.}
\label{Gmat7}
\begin{center}
\begin{ruledtabular}
 \begin{tabular}{l|cc|rrrrrrc|r}
%\hline
      & $T$ & & $^1S_0$ & $^3S_1$ & $^1P_1$ & $^3P_0$ & $^3P_1$ & $^3P_2$ 
      & & $U_\Xi$ \hspace{1.5mm}  \\
 \hline
ESC04a& 0 & &   8.1 & --10.0 &   1.0 & --0.3 & --0.4 & --0.7 & & \\
      & 1 & & --4.5 &   21.8 & --0.7 &   0.7 & --0.1 &   0.3 & &  +15.1 \hspace{0mm}  \\
 \hline
ESC04b& 0 & &   5.9 & --2.4  &   0.7 &   0.7 &   1.0 & --0.4 & & \\
      & 1 & &   0.5 &   27.9 &   0.6 &   0.9 & --0.3 &   1.2 & &  +36.3 \hspace{0mm}  \\
 \hline
ESC04c& 0 & &   5.9 & --15.7 &   1.2 & --0.1 & --1.8 & --1.2 & & \\
      & 1 & &   6.8 &   1.9  & --0.8 &   0.1 & --0.3 & --1.7 & &  --5.5 \hspace{0mm}  \\
 \hline
ESC04d& 0 & &   6.4 & --19.6 &   1.1 &   1.2 & --1.3 & --2.0 & & \\
      & 1 & &   6.4 & --5.0  & --1.0 & --0.6 & --1.4 & --2.8 & & --18.7 \hspace{0mm}  \\
 \hline
ESC04d$^*$& 0 & & 6.3 & --18.4 &   1.2 &   1.5 & --1.3 & --1.9 & & \\
          & 1 & & 7.2 & --1.7  & --0.8 & --0.5 & --1.2 & --2.5 & & --12.1 \hspace{0mm} \\
 \hline
 \hline
NHC-D  & 0 & & --4.5 &   2.6 & --1.8 & --0.2 & --0.6 & --1.7 & & \\
       & 1 & &   0.2 &   5.3 & --2.6 &   0.0 & --2.9 & --5.6 & & --11.9 \hspace{0mm} \\
%\hline
 \end{tabular}
\end{ruledtabular}
\end{center}
\end{table}

In Table~\ref{Gmat7} we show the calculated values of $U_\Xi$
at normal density and their partial-wave contributions for 
ESC04a-d and ESC04d$^*(\alpha_V=0.18)$.
For comparison, the result for NHC-D is also given, where
the hard-core radii are taken as 0.50 fm in all channels,
and the $\Sigma \Sigma$ and $\Lambda \Sigma$ channels are
not taken into account.
Now, the remarkable difference among ESC04a-d is revealed: 
These four versions turn out to give rise to completely 
different values of $U_\Xi$.
It should be noted that the ESC models such as ESC04c/d can 
bring about attractive $\Xi$-nucleus potentials predicting
the existence of $\Xi$ hypernuclei.
It is very interesting that ESC04d$^*$ including the 
medium-induced repulsion leads to the $\Xi$ well depth
similar to the above ``experimental'' value.
%This suggests that our $\alpha_V$ parameter specifying the 
%medium-induced repulsion can be used for fitting 
%$B_\Xi$ values of observed $\Xi$ hypernuclei.
Though the attractive value of $U_\Xi$ is obtained also
in the case of NHC-D, its partial-wave contribution is
completely different from those in the case of ESC04c/d.
In the former case, the attractive $U_\Xi$ is owing to
the strong $P$-state attraction. In the latter case,
however, the strong attraction in the $^{13}S_1$ state
plays an essential role for it.
Because of this reason, various $\Xi$ hypernuclear states
will be predicted even in light $s$-shell systems on the 
basis of ESC04c/d. Level structures of these $\Xi$ states 
have to reflect the peculiar spin- and isospin dependences 
of the underlying $\Xi N$ interactions. 
The detailed analysis will be given in our next paper.

%-----------------------------------------------------------------
% NEW G-MATRIX END   
%-----------------------------------------------------------------

\section{Discussion and Conclusions }
We have shown in this paper that the ESC-approach to the nuclear force 
problem is able to make a connection between on the one hand the at 
present available baryon-baryon data and on the other hand the underlying
quark structure of the baryons and mesons. Namely, a successfull description
of both the {\it N\!N}- and {\it Y\!N}-scattering data is obtained with meson-baryon
coupling parameters which are almost all explained by the QPC-model.
This at the same time in obediance of the strong constraint of no bound states 
in the $S=-1$-systems.
Therefore, 
the ESC04 models of this paper are an important step in the determination of    
the baryon-baryon interactions for low energy scattering and the description
of hypernuclei in the context of broken SU(3)-symmetry.
The values for many parameters, which in previous work were considered to free to 
large extend, are now limited strongly, and tried to be made consistent with the present 
theoretical view on low energy hadron physics. 
This is in particularly the case for the $F/(F+D)$-ratios of the MPE-interactions.
These ratio's for the vector- and scalar-mesons are rather close to the QPC-model
predictions. 
% update -----------------------------------------------------------------------
This holds also for the values of the coupling constants. Here, the introduction 
of a zero in the form factor is important, leading to a sizeable reduction in
the scalar couplings. It is interesting that the features
of $\sigma$-exchange with a zero in the form factor are very similar to those 
obtained in a chiral-unitary-approach \cite{Ose00}.
% update -----------------------------------------------------------------------

The application of the $Q\bar{Q}$-pair creation to baryon-meson couplings using
a $^3S_1$-model \cite{Zha84} for pseudoscalar and vector-meson couplings, and
the nucleon-nucleon interactions has first been attempted by Fujiwara and Hecht 
\cite{Fuj92}. We did not explore this possibility, but it is not unlikely that this
alternative leads to a similar scheme of couplings as the $^3P_0$-model.
% update -----------------------------------------------------------------------

%The values fitted  for the magnetic ratio of the vector mesons is compatible 
%with the estimates from static and non-static SU$(6)$ \cite{Sak65}.
%On the other hand, the fitted value for the $F/(F+D)$-ratio
% $\alpha_{PV}=0.47$ for the pseudoscalar mesons deviates somewhat from the value
%that was favored in the sixties. 
%At the same time, it is not very far from that
%as found in the weak interactions, see e.g. \cite{Nag79b}.

%As compared to the soft-core OBE-models NSC89 \cite{MRS89} and NSC97 \cite{RSY99},
%the ESC04-model G-matrix calculations give worse results for the well depths 
%$U_\Lambda$ and $U_\Sigma$. 
The G-matrix results show that basic features of hypernuclear data are reproduced
nicely by ESC04, improving some weak points of the soft-core OBE-models NSC89 \cite{MRS89}
and NSC97 \cite{RSY99}. In spite of this superiority of ESC04 for hypernuclear data,
perhaps not every aspect of the effective (two-body) 
interactions in hypernuclei can be described
by this model. For example, this could be the case for the well depth $U_\Sigma$.
From the results it is clear that a good fit to the scattering data not necessarily
means success in the G-matrix results. To explain this one can think of two reasons:     
(i) the G-matrix results are sensitive to the two-body interactions below 1 fm, 
whereas the present {\it YN}-scattering data are not, (ii) other than two-body forces play an
important role. 
However, since the ${\it NN}\oplus {\it YN}$-fit is so much superior for ESC04- than for OBE-models, 
we are inclined to look for solutions to the mentioned problems 
outside the two-body forces. 
A natural possibility is the presence of three-body forces (3BF) in hypernuclei
which can be viewed as generating effective two-body forces, 
which could solve the well-depth issues. 
In the case of the $\Delta B_{\Lambda\Lambda}$ also 3BF could be operating. 
This calls for an evaluation of the 3BF's $NNN$, $\Lambda NN$, $\Sigma NN$, 
$\Lambda\Lambda N$, etc. for the soft-core ESC-model, consistent 
with its two-body forces.

The $\Lambda N$ p-waves seem to be better, which is the result of the truly 
simultaneous $NN+{\it YN}$-fitting. This is also reflected in the better $K_\Lambda$-value,
making the well-known small spin-orbit splitting smaller. 

% update -----------------------------------------------------------------
Finally, we mention the extensive work on baryon-baryon using the resonating-group-method
(RGM), exploiting quark-gluon-exchange (QGE) in conjunction with OBE, taking full
account of the antisymmetrization of the six quarks in the two-baryon systems \cite{Fuj02}.
A remarkable difference with the ESC-models is that QGE leads to strong repulsion
in the $\Sigma N(^3S_1,I=3/2)$- and the $\Sigma N(^1S_0, I=1/2)$-channels.
 In contrast, in this paper we have assumed that QGE is very suppressed dynamically.
% update -----------------------------------------------------------------
 
\acknowledgments
 We wish to thank prof. T. Motoba  for many stimulating discussions, 
 and prof. D.E. Lanskoy and dr. A. Dieperinck for useful comments.
 Th.A.R. is very gratefull for the generous hospitality 
 extended to him in the fall of 2004 at the 
 Osaka-EC University, where part of the final work on this paper was done.
Y.Y. thanks IMAPP of the Radboud University Nijmegen for its hospitality.
He is supported by the Japanese Grant-in-Aid for Scientific Research Fund of
Education, Culture, Sports, Science and Technology (No. 16540261).

% \onecolumngrid   ! change i.v.m. equation* etc.

%\begin{references}


\begin{thebibliography}{99}
\bibitem{Rij04a} Th.A.\ Rijken, {\it Extended-soft-core Baryon-Baryon Model,     
    I. Nucleon-Nucleon Scattering }, submitted to Phys. Rev. C.
\bibitem{Rij04c} Th.A.\ Rijken and Y.\ Yamamoto,
 {\it Extended-soft-core Baryon-Baryon Model,     
    III. $S=-2: \Lambda\Lambda, \Xi N$ etc. Scattering }, in preparation.
\bibitem{MRS89} P.M.M.\ Maessen, Th.A.\ Rijken, and J.J.\ de Swart,
         Phys.\ Rev.\ C {\bf 40} (1989) 2226.
\bibitem{RSY99} Th.A.\ Rijken, V.G.J.\ Stoks, and Y.\ Yamamoto, 
         Phys.\ Rev.\ C {\bf 59}, 21, (1999).
\bibitem{Rij85} T.A.\ Rijken, Ann.\ Phys.\ (N.Y.), {\bf 164} (1985) 1,23.
\bibitem{Rij00} T.A.\ Rijken, in {\it Few-Body Problems in Physics `99}, 
Proceedings of the 1st Asian-Pacific Conference, August 23-28, Tokyo,  
p. 317, Editors S.\ Oryu, M.\ Kamimura, and S.\ Ishikawa, Springer-Verlag (2000).
\bibitem{Rij01} T.A.\ Rijken, 
%in  Proceedings of the VII International Conference  
%on Hypernuclear and Strange Particle Physics (HYP2000), Torino, 
Nucl.\ Phys.\ {\bf A\ 691} (2001) 322c-328c.
\bibitem{Mic69} L.\ Micu, Nucl.\ Phys.\ {\bf B10} (1969) 521;          
                R.\ Carlitz and M.\ Kislinger, 
                Phys.\ Rev.\ D {\bf 2} (1970) 336.
\bibitem{Yam90} Y.\ Yamamoto and H.\ Band\={o},
         Progr.\ Theor.\ Phys.\ 83 (1990) 254.
\bibitem{Mot95} T.\ Motoba and Y.\ Yamamoto, Nucl. Phys. A 585 (1995) 29c.
\bibitem{Has95} T.\ Hasegawa {\it et al.}, Phys.\ Rev.\ Lett.\ 74 (1995) 224.
\bibitem{Dal64} R.H.\ Dalitz and F.\ von Hippel,
         Phys.\ Lett.\ {\bf 10}, 153 (1964).
\bibitem{Nag73} M.M.\ Nagels, T.A.\ Rijken, and J.J.\ de Swart,
         Ann.\ Phys.\ (N.Y.) {\bf 79}, 338 (1973).
\bibitem{Swa63} J.J.\ de Swart, Rev.\ Mod.\ Phys.\ {\bf 35} (1963) 916;
         {\it ibid.} {\bf 37} (1965) 326(E).
\bibitem{Car66} P.A.\ Carruthers, 'Introduction to unitary symmetry',      
         John Wiley \& Sons Inc., New York , 1966.
\bibitem{Mar69} R.E.\ Marshak, Riazuddin, and C.P.\ Ryan,                
         'Theory of weak interactions in particle physics',
         John Wiley \& Sons Inc., New York , 1969.
\bibitem{LeY73} A.\ Le\ Yaouanc, L.\ Oliver, O.\ P\'{e}ne, and J.-C.
                Raynal, Phys.\ Rev.\ D {\bf 8} (1973) 2223;
                Phys.\ Rev.\ D {\bf 11} (1975) 1272.
\bibitem{Cha80} M.\ Chaichain and R.\ K\"{o}gerler,
                Ann.\ Phys.\ (N.Y.)\ {\bf 124} (1980) 61.
\bibitem{NRS78} M.M.\ Nagels, T.A.\ Rijken, and J.J.\ de Swart,
         Phys.\ Rev.\ D {\bf 17} (1978) 768.
\bibitem{Bry72} R.A.\ Bryan and A.\ Gersten, Phys.\ Rev.\ D {\bf 6} (1972) 341.
\bibitem{Sto93}
 V.G.J.\ Stoks, R.A.M.\ Klomp, M.C.M.\ Rentmeester, and J.J.\ de Swart,
 Phys.\ Rev.\ C 48 (1993) 792.
\bibitem{Klo93} R.A.M.\ Klomp, private communication (unpublished).
%\bibitem{SKTS94}
% V.G.J.\ Stoks, R.A.M.\ Klomp, M.C.M.\ Rentmeester, C.P.F.\ Terheggen,
% and J.J.\ de Swart, Phys.\ Rev.\ C 49 (1994) 2950.

\bibitem{Kanda05}
 J.K.\ Ahn\ et\ al, Nucl.\ Phys.\ {\bf A761} (2005) 41.
 H.\ Kanda, Tohoku University Sendai, KEK-experiment E289, 
 private communication 2005;          
 J.K.\ Ahn\ et\ al, Nucl.\ Phys.\ {\bf A761} (2005) 41.

\bibitem{RS96ab} Th.A.\ Rijken and V.G.J.\ Stoks,
                Phys.\ Rev.\ C 54 (1996) 2869; 
                ibid.\ C 54 (1996) 2869.
\bibitem{SR97} V.G.J.\ Stoks and Th.A.\ Rijken,
                Nucl.\ Phys.\ {\bf A 613} (1997) 311. 

\bibitem{Ale68} G.\ Alexander, U.\ Karshon, A.\ Shapira,
         G.\ Yekutieli, R.\ Engelmann, H.\ Filthuth, and W.\ Lughofer,
         Phys.\ Rev.\ {\bf 173}, 1452 (1968).
\bibitem{Sec68} B.\ Sechi-Zorn, B.\ Kehoe, J.\ Twitty, and
         R.A.\ Burnstein, Phys.\ Rev.\ {\bf 175}, 1735 (1968).
\bibitem{Eis71} F.\ Eisele, H.\ Filthuth, W.\ F\"olisch, V.\ Hepp,
         E.\ Leitner, and G.\ Zech,
         Phys.\ Lett.\ {\bf 37B}, 204 (1971).
\bibitem{Eng66} R.\ Engelmann, H.\ Filthuth, V.\ Hepp, and E.\ Kluge,
         Phys.\ Lett.\ {\bf 21}, 587 (1966).
\bibitem{Hep68} V.\ Hepp and M.\ Schleich,
         Z.\ Phys.\ {\bf 214}, 71 (1968).
\bibitem{Ste70} D.\ Stephen, Ph.D.\ thesis, University of Massachusetts, 1970.
         Z.\ Phys.\ {\bf 214}, 71 (1968).

%\bibitem{Kad71} J.A.\ Kadyk, G.\ Alexander, J.H.\ Chan,
%         P.\ Gaposchkin, and G.H.\ Trilling,
%         Nucl.\ Phys.\ {\bf B27}, 13 (1971).
%\bibitem{Hau77} J.M.\ Hauptman, J.A.\ Kadyk, and G.H.\ Trilling,
%         Nucl.\ Phys.\ {\bf B125}, 29 (1977).

% update ----------------------------------------------------------
\bibitem{Swa62} J.J.\ de Swart and C.\ Dullemond, Ann. of Phys.\ {\bf 19} (1962) 458. 
\bibitem{Swa71} J.J.\ de Swart, M.M.\ Nagels, T.A.\ Rijken, and P.A.\ Verhoeven, 
 Springer Tracts in Modern Physics, {\bf 60} (1971) 138.
\bibitem{Fuj98}   
Y.\ Fujiwara, T.\ Fujita, C.\ Nakamoto, and Y.\ Suzuki,  
Progr.\ Theor.\ Phys.\ {\bf 100} (1998) 957.          

\bibitem{ynonline} ESC04 {\it YN}-potentials, see http://nn-online.org.                        
% update ----------------------------------------------------------

\bibitem{Sak65} B.\ Sakita and K.C.\ Wali, 
         Phys.\ Rev.\ {\bf 139} (1965) B1355.
%-----------------------------------------------------------------------------

%---------------------------------------------------------------------
%NEW BEGIN   YAMAMOTO references.
%\bibitem{NSC97} Th.A.\ Rijken, V.G.J.\ Stoks, and Y.\ Yamamoto, 
%Phys.\ Rev.\ C 59 (1999) 21.

\bibitem{Hiya01}
E.\ Hiyama, M.\ Kamimura, T.\ Motoba, T.\ Yamada and Y.\ Yamamoto,
Phys.\ Rev.\ C 65 (2001) 011301(R).

\bibitem{Nogga02}
A.\ Nogga, H.\ Kamada, and W.\ Gl\"{o}ckle,
Phys.\ Rev.\ Lett.\ 88 (2002) 172501.

\bibitem{Nemu02}
H.\ Nemura, Y.\ Akaishi, and Y.\ Suzuki,
Phys.\ Rev.\ Lett.\  89 (2002) 142504.

\bibitem{Mil01}
D.J.\ Millener,
Nucl.\ Phys.\ A691 (2001) 93c.

%\bibitem{Yam94} Y.\ Yamamoto, T.\ Motoba, H.\ Himeno, K.\ Ikeda, and  
%                  S.\ Nagata, Progr.\ Theor.\ Phys.\ Suppl.\ 117 (1994) 361.
\bibitem{Sch76} R.R.\ Scheerbaum,
         Nucl.\ Phys.\ A257 (1976) 77.

\bibitem{Hiya00}
E.\ Hiyama, M.\ Kamimura, T.\ Motoba, T.\ Yamada, and Y.\ Yamamoto,
Phys.\ Rev.\ Lett.\  85 (2000) 270.

% update --------------------------------------------------------------
\bibitem{Fuj04} 
Y.\ Fujiwara, M.\ Kohno, K.\ Miyagawa, and Y.\ Suzuki,
Phys.\ Rev.\ {\bf C70} (2004) 047002.
% update --------------------------------------------------------------

\bibitem{Yam00}
Y.\ Yamamoto, S.\ Nishizaki, and T.\ Takatsuka, 
Prog.\ Theor.\ Phys.\ 103 (2000) 981.

\bibitem{BFG94}
C.J.\ Batty, E.\ Friedman, and A.\ Gal,
Prog.\ Theor.\ Phys.\ Suppl.\ No.117 (1994) 227.

\bibitem{Dab99} J.\ Dabrowski, 
Phys.\ Rev.\ C 60 (1999) 025205.

\bibitem{Noumi}  H.\ Noumi {\it et al.},
Phys.\ Rev.\ Lett.\ 89 (2002) 072301.

\bibitem{Kohno04} 
M.\ Kohno, Y.\ Fujiwara, Y.\ Watanabe, K.\ Ogata, and M.\ Kawai,
Prog.\ Theor.\ Phys.\ 112 (2004) 895.

\bibitem{Jack83} 
A.D.\ Jackson, Ann.\ Rev.\ Nucl.\ Part.\ Sci.\ {\bf 33} 105 (1983);
A.D.\ Jackson, E.\ Krotcheck, and M.\ Rho, Nucl.\ Phys.\ A407 (1983) 495.

\bibitem{LaPa81}
I.E.\ Lagaris and V.R.\ Pandharipande,
Nucl.\ Phys.\ A359 (1981) 349.

\bibitem{Baldo}
M.\ Baldo, A.\ Fiasconaro, H.Q.\ Song, G.\ Giansiracusa, and U.\ Lombardo,
Phys.\ Rev.\ C 65 (2002) 017303.

\bibitem{TNY02}
T.\ Takatsuka, S.\ Nishizaki, and Y.\ Yamamoto, Proc.\ Int.\ Symp.\ on\
Perspectives\ in\ Physics\ with\ Radiactive\ Isotopes,
Hayama, Kanagawa, Japan, Nov. 13-16, 2000 ; Eur.\ Phys.\ J. A. 13 (2002) 213.

\bibitem{NYT01}
S.\ Nishizaki, and Y.\ Yamamoto, T.\ Takatsuka, Prog.\ Theor.\ Phys.\ 105 (2001) 607; 
ibid\ 108 (2002) 703.

\bibitem{FM} 
J.\ Fujita and H.\ Miyazawa,
Prog.\ Theor.\ Phys.\ 17 (1957) 360; 17 (1957) 366.

\bibitem{NRS77}  M.M.\ Nagels, T.A.\ Rijken, and J.J.\ deSwart, 
Phys.\ Rev.\ D 15 (1977) 2547.

\bibitem{Tak01}  H.\ Takahashi {\it et al.}, 
Phys.\ Rev.\ Lett.\ 87 (2001) 212502.

\bibitem{Lans04} 
D.E.\ Lanskoy and Y.\ Yamamoto, 
Phys.\ Rev.\ C 69 (2004) 014303.

\bibitem{Yam85} 
Y.\ Yamamoto and H.\ Band\=o,
Prog.\ Theor.\ Phys.\ Suppl.\ No.81 (1985) 9.

\bibitem{Yama89} 
Y.\ Yamamoto, H.\ Takaki, and K.\ Ikeda,
Prog.\ Theor.\ Phys.\ 82 (1989) 13.

\bibitem{Kanada}
H.\ Kanada, T.\ Kaneko, S.\ Nagata, and M.\ Nomoto,
Prog.\ Theor.\ Phys.\ 61 (1979) 1327.

\bibitem{Yamada04} T.\ Yamada, 
Phys.\ Rev.\ C69 (2004) 044301.

\bibitem{Vidana} 
I.\ Vida\~{n}a, A.\ Ramos, and A.\ Polls,
Phys.\ Rev.\ C70 (2004) 024306.

% update ---------------------------------------------------
\bibitem{Kohno03} 
M.\ Kohno, Y.\ Fujiwara, and Y.\ Akaishi, 
Phys.\ Rev.\ C68 (2003) 034302.

\bibitem{Usmani04} 
Q.N.\ Usmani, A.R.\ Bodmer, and Bhupali Sharma,
Phys.\ Rev.\ C70 (2004) 061001(R).

\bibitem{Nemu05}
H.\ Nemura, S.\ Shinmura, Y.\ Akaishi, and Khin Swe Myint,
Phys.\ Rev.\ Lett.\  94 (2005) 202502.

% update ---------------------------------------------------

\bibitem{E885}  
P.\ Khaustov {\it et al.}, Phys.\ Rev.\ C 61 (2000) 054603.

%NEW END   YAMAMOTO references.

% update -----------------------------------------------------------
\bibitem{Ose00}   
E.\ Oset, H.\ Toki, M.\ Mizobe, and T.T.\ Takahashi,                 
Progr.\ Theor.\ Phys.\ {\bf 103} (2000) 351.          
\bibitem{Zha84}   
Y.W.\ Yu and Z.Y.\ Zhang, Nucl.\ Phys., A426 (1984) 557.       
\bibitem{Fuj92}   
Y.\ Fujiwara,   
Progr.\ Theor.\ Phys.\ {\bf 88} (1992) 933, and the references cited here.

\bibitem{Fuj02}   
Y.\ Fujiwara, T.\ Fujita, M.\ Kohno, C.\ Nakamoto, and Y.\ Suzuki,  
Phys.\ Rev.\ C 65 (2002) 014002, and cited references.
%---------------------------------------------------------------------

\end{thebibliography}
\end{document}